\RequirePackage{ifpdf}
\documentclass[hyper,letterpaper]{JHEP3}
\usepackage{amssymb,amsfonts,bm,amsmath,empheq}
\usepackage{cite}
\usepackage{graphicx}
\usepackage{multirow}
\usepackage{verbatim}
\usepackage{appendix}

\usepackage{url}
\usepackage{float}
\usepackage[T1]{fontenc}

\newcommand{\bea}{\begin{eqnarray}}
\newcommand{\eea}{\end{eqnarray}}
\newcommand{\be}{\begin{equation}}
\newcommand{\ee}{\end{equation}}

\def\IN{\mathbb {N}}
\def\IZ{\mathbb {Z}}

\def\IR{\mathbb {R}}
\def\IC{\mathbb {C}}
\def\IP{\mathbb {P}}

\newcommand{\D}{{\frak D}}

\renewcommand{\hat}{\widehat}

\title{Quantum curves and conformal field theory}

\author{Masahide Manabe$^{1}$ and Piotr Su{\l}kowski$^{1,2}$
\\
$^1$ Faculty of Physics, University of Warsaw, ul. Pasteura 5, 02-093 Warsaw, Poland  \\
$^2$ Walter Burke Institute for Theoretical Physics, California Institute of Technology, Pasadena, CA 91125, USA}

\abstract{To a given algebraic curve we assign an infinite family of quantum curves (Schr{\"o}dinger equations), which are in one-to-one correspondence with, and have the structure of, Virasoro singular vectors. For a spectral curve of a matrix model we build such quantum curves out of an appropriate representation of the Virasoro algebra, encoded in the structure of the $\alpha/\beta$-deformed matrix integral and its loop equation. We generalize this construction to a large class of algebraic curves by means of a refined topological recursion. We also specialize this construction to various specific matrix models with polynomial and logarithmic potentials, and among other results, show that various ingredients familiar in the study of conformal field theory (Ward identities, correlation functions and a representation of Virasoro operators acting thereon, BPZ equations) arise upon specialization of our formalism to the multi-Penner matrix model. 
\\
\\
\\
\\
\\
\\
\\
\\
\\
{\tt CALT-2015-061}}

\begin{document}


\newpage

\section{Introduction}  \label{sec-intro}

In this paper we consider a quantization of plane algebraic curves. Such curves can be thought of as ``classical'' objects defined by a polynomial equation
\be  
A(x,y) = 0     \label{Axy}
\ee
in two complex variables $x$ and $y$. By a quantum curve associated to the above algebraic curve we understand an operator $\widehat{A}(\hat x, \hat y)$ that imposes a Schr{\"o}dinger equation of the form
\be
\widehat{A}(\hat x, \hat y)\psi(x) = 0,      \label{Ahatxy}
\ee
where $\hat x$ and $\hat y$ are operators that satisfy a commutation relation
\be
[\hat  y, \hat x] = g_s,
\ee
and in the $g_s\to 0$ limit $\widehat{A}(\hat x, \hat y)$ reduces to $A(x,y)$, or contains this polynomial as a factor. The solution $\psi(x)$ of the equation (\ref{Ahatxy}) is often referred to as a wave-function and it has specific interpretations in various actual realizations. 

Quantum curves arise in various contexts in modern mathematical physics: in matrix models as quantization of spectral curves \cite{abmodel,ACDKV}, in topological string theory as quantization of mirror curves \cite{ADKMV}, in systems of intersecting branes and in Seiberg-Witten theory  \cite{DHSV,DHS}, as quantum A-polynomials and their generalizations in knot theory and its physical realizations \cite{DijkgraafFuji-1,DijkgraafFuji-2}, in various enumerative problems related to moduli spaces of Riemann surfaces \cite{Norbury-quantum,Dumitrescu:2015mpa}. In all these cases it is usually claimed that a quantum curve can be uniquely assigned to a given classical curve. It has been postulated that for a given (classical) algebraic curve the corresponding wave-function, as well as the quantum curve, can be reconstructed in a universal way, by means of the topological recursion. Recall that the topological recursion was originally formulated in the context of matrix models as a redefinition of loop equations, and in this case the spectral curve of a matrix model plays a role of the initial condition for this recursion \cite{Eynard:2004mh,Chekhov:2005rr,Chekhov:2006rq}. The topological recursion was subsequently reformulated more abstractly, as a tool that assigns various new invariants to a large class of algebraic curves (not necessarily matrix model spectral curves) \cite{eyn-or}. It is this more general formulation that turned out to be surprisingly powerful in various physical and mathematical contexts mentioned above, whenever algebraic curves play an important role, and even if the corresponding matrix model formulation does not exist.

In this paper we show that to a given classical algebraic curve one can naturally assign not just one quantum curve, but an infinite family of quantum curves. These curves are in one-to-one correspondence with, and have the structure of, Virasoro singular vectors (also known as null vectors) in two-dimensional conformal field theory. Quantum curves that have been considered in literature so far, from our perspective correspond to Virasoro singular vectors at level 2. Other quantum curves that we identify in this work correspond to singular vectors at higher levels, and we refer to them as higher level quantum curves. 

In this paper we provide a construction of the above mentioned family of quantum curves in the formalism of $\beta$-deformed matrix models. Moreover, we postulate that such quantum curves exist in more general context and can be assigned to a large class of algebraic curves. This can be argued simply as follows. Our construction is based on the analysis of loop equations and the corresponding Virasoro constraints in a $\beta$-deformed matrix model. As mentioned above, the topological recursion, being also a manifestation of matrix model loop equations, can be formulated more generally and independently of the existence of a matrix model, and applied to a large class of algebraic curves. In the same spirit, more general wave-functions and corresponding higher level quantum curves that we consider in this paper can be associated to a more general class of algebraic curves and constructed by means of the topological recursion, presumably even if no corresponding matrix model is known. We provide an explicit construction, based on the topological recursion, of a large family of more general wave-functions and higher level quantum curves, and postulate that such construction exists for all higher level quantum curves.

Our results could be expressed also in some other language, and they should have applications to other systems related to matrix models by various dualities. In particular, $\beta$-deformed matrix models provide one formulation of the refined topological string theory, and our results could be equivalently pronounced in this context. Exploiting this link further, they can be related to surface operators in supersymmetric gauge theories, and then to all other systems connected via Alday-Gaiotto-Tachikawa (AGT) duality, Nekrasov-Shatashvili correspondence, etc. We briefly discuss those other systems and corresponding dualities in section \ref{sec-top-strings}. We leave unraveling the full role of higher level quantum curves in these systems for future work, and present the results in this paper primarily in the language of matrix models.

As has been already mentioned, quantum curves identified in this paper are in one-to-one correspondence with Virasoro singular vectors. This fact is a manifestation of the underlying conformal invariance of matrix models. Conformal invariance of matrix models was discovered around 1990, when it was found that loop equations of a hermitian matrix model can be rephrased as Virasoro constraints on its partition function \cite{Ambjorn:1990ji,David:1990ge,Mironov:1990im,Fukuma:1990jw,Itoyama:1991hz,Dijkgraaf:1990rs}. Subsequently it was shown that conformal invariance is preserved upon the $\beta$-deformation \cite{Kharchev:1992iv,Awata:1994xd}, and more recently it was realized that certain quantum curves can be written as the BPZ-like equation at level 2 \cite{ACDKV} (for related ideas see also \cite{GatoRivera:1992in,TakiMaruyoshi,Marshakov:2010fx}). Our results provide a generalization of all those statements to arbitrary singular vectors in Virasoro algebra, or equivalently to BPZ-like equations at arbitrary levels. In particular Virasoro constraints for the matrix model partition function found in \cite{Ambjorn:1990ji,Awata:1994xd}, as well as BPZ-like equations at level 2 identified in \cite{ACDKV}, correspond respectively to level 1 and level 2 quantum curves from our perspective. Apart from these two lowest levels, we identify and present explicitly the structure of Virasoro constraints and quantum curves at all higher levels.

As asserted above, we conduct our analysis in the formalism of matrix models. As the main ingredient in this analysis we consider some particular matrix integral, that we refer to as the $\alpha/\beta$-deformed matrix integral, or $\alpha/\beta$-deformed matrix model. We define it as the $\beta$-deformed expectation value of an $x$-dependent determinant-like expression, raised to a power parameterized by a parameter $\alpha$. Despite this simple definition we believe that this object, due to its special properties, deserves a special name. The crucial property of $\alpha/\beta$-deformed integrals is that they satisfy finite order differential equations in $x$ only for special, discrete values of parameters $\alpha$. For this reason we refer to these integrals as wave-functions, and identify differential operators that annihilate these wave-functions as quantum curves. One of the main aims of this paper is to show that these special values of parameters $\alpha$ are precisely the values of degenerate momenta in conformal field theory, and that in this case the corresponding differential operators have the structure of Virasoro singular vectors.


\subsection{Summary of main results -- quantum curves and their structure}

Let us present the above statements and main results of this paper in more detail. A crucial role in our analysis is played by the following expression
\be
\widehat{\psi}_{\alpha}(x) = \frac{e^{-\frac{2\alpha}{\epsilon_1 \epsilon_2}V(x)}}{(2\pi)^NN!}  \int_{{\IR}^N}   \Delta(z)^{2\beta} 
\, \Big( \prod_{a=1}^N(x-z_a)^{-\frac{2\alpha}{\epsilon_{2}}} \Big)\, 
e^{-\frac{\sqrt{\beta}}{\hbar}\sum_{a=1}^NV(z_a)} \prod_{a=1}^Ndz_a,    \label{Psi-alpha}
\ee 
which we refer to as an $\alpha/\beta$-deformed matrix integral. In this expression $(2\pi)^NN!$ is an overall normalization, $\Delta(z)=\prod_{1\le a<b\le N}(z_a-z_b)$ denotes the Vandermonde determinant, and 
\begin{equation}
\epsilon_1=-\beta^{1/2}g_s,\qquad \epsilon_2=\beta^{-1/2}g_s,\qquad g_s=2\hbar,\qquad b^2 = -\beta  = \frac{\epsilon_1}{\epsilon_2}.
\label{matparam-intro}
\end{equation}
Without the prefactor $e^{-\frac{2\alpha}{\epsilon_1 \epsilon_2}V(x)}$ and without the term in the bracket in the integrand (or simply for $\alpha=0$), (\ref{Psi-alpha}) reduces to the expression for the partition function $Z\equiv\widehat{\psi}_{\alpha=0}(x)$ of the standard $\beta$-deformed matrix model. For  $\beta=1,\frac{1}{2},2$ the above expression is an eigenvalue representation of an integral over hermitian, orthogonal, and symplectic matrices $M$ respectively, and in those cases the term in the bracket in the integrand is the eigenvalue representation of the determinant $\textrm{det}(x-M)$, raised to some power parametrized by a parameter $\alpha$. The exponential term in the integrand is the eigenvalue representation of  $e^{-\frac{\sqrt{\beta}}{\hbar}\textrm{Tr}V(M)}$, where $V=V(x)$ is called a potential, and we consider it to be of the form 
\be
V(x) = \sum_{n=0}^{\infty}  t_n x^n, \label{V}
\ee
with parameters $t_n$ referred to as times. Matrix integrals with $\beta=1$ are often called unrefined matrix models. For values of $\beta$ other than $1,\frac{1}{2}$ and 2 the corresponding matrix ensemble does not exist, however it still makes sense to consider integrals of the form (\ref{Psi-alpha}), customarily (and typically for $\alpha=0$) referred to as $\beta$-deformed matrix integrals.

While at first sight the expression (\ref{Psi-alpha}) might seem complicated, it is in fact very natural and arises as the expectation value of the exponent $e^{\frac{2\alpha}{g_s}\phi(x)}$ of the ($\beta$-deformed) matrix model realization of the chiral boson field $\phi(x)$, with the background charge $Q = i\big(b+\frac{1}{b}\big) = \frac{\epsilon_1+\epsilon_2}{g_s}$. While sometimes it is useful to consider other normalizations of (\ref{Psi-alpha}), in particular without the prefactor $e^{-\frac{2\alpha}{\epsilon_1 \epsilon_2}V(x)}$, one should keep in mind that this prefactor has the origin in the exponent of the chiral boson field. All our results, in particular the correspondence between quantum curves and Virasoro singular vectors, arise in consequence of the conformal invariance of such chiral boson theory in two dimensions and the structure of the associated Virasoro algebra, which in this case has central charge $c=1-6Q^2$. 

Having written down the expression (\ref{Psi-alpha}), we ask whether it satisfies a finite order differential equation equation in parameter $x$. We find a beautiful answer to this question: it turns out that this is so only for discrete values of $\alpha$, which are of the form
\be
\alpha = \alpha_{r,s} = -\frac{r-1}{2}\epsilon_1 - \frac{s-1}{2}\epsilon_2,  \qquad \textrm{for}\ r,s=1,2,3,\ldots  \label{alpha-rs}
\ee
Up to a normalization by $ig_s$ (which we find convenient not to include in the definition of $\alpha_{r,s}$) these values can be written as $(1-r)b+(1-s)b^{-1}$, which are immediately recognized as the degenerate momenta of the chiral boson in presence of the background charge. We find that for a particular value of $\alpha=\alpha_{r,s}$ the differential equation for $\widehat{\psi}_{\alpha}(x)$ has order $n=rs$, and we write it as
\be
\widehat{A}^{\alpha}_n \widehat{\psi}_{\alpha}(x) = 0.  \label{Aalpha-psialpha-intro}
\ee
We refer to such equations as higher level quantum curve or Schr{\"o}dinger equations, and we call operators $\widehat{A}^{\alpha}_n$ as (higher level) quantum curves, and  $\widehat{\psi}_{\alpha}(x)$ defined in (\ref{Psi-alpha}) as wave-functions. Furthermore, we show that these quantum curves can be written as
\be
\widehat{A}^{\alpha}_n = \sum_{p_1+p_2+\ldots+p_k=n}  \widehat{c}_{p_1,p_2,\ldots,p_k}(\alpha)\,   \widehat{L}_{-p_1} \widehat{L}_{-p_2}\cdots \widehat{L}_{-p_k}.       \label{chat-Lhat-intro}
\ee
This expression has the same structure as an operator that acting on a primary state creates a Virasoro singular vector corresponding to a given value $\alpha_{r,s}$. In particular $\widehat{c}_{p_1,p_2,\ldots,p_k}(\alpha)$ are appropriate constants that appear in expressions for such singular vectors, while in our case operators $\widehat{L}_{-p}$ with $p\geq 0$ form a representation of the  Virasoro algebra on a space of functions in $x$ and times $t_k$. We find that this representation takes form
\be
\begin{split}
\widehat{L}_0\,  &=\Delta_{\alpha}\equiv \frac{\alpha}{g_s}\Big(\frac{\alpha}{g_s} - Q \Big) ,  \qquad  \qquad   \widehat{L}_{-1} =\partial_x ,  \label{Lm_fer_op-intro}  \\ 
\, \widehat{L}_{-n} = - \frac{1}{\epsilon_1\epsilon_2 (n-2)!} \Big( &  \partial_x^{n-2} \big(  V'(x)^2  \big)  + (\epsilon_1+\epsilon_2) \partial_x^n V(x)  + \partial_x ^{n-2}\hat{f}(x) \Big),\  \textrm{for}\ n\geq 2, \,
\end{split}
\ee
where $\partial_x^n\widehat{f}(x) \equiv [\partial_x, \partial_x^{n-1}\widehat{f}(x)]$ and
\begin{equation}
\hat{f}(x)=-\epsilon_1\epsilon_2\sum_{m=0}^{\infty}x^m\partial_{(m)},\ \ \ \ 
\partial_{(m)}=\sum_{k=m+2}^{\infty}kt_k\frac{\partial}{\partial t_{k-m-2}}.
\label{sp_hat_f_op-intro}
\end{equation}
We find the representation (\ref{Lm_fer_op-intro}) by interpreting (\ref{Psi-alpha}) in conformal field theory terms, as an insertion of an operator $\prod_a (x-z_a)^{-\frac{2\alpha}{\epsilon_{2}}}$ at a position $x$, and determining modes of the corresponding energy-momentum tensor. Note that $\widehat{L}_{-n}$ involves derivatives with respect to times, which are encoded in the term $\partial_x ^{n-2}\hat{f}(x)$. For this reason, in general, operators (\ref{chat-Lhat-intro}) are time-dependent quantum curves, and they impose partial differential equations in $x$ and times $t_k$. However, as we discuss in what follows, in certain specific situations quantum curves become time-independent and impose ordinary differential equations in $x$ for $\widehat{\psi}_{\alpha}(x)$.

As this is one of the main results of this paper, let us comment a little more on the structure of $\widehat{A}^{\alpha}_n$. Recall that, in an abstract formulation, singular vectors of Virasoro algebra $|\phi_{r,s} \rangle$ are parametrized by two positive integers $r$ and $s$ (that in the free boson realization label the momenta (\ref{alpha-rs})), and can be written as $|\phi_{r,s} \rangle = A_{r,s} | \Delta_{r,s}\rangle$, where $|\Delta_{r,s} \rangle$ is a primary state of appropriate weight $\Delta_{r,s}$, and $A_{r,s}$ is an operator that is also of the form (\ref{chat-Lhat-intro}), however with $\widehat{L}_{-p}$ being abstract Virasoro generators. Determining an explicit form of coefficients $\widehat{c}_{p_1,p_2,\ldots,p_k}(\alpha)$ is an outstanding problem in conformal field theory, which has been solved only in some particular cases (for example in case $r=1$ or $s=1$ \cite{BenoitSaintAubin}). We show that coefficients $\widehat{c}_{p_1,p_2,\ldots,p_k}(\alpha)$ encoded in differential equations for $\widehat{\psi}_{\alpha}(x)$ indeed agree with the values expected for Virasoro singular vectors; moreover, in our approach we find an interesting formulas for these coefficients, which do not seem to have been known before. In principle, following the algorithm that we propose to determine differential equations for $\widehat{\psi}_{\alpha}(x)$ for arbitrary $\alpha=\alpha_{r,s}$, one can determine explicit form of coefficients $\widehat{c}_{p_1,p_2,\ldots,p_k}(\alpha)$ at arbitrary level. This can be regarded as a new way of determining Virasoro singular vector.  

Moreover, we find yet another representation $\ell_n^{\alpha}(x)$ of Virasoro operators associated to (\ref{Psi-alpha}), by considering the energy-momentum tensor of a generalized chiral boson field, modified in a way that represents an insertion of $\prod_a (x-z_a)^{-\frac{2\alpha}{\epsilon_{2}}}$. This second representation leads to Virasoro constraints for the integral (\ref{Psi-alpha}), which are analogous to Virasoro constraints for a matrix model partition function without the insertion of $\prod_a (x-z_a)^{-\frac{2\alpha}{\epsilon_{2}}}$ \cite{Ambjorn:1990ji,Fukuma:1990jw,Itoyama:1991hz,Dijkgraaf:1990rs,Awata:1994xd}. Explicit expression for operators $\ell_n^{\alpha}(x)$ is given in (\ref{twi_vira_g}); while it is quite lengthy, Virasoro constraints imposed by $\ell_n^{\alpha}(x)$ are very useful in the analysis of quantum curves and in particular, for some specific models, enable to turn time-dependent quantum curves into time-independent ones. 

Let us illustrate our results with the following examples. Demanding that $\widehat{\psi}_{\alpha}(x)$ satisfies a second order differential equation in $x$, we find quantum curves at level 2 of the form
\be
\widehat{A}^{\alpha}_2 \widehat{\psi}_{\alpha}(x) \equiv \Big(\widehat{L}_{-1}^2+\frac{4\alpha^2}{\epsilon_1\epsilon_2}\widehat{L}_{-2}\Big) \widehat{\psi}_{\alpha}(x)=0,\ \ \  
\textrm{for}\ \alpha=-\frac{\epsilon_1}{2}, -\frac{\epsilon_2}{2}.       \label{BPZ-level2-intro}
\ee
Substituting values $\alpha=\alpha_{2,1}=-\frac{\epsilon_1}{2}$ or $\alpha=\alpha_{1,2}=-\frac{\epsilon_2}{2}$, the differential operator in this equation takes form $\widehat{A}^{\alpha}_2=\widehat{L}_{-1}^2+b^{\pm 2} \widehat{L}_{-2}$ respectively, with $b^2$ defined in (\ref{matparam-intro}), which we indeed recognize as standard operators that acting on relevant primary states create singular vectors of Virasoro algebra at level 2. A variant of this calculation was first presented in \cite{ACDKV}. It is important to note that the form (\ref{BPZ-level2-intro}) is universal for both singular vectors at level 2, and specializes to an expression for a particular singular vector simply upon a choice of $\alpha$. Furthermore, invoking the representation (\ref{Lm_fer_op-intro}), quantum curves in (\ref{BPZ-level2-intro}) can be written explicitly as
\begin{equation}
\widehat{A}^{\alpha}_2 = \partial_x^2-\frac{4\alpha^2}{\epsilon_1^2\epsilon_2^2}V'(x)^2-\frac{4\alpha^2}{\epsilon_1^2\epsilon_2^2}(\epsilon_1+\epsilon_2)V''(x)
-\frac{4\alpha^2}{\epsilon_1^2\epsilon_2^2}\widehat{f}(x).\ \ \
\label{BPZ-level-2bis-intro}
\end{equation}
One can also rewrite (\ref{BPZ-level2-intro}) as equations for the normalized wave-function $\Psi_{\alpha}(x) = \widehat{\psi}_{\alpha}(x) / Z$, where $Z$ is identified as the value of (\ref{Psi-alpha}) for $\alpha=0$. Then, for $\beta=1$ and in the classical limit $g_s\to 0$, after fixing the potential and adjusting filling fractions, quantum curves $\widehat{A}^{\alpha}_2$ reduce to the classical equation
\be
y^2 - V'(x)^2 - f_{cl}(x) = 0,   \label{Ahat-class-limit-level2-intro}
\ee
where $y$ is identified with the classical limit of $g_s\partial_x$, and $f_{cl}(x)$ denotes the classical limit of expression that originates from the action of $\widehat{f}(x)$. This last equation coincides with the spectral curve of the matrix model, and for this reason the quantum curves $\widehat{A}^{\alpha}_2$ can be regarded as quantizations of this spectral curve. 

We determine quantum curves at higher levels analogously, by demanding that  $\widehat{\psi}_{\alpha}(x)$ satisfies a finite order differential equation in $x$. We stress that for a given level we obtain a universal formula for quantum curves, which specializes to an expression corresponding to a particular singular vector simply upon a choice of the relevant value of $\alpha$. Moreover, such a formula encodes expressions for singular vectors not only at a given level, but also at all lower levels. For example, we find that quantum curves at level 3 take form
\be
\widehat{A}_3^{\alpha}  = \widehat{L}_{-1}\widehat{A}_2^{\alpha}
+\frac{2\alpha^2}{\epsilon_1^2\epsilon_2^2}(2\alpha+\epsilon_1)(2\alpha+\epsilon_2)\widehat{L}_{-3},    \label{BPZ-level-3-intro} \\ 
\ee
and choosing $\alpha=\alpha_{3,1}=-\epsilon_1$ or $\alpha=\alpha_{1,3}=-\epsilon_2$ the above operator takes form that encodes singular vectors at level 3. Using the representation (\ref{Lm_fer_op-intro}) it is straightforward to write down explicit form of these operators.
Moreover, substituting values of $\alpha$ for level 2, i.e. $\alpha_{2,1}=-\frac{\epsilon_1}{2}$ or $\alpha_{1,2}=-\frac{\epsilon_2}{2}$, the second term in (\ref{BPZ-level-3-intro}) drops out and $\widehat{A}_3^{\alpha}$ reduces essentially to quantum curves $\widehat{A}_2^{\alpha}$ at level 2. In this paper we determine analogous explicit form of quantum curves up to level 5, with final expressions given in (\ref{BPZ_level_5c}) and (\ref{BPZ-levels-234}), and in section \ref{ssec-general-construction} we propose a general algorithm that enables to determine such expressions at arbitrary level. It is an important task for future work to determine such explicit expressions for arbitrary level. We also note that in the classical limit quantum curves at higher levels factorize and reduce simply to multiple copies of the underlying spectral curve (\ref{Ahat-class-limit-level2-intro}).


\subsection{Some comments and summary of other results}

Let us add a few more comments and briefly summarize other results presented in this paper.

First, recall that via the state-operator correspondence in conformal field theory, to a singular vector $|\phi_{r,s} \rangle$ one can assign a degenerate field that is a descendant of a primary field $\Phi_{r,s}(x)$. Correlation functions $\langle \Phi_{r,s}(x)\prod_i \Phi(x_i)\rangle$ of this field with other primary fields $\Phi(x_i)$ inserted at positions $x_i$ and of weights $\Delta_i$ satisfy Belavin-Polyakov-Zamolodchikov (BPZ) equations \cite{Belavin:1984vu}, i.e. they are annihilated by an operator which also has the same structure as (\ref{chat-Lhat-intro}), however with Virasoro operators represented as $\widetilde{L}_{-1}=\partial_x$ and 
\be
\widetilde{L}_{-n} = \sum_i\Big(\frac{(n-1)\Delta_i}{(x_i-x)^n} - \frac{1}{(x_i-x)^{n-1}}\partial_{x_i}  \Big), \qquad \textrm{for}\ n\geq 2.   \label{tildeLp-intro}
\ee
Therefore quantum curve equations (\ref{Aalpha-psialpha-intro}) can be thought of as being analogous to BPZ equations, with the role of derivatives with respect to $x_i$ played by time derivatives encoded in the term $\partial_x ^{n-2}\hat{f}(x)$.  

Second, it is interesting to note that quantum curves (\ref{chat-Lhat-intro}) have a double quantum structure: on one hand they are quantum in the usual 't Hooft sense, with the parameter $g_s$ or equivalently $1/N$ playing a role of the Planck constant; when this parameter vanishes, quantum curves reduce to classical algebraic curves. Interestingly, from the viewpoint of Virasoro algebra this limit corresponds to setting all $\widehat{L}_{-n}$ to zero for $n>2$, and treating $\widehat{L}_{-1}$ and $\widehat{L}_{-2}$ as commuting objects; precisely such a limit was considered in \cite{Feigin:1988se,Kent:1991wm}. On the other hand, quantum curves (\ref{chat-Lhat-intro}) are quantum in the sense of conformal field theory; from this perspective classical limit corresponds to an infinite central charge $c=1-6Q^2$, which is achieved by taking $b$ or equivalently $\beta$ to zero or infinity. In particular in the context of Liouville theory this is a very interesting limit, whereupon singular vectors reduce to differential equations that represent equations of motion for certain fields in classical Liouville theory \cite{Zamolodchikov:2003yb}. We also reproduce these differential equations in our formalism, and furthermore show that this limit is equivalent to Nekrasov-Shatashvili limit, i.e. it corresponds to setting either $\epsilon_1$ or $\epsilon_2$ to zero, and keeping the value of the other of these parameters constant \cite{NS}. To sum up, two quantum structures built into quantum curves are related to the presence of two parameters $g_s$ and $\beta$, or equivalently $\epsilon_1$ and $\epsilon_2$, and corresponding classical limits arise for appropriate choices of these parameters. 

Another important aspect of this work is the application of the topological recursion \cite{Eynard:2004mh,Chekhov:2005rr,eyn-or}, and in particular its refined version \cite{Chekhov:2006rq,Eynard:2008mz,Chekhov:2009mm,Chekhov:2010zg,Chekhov:2010xj,Brini:2010fc,Marchal:2011iu,Manabe:2012bq},  in the context of quantum curves. As we already mentioned, the topological recursion can be regarded as a reformulation of loop equations of a matrix model, and can be formulated more generally as an algorithm that assigns to a given algebraic curve an infinite set of symplectic invariants encoded in the free energy $F=\log Z$, where $Z=\widehat{\psi}_{\alpha=0}(x)$, and various multi-differentials $W_h(x_1,\ldots,x_h)$. In the context of matrix models the role of an algebraic curve is played by the spectral curve (\ref{Ahat-class-limit-level2-intro}). Note that to consider some particular spectral curve one needs to fix the potential, i.e. the values of times $t_n$ in (\ref{V}) to some particular value, as well as filling fractions in case of a multi-cut solution. In such a setup, it was proposed in \cite{abmodel} that the topological recursion can be used to determine perturbatively, in $g_s$ expansion, the form of quantum curves, that in our present context correspond to unrefined quantum curves at level 2. Generalizing this claim, in this work we propose that the topological recursion, and in particular its refined version, can be used to determine all higher level quantum curves. To this end we determine first the wave-function normalized by the partition function $Z$ in terms of multi-differentials $W_h(x_1,\ldots,x_h)$ as (for details see section \ref{ssec-det})
\be
\log \frac{\widehat{\psi}_{\alpha}(x)}{Z} =
-\frac{2\alpha}{\epsilon_1 \epsilon_2}V(x)+
\sum_{h=1}^{\infty}\frac{1}{h!}\Big(-\frac{2\alpha}{g_s}\Big)^h \int^{x}_{\infty}\cdots\int^{x}_{\infty}W_h(x_1',\ldots,x_h'),  \label{Psi-pert-intro}
\ee
and then reconstruct the corresponding quantum curve perturbatively, as explained in section \ref{ssec-x-expansion}. Nonetheless, there are some important subtleties related to the above formula, and in particular to precise definition of integrals of $W_h(x_1,\ldots,x_h)$. In this work we mainly use (\ref{Psi-pert-intro}) to determine the form of wave-functions and quantum curves that correspond to momenta $\alpha_{r,1}$ or $\alpha_{1,s}$, and in various examples we confirm that in this way we obtain the same results as from the earlier analysis of Virasoro constraints. However, it is not obvious how to evaluate the integrals for wave-functions corresponding to more general singular vectors, and more generally, how to identify the structure of singular vectors or BPZ equations in the perturbative expansion (based on the topological recursion) of quantum curves. We leave such an analysis for future work. 

Yet another subtlety has to do with a choice of reference points or limits in integrals in (\ref{Psi-pert-intro}). As we just mentioned, from the integrals in the limits $\int_{\infty}^x$ one can rederive quantum curves that we constructed in (\ref{chat-Lhat-intro}). However, in literature another definition of wave-functions was proposed \cite{Mulase:2012tm,Dumitrescu:2012dka,Norbury-quantum}, also based on expression (\ref{Psi-pert-intro}) but with integrals in the limits $\int_{\overline{x}}^x$, where $\overline{x}$ is a point conjugate to $x$. This different definition leads in consequence to a different form of quantum curves. In this work we construct both types of wave-functions and corresponding quantum curves (based on two choices of reference points), and find that they are related by a simple transformation, given in (\ref{psi-refpts-relation}) (see also a discussion below this formula). Undoubtedly, especially in view of a recent mathematical work on quantum curves defined via $\overline{x}$ reference point, it is desirable to understand the relation between these two approaches -- and possibly with yet more general quantum curves, based on other reference points -- in more detail.

As follows from the above remarks, the topological recursion, and in particular its refined (or $\beta$-deformed) version, plays an important role in our analysis. It is worth stressing, that apart from the original formulation and several other papers \cite{Eynard:2008mz,Chekhov:2009mm,Chekhov:2010zg,Brini:2010fc,Marchal:2011iu,Manabe:2012bq}, not much work has been done on the refined version of the topological recursion. To make this paper complete we collect various known results, and also present some new results concerning this recursion. In particular we provide a detailed treatment and various explicit formulas relevant for curves of genus zero, which in the $\beta$-deformed case still pose some technical problems. Even though to determine wave-functions and quantum curves one needs to consider only multi-differentials $W_h(x_1,\ldots,x_h)$, for completeness we also provide detailed formulas, and explicit computations in various specific models, of free energies $F=\log Z$, both in the stable and the unstable range. In view of many applications of $\beta$-deformed matrix models in various contexts \cite{Dijkgraaf:2009pc,ACDKV,Marshakov:2010fx,Maruyoshi:2014eja,Witte:2013cea,Sulkowski:2010ux,Sulkowski:2009ne}, we hope that our presentation will be useful for everyone interested in this formalism.

In fact, the analysis of the free energy $F=\log Z$ is not completely unrelated to our analysis of quantum curves. This is so because the wave-function (\ref{Psi-alpha}) contains free energy contributions that -- if needed -- can be factored out by including the normalization  by $Z$, as follows from (\ref{Psi-pert-intro}). We show that in consequence it is possible to extract free energy contributions encoded in (the unnormalized) $\widehat{\psi}_{\alpha}(x)$, or more precisely their time-dependent parts, by the analysis of quantum curves in the limit of large $x$.

From the above discussion it follows that in this paper we analyze matrix models from two different perspectives. On one hand, we consider a matrix model with a potential (\ref{V}), with general and independent times $t_n$. This form of the potential is necessary in order to be able to take partial derivatives with respect to times; such time derivatives enter the Virasoro operators (\ref{Lm_fer_op-intro}) via the dependence on $\hat{f}(x)$, and in consequence quantum curves impose time-dependent, partial differential equations for $\widehat{\psi}_{\alpha}(x)$. On the other hand, the formalism of the topological recursion involves a fixed a spectral curve, which requires fixing the matrix model potential (and so all times $t_n$) to some specific form (which leads to the spectral curve in question). While these two viewpoints might seem contradictory, they can be made consistent, at least for some specific matrix models. 

To illustrate our formalism, and in particular its features just mentioned, we conclude the paper by considering various specific models, for which we construct and analyze quantum curves and corresponding wave-functions. We consider models with polynomial and logarithmic potentials, in particular Gaussian, Penner and multi-Penner model, which are not only interesting in themselves, but also arise in various dualities and compute quantities relevant for completely different systems.  Our formalism reveals interesting features of all these models. For example, while the Gaussian model with the potential $V(x)=\frac{1}{2}x^2$ is the simplest matrix model that has been analyzed in depth from many viewpoints, it appears that differential equations imposed by its higher level quantum curves have not been considered before. Then, an interesting model with infinitely many times $t_n$ fixed to specific values is the Penner model with a logarithmic potential $V(x)=-x-\log(1-x)$; we show that in this case Virasoro constraints can be used to replace derivatives with respect to times by derivatives with respect to $x$, and so to rewrite all quantum curves as ordinary differential operators in $x$. We also find and analyze an interesting generalization of this potential, that depends on one additional parameter. Furthermore, we note that the quantum curve equations at level 2 are essentially equations that define orthogonal polynomials for a given model, in particular Hermite and Laguerre polynomials respectively for the Gaussian and the Penner model; it would be interesting to find analogous interpretation of solutions of higher level quantum curve equations and its interpretation in conformal field theory.

Finally, perhaps the most interesting example is the multi-Penner model with the potential of the form $V(x)=\sum_{i=1}^M\alpha_i\log(x-x_i)$. It is amusing to realize that in this case the operators (\ref{Lm_fer_op-intro}) reduce precisely to Virasoro operators familiar in conformal field theory (\ref{tildeLp-intro}). In consequence quantum curve are not just analogous, but take form identical to BPZ equations, while wave-functions $\widehat{\psi}_{\alpha}(x)$ are identified with conformal field theory correlation functions. Moreover, using Virasoro constraints, in case $M\leq 3$  quantum curve equations can be reduced to a time-independent form; in particular at level 2 they become then identical to hypergeometric equations for four-point functions also familiar in conformal field theory.


\subsection{Matrix models, dualities and related systems}    \label{sec-top-strings}

Over the years matrix models attracted immense attention due to their applications in other systems, which we now briefly review. Higher level quantum curves, which we introduce in this paper, should have an interesting interpretation in all those other systems too. 


{\bf Two-dimensional gravity, intersection theory, integrable hierarchies, etc.} 
First, it is natural to ask what is the role of (higher level) quantum curves in an intricate web of dualities -- uncovered since the end of 1980's -- relating matrix models to two-dimensional quantum gravity, random surfaces, intersection theory on the moduli space of Riemann surfaces, integrable hierarchies, soliton equations, free fermion formalism, etc.   \cite{P2,Moore:1990mg,Witten:1990hr,Kontsevich:1992ti,Dijkgraaf:1990rs,Fukuma:1990jw,Dijkgraaf:1991qh,Kharchev:1992iv,GatoRivera:1992in,DFGZJ}.  


{\bf Topological strings and Dijkgraaf-Vafa theory.} 
Second, the Dijkgraaf-Vafa theory revived the interest in matrix models, relating them to supersymmetric gauge theories and topological strings  \cite{Dijkgraaf:2002fc,Dijkgraaf:2002vw,Dijkgraaf:2002dh}. In particular the B-model topological strings on a Calabi-Yau 
\be
uv = A(x,y),   \label{uv-Axy}
\ee
with $u,v,x,y\in \mathbb{C}$, and for $A(x,y)$ taking form of the spectral curve (\ref{Ahat-class-limit-level2-intro}), turned out to be described by the unrefined ($\beta=1$) matrix model with the potential $V(x)$: the topological string partition function equals the matrix model partition function, which at the same time computes an effective superpotential of $\mathcal{N}=1$ gauge theory with a tree level superpotential $V(x)$. 
The curve (\ref{Ahat-class-limit-level2-intro}) plays role of the moduli space of B-branes characterized by $u=0$ or $v=0$, with partition functions given by (\ref{Psi-alpha}) with $\beta=1$ and $\alpha=-\frac{g_s}{2}$. Our work naturally identifies (\ref{Psi-alpha}), for $\beta=1$ and $\alpha$ being a multiplicity of $-\frac{g_s}{2}$, with a wave-function of a stack of coinciding B-branes, and the corresponding quantum curve (\ref{Aalpha-psialpha-intro}) with a Schr{\"o}dinger equation for those B-branes. This picture is even more interesting for $\beta\neq 1$, which we now turn to.


{\bf M-theory and refined topological string theory.}
The matrix model plays yet more profound role: its $\beta$-deformation defines refined topological string theory \cite{Dijkgraaf:2009pc,ACDKV}. Despite the lack of a worldsheet definition, it is postulated that refined topological strings have other descriptions too, via M-theory \cite{Hollowood:2003cv}, supersymmetric gauge theories \cite{Nekrasov}, or refined topological vertex \cite{Awata:2005fa,Iqbal:2007ii}. These descriptions are interrelated, as follows from M-theory formulation on 
\be
S^1\times \textrm{Taub-NUT}\times X,   \label{Mtheory}
\ee
where $X$ is a Calabi-Yau manifold mirror to (\ref{uv-Axy}), and the Taub-NUT space parametrized by $z_1,z_2\in\mathbb{C}$ is twisted such that a rotation along the circle $S^1$ induces a rotation $z_1\mapsto e^{i\epsilon_1}z_1, z_2\mapsto e^{i\epsilon_2}z_2$. This M-theory picture captures indices of refined BPS states of M2-branes wrapping two-cycles of $X$ \cite{Hollowood:2003cv}, and can be also described in terms of a supersymmetric gauge theory (with $\epsilon_1,\epsilon_2$ encoding the $\Omega$-background \cite{Nekrasov}), or via the B-model topological strings on a manifold mirror to $X$; if this mirror is of the form (\ref{uv-Axy}) with $A(x,y)$ as in (\ref{Ahat-class-limit-level2-intro}), a description via a $\beta$-deformed matrix models arises, with $\hbar$ and $\beta$ related to $\epsilon_1$ and $\epsilon_2$ via (\ref{matparam-intro}). 

In the M-theory system (\ref{Mtheory}) one can include M5-branes, referred to as $\epsilon_1$- or $\epsilon_2$-branes respectively, that wrap $S^1$, a lagrangian subspace of $X$, and a complex line in the Taub-NUT space parametrized by $z_1$ or $z_2$. Reducing this system to the internal space two types of B-branes arise, which in the mirror description are described by the $\beta$-deformed matrix model with a determinant insertion. Such a system with a single $\epsilon_1$- or a single $\epsilon_2$-brane was analyzed in \cite{ACDKV}; its partition function equals (\ref{Psi-alpha}), with either $\alpha=\alpha_{1,2}=-\frac{1}{2}\epsilon_2$ or $\alpha=\alpha_{2,1}=-\frac{1}{2}\epsilon_1$.

From the above description, the interpretation of wave-functions that we introduce in (\ref{Psi-alpha}), with values of $\alpha=\alpha_{r,s}$ given in (\ref{alpha-rs}), is straightforward: a wave-function $\widehat{\psi}_{\alpha_{r,s}}(x)$ represents a refined topological string amplitude for a stack that consists of $(s-1)$ overlapping $\epsilon_1$-branes and $(r-1)$ overlapping $\epsilon_2$-branes. The fact that only integer multiplicities of these branes are allowed (i.e. that $r,s=1,2,3,\ldots$) is automatically encoded in the matrix model, or equivalently in refined topological strings, and it follows from the postulate that $\widehat{\psi}_{\alpha}(x)$ satisfy differential equations of a finite order in $x$ (i.e. Schr{\"o}dinger equations for stacks of branes, which take form of higher level quantum curves (\ref{Aalpha-psialpha-intro})). While we present our results in terms of matrix models, we could also express them in the language of refined topological strings, regarding (\ref{Psi-alpha}) as brane amplitudes, the spectral curve (\ref{Ahat-class-limit-level2-intro}) as part of (\ref{uv-Axy}), etc.


{\bf Supersymmetric gauge theories and surface operators.}
As already mentioned, a duality between matrix models and $\mathcal{N}=1$ gauge theories follows from the Dijkgraaf-Vafa theory \cite{Dijkgraaf:2002fc,Dijkgraaf:2002vw,Dijkgraaf:2002dh}. Reducing the M-theory setup (\ref{Mtheory}) one can also relate matrix models to $\mathcal{N}=2$ four-dimensional gauge theories; this viewpoint is also linked to the AGT duality \cite{Alday:2009aq} (between $\mathcal{N}=2$ gauge theories to the Liouville theory on a Riemann surface $\Sigma$, that encodes the Seiberg-Witten curve). In \cite{Dijkgraaf:2009pc} this duality was explained by means of multi-Penner matrix models, whose potentials lead to Seiberg-Witten curves as spectral curves. In \cite{ACDKV} the system (\ref{Mtheory}) was generalized to include an additional M5-brane, described above, which wraps $S^1$, a lagrangian submanifold of $X$, and a complex $z_1$- or $z_2$-plane in the Taub-NUT space. Reducing (\ref{Mtheory}) to type IIA string theory, and then to four dimensional gauge theory, these branes reduce to two types of surface operators, extended along planes parameterized by $z_1$ or $z_2$. Moreover, in the Nekrasov-Shatashvili limit \cite{NS,ACDKV} one can reduce this system further to an effective theory on a surface operator by considering. Links between the AGT conjecture, $\beta$-deformed matrix models, Nekrasov-Shatashvili limit, and Hitchin systems are discussed in \cite{Bonelli:2011na,Choi:2015idw}. 

The AGT conjecture was also explained by considering M5-branes wrapping spacetime and the Riemann surface $\Sigma$, with surface operators arising from extra M2-branes 
 \cite{AGGTV,DGH}. For relations between surface operators, topological recursion, and Liouville theory see \cite{Kozcaz:2010af,TakiMaruyoshi,Awata:2010bz,Marshakov:2010fx}.

We see from the above perspectives that surface operators in $\mathcal{N}=2$ theories are related to correlation functions in Liouville theory that involve degenerate fields. Moreover, amplitudes in Liouville theory can be described by the multi-Penner matrix model. We find that we can also relate the formalism in this paper to this picture, and generalize it to surface operators corresponding to degenerate fields at arbitrary levels. Similarly as above, $\widehat{\psi}_{\alpha}(x)$ with $\alpha=\alpha_{r,s}$ represents then a stack of $(s-1)$ surface operators wrapping a subspace parameterized by $z_1$, and $(r-1)$ surface operators parametrized by $z_2$. In section \ref{ssec-multiPenner} we show that specializing matrix model potential to the multi-Penner form, wave-functions $\widehat{\psi}_{\alpha}(x)$ are identified with Liouville correlation functions that involve degenerate fields with momenta $\alpha_{r,s}$, and differential equations that they satisfy are quantum curve equations (\ref{Aalpha-psialpha-intro}). The Nekrasov-Shatashvili limit can also be considered, following section \ref{ssec-NSlimit}. While we focus on relations between multi-Penner model and the Liouville theory, it is also desirable to understand the gauge theory construction of surface operators corresponding to degenerate fields at higher levels.


{\bf Topological recursion: remodeling conjecture, knot theory, and more.}
We also briefly review applications of the topological recursion to other systems that involve algebraic curves. Recall that the B-model topological strings on the Calabi-Yau manifold (\ref{uv-Axy}) reduce to the Kodaira-Spencer theory on the curve $A(x,y)=0$, with Ward identities taking form of the topological recursion \cite{Dijkgraaf:2007sx}. This explains the remodeling conjecture \cite{Marino:2006hs,BKMP,Fang:2013dna}, which concerns B-model topological strings also on (\ref{uv-Axy}), however now with $\mathbb{C}^*$ variables $x=e^s$ and $y=e^t$. In this case the mirror manifold is a non-compact, toric Calabi-Yau threefold, and $A(x,y)=0$ is called the mirror curve. The remodeling conjecture states that closed B-model topological amplitudes agree with free energies computed by the topological recursion, and open (brane) amplitudes are captured by multi-differentials $W_h(x_1,\ldots,x_h)$. In particular the topological recursion can produce wave-functions for a single B-brane, which satisfy certain quantum curve equations \cite{ADKMV}, corresponding to quantum curves at level 2 in our formalism. Generalizing this statement, we identify partition functions of multiple coinciding B-branes on a mirror curve, and corresponding quantum curves, respectively with wave-functions (\ref{Psi-alpha}) and quantum curves (\ref{Aalpha-psialpha-intro}) at higher levels. 
 To identify explicitly the underlying structure of singular vectors for such higher level quantum curves, one might consider matrix models that encode Nekrasov partition functions \cite{Klemm:2008yu,Sulkowski:2009br,Sulkowski:2009ne} or topological strings on toric manifolds \cite{Eynard:2010dh,Eynard:2010vd}, and construct wave-functions (\ref{Psi-alpha}) and quantum curves (\ref{Aalpha-psialpha-intro}) associated to those models. 

Algebraic curves $A(x,y)=0$ in $\mathbb{C}^*$ variables $x=e^s$ and $y=e^t$ arise also in knot theory. Such curves are known as A-polynomials \cite{CCGLS}, augmentation polynomials and super-A-polynomials \cite{Ng,AVqdef,superA}. Quantization of these curves $\widehat{A}(\widehat{x}, \widehat{y})$ imposes recursion relations for colored Jones and HOMFLY polynomials, or superpolynomials. It was shown in \cite{DijkgraafFuji-1,DijkgraafFuji-2,Borot:2012cw,Gu:2014yba} that A-polynomials and their generalizations can be identified as mirror curves underlying (\ref{uv-Axy}), and in consequence various objects in knot theory are computed by B-model topological strings. Moreover, following the above discussion, computations in this topological string theory reduce to the topological recursion. Ultimately various objects in knot theory are computable by this recursion. It is desirable to understand the role of higher level quantum curves and corresponding wave-functions presented in this paper, in knot theory too.

Other applications of the topological recursion make contact with moduli spaces of Riemann surfaces and related systems \cite{Andersen:2012yb,Dumitrescu:2012dka,Mulase:2012tm,Bouchard:2013xd,Mulase:2013pa,Liu:2013lra,Dumitrescu:2013tca,Dunin-Barkowski:2013wca,Do:2013wda,Do:2014gda,Dumitrescu:2014dqa,Schwarz:2014hfa,Luu:2015fga,2015arXiv150706557I,Dumitrescu:2015mpa,Norbury-quantum}. The role of higher level quantum curves identified in (\ref{Aalpha-psialpha-intro}) in all those systems is worth analysis too.






\subsection{Plan of the paper}

Having summarized our results and their relations to other systems, let us also summarize the contents of this paper. In section \ref{sec-virasoro} we review the construction and properties of Virasoro singular vectors from conformal field theory perspective. In section \ref{sec-loop} we introduce the notion of $\alpha/\beta$-deformed matrix integrals, analyze their properties, and derive corresponding representations of Virasoro algebra; in particular in section \ref{ssec-interlude} we present operator expressions and various technical tools, on which computations throughout the paper are based. In section \ref{sec-quantum} we present a general algorithm that enables to determine quantum curves at higher levels, and determine such curves explicitly up to level 5. In section \ref{sec-double} we discuss a double quantum structure of quantum curves, their properties in various limits, and various perturbative expansions. In section \ref{sec-refined} we provide a concise summary, including a few new results, of the refined version of the topological recursion, and discuss how this recursion can be used to reconstruct wave-functions and quantum curves; in section \ref{ssec-det} we also discuss quantum curves for various reference points. Finally, in section \ref{sec-examples} our general formalism is applied in several examples of matrix models with polynomial and logarithmic potentials. Appendices \ref{sec-Funstable}, \ref{app:integrands_rec} and \ref{ssec-compPenner} contain technical details related to the refined topological recursion, while appendix \ref{sec-GP} summarizes known formulas for free energies in Gaussian and Penner matrix models, which we rederive independently in section \ref{sec-examples} by means of the refined topological recursion.


\newpage 

\section{Virasoro singular vectors}     \label{sec-virasoro}


In this section we briefly review properties of singular vectors of the Virasoro algebra \cite{Belavin:1984vu,Bauer-singvect-1}. 

Consider the Virasoro algebra
\begin{equation}
[L_{m}, L_{n}]=(m-n)L_{m+n}+\frac{c}{12}(m^3-m)\delta_{m+n,0}   \label{L-Virasoro}
\end{equation}
and a Verma module $V(\Delta,c)$ generated by the highest weight vector $| \Delta\rangle$ that satisfies $L_n| \Delta\rangle=0$ for $n\geq 1$ and $L_0|\Delta\rangle = \Delta|\Delta\rangle$. This Verma module can be decomposed into subspaces $V_n(\Delta,c)$ at level $n$, for $n=0,1,2,\ldots$, defined as eigenspaces of $L_0$ with eigenvalues $n+\Delta$. Singular vectors are defined as vectors annihilated by all Virasoro generators $L_m$ with positive $m$. It can be shown that at level $n$, in the subspace $V_n(\Delta,c)$, there exists a unique (up to an overall normalization) singular vector, if and only if $\Delta$ and $c$ can be parametrized by a complex parameter $b$  as follows
\begin{equation}
c  = 13+6b^2+6b^{-2},\qquad   \Delta=\Delta_{r,s} = \frac{1-r^2}{4}b^2 + \frac{1-s^2}{4}b^{-2} + \frac{1-rs}{2},   \label{c-Delta-null}
\end{equation}
where $r,s$ are two positive integers satisfying $n=rs$. We denote such singular vector by $|\phi_{r,s}\rangle$; by the above definition it satisfies
\be
L_m |\phi_{r,s}\rangle = 0,\qquad \textrm{for}\ m\ge 1.    \label{singvect}
\ee
In general such a singular vector can be represented as 
\be
|\phi_{r,s} \rangle = A_{r,s} | \Delta_{r,s}\rangle,   \label{singular-vect}
\ee
where
\be
A_{r,s} = \sum_{p_1+p_2+\ldots+p_k=n}  c^{r,s}_{p_1,p_2,\ldots,p_k}(b)\,   L_{-p_1} L_{-p_2}\cdots L_{-p_k},   \label{Lrs}
\ee
and where $p_1,p_2,\ldots>0$, and coefficients $c^{r,s}_{p_1,p_2,\ldots,p_k}(b)$ depend on $r$ and $s$ and are functions of $b$. Typically one chooses a normalization such that $c^{r,s}_{1,1,\ldots,1}(b)=1$.

Finding an explicit form of operators $A_{r,s}$ is an outstanding problem in conformal field theory. For $r=1$ or $s=1$ it was solved by Benoit and Saint-Aubin, who showed in \cite{BenoitSaintAubin} that 
\be
A_{r,1} = \sum_{p_1+p_2+\ldots+p_k=r} \frac{(r-1)!^2}{\prod_{i=1}^{k-1} \big( \sum_{j=1}^i p_j \big)\big( r - \sum_{j=1}^i p_j   \big)  }  b^{2(r-k)}   L_{-p_1} L_{-p_2}\cdots L_{-p_k},  \label{Lnull}
\ee
and for $A_{1,s}$ an analogous formula holds, however with $b$ replaced by $b^{-1}$. Note that in the above formula one needs to include all possible combinations of positive $p_1,p_2,\ldots$ (i.e. $(p_1,p_2,\ldots,p_k)$ does not represent an ordinary two-dimensional partition of $n$, but rather an \emph{ordered} partition). 
From the formula (\ref{Lnull}) and Virasoro relations $[L_{-1},L_{n}]=-(n+1)L_{n-1}$, the following form of operators $A_{r,1}$ encoding singular vectors at first few levels can be found:
\begin{equation}
\begin{split}
A_{2,1} &= L_{-1}^2 + b^2 L_{-2}  \label{null-level234} \\
A_{3,1} &= L_{-1}^3  + 2b^2 L_{-1}L_{-2} + 2b^2 L_{-2}L_{-1} + 4b^4 L_{-3}
= L_{-1}^3  + 4b^2 L_{-2}L_{-1} + (2b^2 + 4b^4 )L_{-3}    \\
A_{4,1} & = L_{-1}^4 + 10b^2 L_{-1}^2 L_{-2} + 9b^4L_{-2}^2 + (24b^4-10b^2)L_{-1}L_{-3} + (36b^6 - 24b^4 + 6b^2)L_{-4}.   
\end{split}
\end{equation}
Even though for $r,s\neq 1$ a general and explicit formula for $A_{r,s}$ is not known, there are various methods to determine other singular vectors, such as a recursive algorithm by Bauer \emph{et al.} \cite{Bauer-singvect-1,Bauer-singvect-2}, a method by Kent based on the extension of the enveloping algebra of the Virasoro algebra by complex powers of $L_{-1}$ operator \cite{Kent}, or expressions in terms of Jack polynomials \cite{mimachi1995}. These methods enable to determine explicitly various specific operators (\ref{Lrs}) that encode singular vectors; for example, one can show that the first singular vector not of the form (\ref{Lnull}), which belongs to level 4, is determined by
\be
A_{2,2}=L_{-1}^4 + \big(b^2-b^{-2}\big)^2 L_{-2}^2 + \frac{3}{2}\big(b+b^{-1}  \big)^2 L_{-1}^2L_{-2} + 
\frac{3}{2}\big(b-b^{-1}  \big)^2 L_{-2}  L_{-1}^2 - \big( b^2 + b^{-2}\big) L_{-1}L_{-2}L_{-2}. \label{null-level4-mixed} 
\ee

By the state-operator correspondence, to a singular vector $|\phi_{r,s} \rangle$ we can associate a degenerate field that is a descendant of a primary field $\Phi_{r,s}(x)$. In consequence of the singular vector condition (\ref{singvect}), correlation functions of this primary field with other local fields $\Phi(x_i)$ satisfy differential equations called BPZ equations 
\be
\widetilde{A}_{r,s} \big\langle \Phi_{r,s}(x) \prod_i \Phi(x_i)\big\rangle = 0,   \label{BPZ-cft}
\ee
where $\widetilde{A}_{r,s}$ takes the same form as (\ref{Lrs}), however with $L_{-p}$ replaced by operators
\be
\widetilde{L}_{-p} = \sum_i\Big(\frac{(p-1)\Delta_i}{(x_i-x)^p} - \frac{1}{(x_i-x)^{p-1}}\partial_{x_i}  \Big),   \label{tildeLp}
\ee
where $\Delta_i$ denotes a conformal weight of the field $\Phi(x_i)$. 

As we discuss in the next section, ($\beta$-deformed) matrix model formalism is closely related to the Coulomb gas realization of conformal field theory, which is a theory of a free chiral boson field $\varphi(x)$ in the presence of a background charge $Q=i(b+1/b)$. The energy-momentum tensor in this theory takes form $T = :(\partial \varphi)^2: + Q\partial^2 \varphi$ and the central charge  $c=1-6Q^2$. One class of primary fields in this theory are vertex operators
$e^{\frac{2\alpha}{g_s}\varphi(x)}$ with momenta that we denote $\frac{\alpha}{g_s}$, whose conformal dimension reads $\Delta_{\alpha}=\frac{\alpha}{g_s}(\frac{\alpha}{g_s}-Q)$ (normalization of the momentum by $g_s$, as well as the factor of $i$ in the background charge $Q$ and $\alpha_{r,s}$ below, are chosen for consistency with matrix model conventions which will be introduced in section \ref{sec-loop}). Degenerate fields corresponding to singular vectors $|\phi_{r,s}\rangle$ can also be realized as vertex operators $\phi_{r,s}(x)=e^{\frac{2\alpha_{r,s}}{g_s}\varphi(x)}$, with particular, discrete values of momenta 
\be
\frac{\alpha_{r,s}}{g_s} = -i\Big( \frac{b}{2}(r-1) + \frac{1}{2b}(s-1)\Big),\qquad r,s = 1,2,3\ldots    \label{alpha-rs-cft}
\ee
For these 
momenta the conformal dimension $\Delta_{{\alpha}_{r,s}}=\frac{\alpha_{r,s}}{g_s}(\frac{\alpha_{r,s}}{g_s}-Q)$ indeed takes form (\ref{c-Delta-null}). 

Our presentation in the following sections can be regarded as the analysis of the matrix model realization of the Coulomb gas formalism. Amusingly, this analysis will reveal some new answers to the old questions in conformal field theory. In particular, we will find explicit expressions that capture all operators (\ref{Lrs}) up to a given level $n=rs$, which take form of $\alpha$-dependent operator expressions that specialize to $A_{r,s}$ upon the specialization of the momenta to the values (\ref{alpha-rs-cft}).


\newpage 

\section{$\alpha/\beta$-deformed matrix integrals and Virasoro constraints}  \label{sec-loop}

In this section we introduce a certain deformation of a determinant expectation value in a matrix model. We call this deformation the $\alpha/\beta$-deformed matrix integral, as it is parametrized by a parameter $\alpha$, and in general we consider it together with the $\beta$-deformation of the matrix model integration measure. We also call such an object a wave-function, or a brane partition function in the context of topological string theory. In this section we derive loop equations for the $\alpha/\beta$-deformed matrix integral and show that they are equivalent to an interesting representation of Virasoro constraints. Furthermore, by interpreting the wave-function as an operator insertion in conformal field theory, we derive the corresponding representation of Virasoro operators acting on such an insertion. As we discuss in the next section, Virasoro operators in this latter representation form building blocks of higher level quantum curves.


\subsection{Matrix models, loop equations and spectral curves}  \label{ssec-loopeqs}

Consider the partition function of the $\beta$-deformed ensemble
\begin{equation}
\boxed{ \ Z=\frac{1}{(2\pi)^NN!}\int_{{\IR}^N}  \Delta(z)^{2\beta} e^{-\frac{\sqrt{\beta}}{\hbar}\sum_{a=1}^NV(z_a)}  \prod_{a=1}^Ndz_a  \ }
\label{matrix_def}
\end{equation}
where $\Delta(z)=\prod_{1\le a<b\le N}(z_a-z_b)$ denotes the Vandermonde determinant. For $\beta=1,\frac{1}{2},2$ the above expression is an eigenvalue representation of, respectively, hermitian, orthogonal, and symplectic matrix model. In this paper we are however interested in an analytic continuation of those models to arbitrary values of $\beta$. In what follows we use the notation
\begin{equation}
\epsilon_1=-\beta^{1/2}g_s,\qquad \epsilon_2=\beta^{-1/2}g_s,\qquad g_s=2\hbar,
\label{matparam}
\end{equation}
and to make contact with conformal field theory notation we also denote
\be
b^2 = -\beta  = \frac{\epsilon_1}{\epsilon_2},\qquad
Q = i\big(b+\frac{1}{b}\big) = \frac{\epsilon_1+\epsilon_2}{g_s} .   \label{Qdef}
\ee
Note that using both $\hbar$ and $g_s=2\hbar$  may seem superfluous, and we introduce these two parameters mainly to remind and to comply with two different conventions commonly used in literature; nonetheless, in most of this paper only $g_s$ is used. We also introduce the 't Hooft parameter $\mu = \beta^{1/2}\hbar N$  which is useful, among the others, in the analysis of the large $N$ limit, defined as  
\be
N\to \infty,\qquad \hbar\to 0, \qquad {\rm with}\quad  \mu=\beta^{1/2}\hbar N=const. \label{tHooft}
\ee

As is well known, the invariance of the partition function $Z$ under the infinitesimal transformation
\begin{equation}
z_a\ \to\ z_a+\frac{\varepsilon}{x-z_a},\qquad
x\neq z_a,     \label{mm-variation}
\end{equation}
gives rise to the Ward identity, often referred to as the loop equation
\be
\begin{split}
0=\int_{{\IR}^N} &
\Delta(z)^{2\beta}
e^{-\frac{\sqrt{\beta}}{\hbar}\sum_{a=1}^NV(z_a)}  \prod_{a=1}^Ndz_a
\label{ward_id} \\
&\times
\Big(\sum_{a=1}^N\frac{1}{(x-z_a)^2}-\frac{2\epsilon_1}{\epsilon_2}\sum_{1\le a<b\le N}\frac{1}{(x-z_a)(x-z_b)}-\frac{2}{\epsilon_2}\sum_{a=1}^N\frac{V'(z_a)}{x-z_a}\Big).
\end{split}
\ee
The loop equation is of crucial significance for the whole matrix model analysis. On one hand, in the large $N$ limit it reduces to the spectral curve of the matrix model, i.e. an algebraic curve that encodes an equilibrium distribution of eigenvalues, and in fact also all order asymptotic $\hbar$-expansion of the partition function $Z$, as we will discuss in section \ref{sec-refined}. On the other hand, the loop equation can be reformulated as a system of Virasoro constraints, which also play a prominent role in our analysis. Let us briefly review these two aspects.

To derive the spectral curve, we consider first the planar part of the resolvent
\be
\omega(x) = \lim_{N\to\infty}\frac{\mu}{N}\frac{1}{Z}\Big\langle \sum_{a=1}^N \frac{1}{x-z_a}\Big\rangle.  \label{resolvent}
\ee
Here and in what follows $\langle\mathcal{O}\rangle$ denotes an unnormalized expectation value for an operator $\mathcal{O}$,
\begin{equation}
\left<\mathcal{O}\right>
=\frac{1}{(2\pi)^N N!}\int_{{\IR}^N}  \Delta(z)^{2\beta}\mathcal{O}\, e^{-\frac{\sqrt{\beta}}{\hbar}\sum_{a=1}^NV(z_a)}  \prod_{a=1}^N dz_a.
\end{equation}
For $\beta=1/2,1,2$, when $Z$ is given by an actual matrix integral, the resolvent is simply an expectation value of appropriately normalized $\textrm{Tr}\frac{1}{x-M}$. We also introduce the quantity $f(x)$ and its normalized expectation value in the large $N$ limit
\be
f(x) = 2\epsilon_1\sum_{a=1}^N \frac{V'(x) - V'(z_a)}{x-z_a}, \qquad\quad f_{cl}(x) = \lim_{N\to\infty}  \frac{\langle f(x) \rangle}{Z}.   \label{f-cl}
\ee 
In particular, if $V(x)$ is a potential of degree ${\rm deg}\, V(x)$, then $f_{cl}(x)$ is a polynomial of degree $({\rm deg}\, V(x)-2)$, whose coefficients are determined by certain asymptotic conditions and filling fractions specifying distribution of eigenvalues between cuts. Setting $\beta=1$, dividing the loop equation (\ref{ward_id}) by $Z$ and taking the large $N$ limit (whereupon expectation values factorize) we obtain the following equation
\be
\omega(x)^2 - \omega(x)V'(x) - \frac{1}{4} f_{cl}(x) = 0,   \label{spcurve-pre}
\ee
which in terms of
\be
y(x) = V'(x) - 2\omega(x),   \label{yVomega}
\ee
is written as
\be
\boxed{ \ A(x,y) = y^2 - V'(x)^2 - f_{cl}(x) = 0  \ }  \label{spcurve}
\ee
where we introduced an algebraic function $A(x,y)$. This algebraic equation, which relates complex variables $x$ and $y=y(x)$, represents the spectral curve that we have been after. 

The spectral curve can also be derived by writing the Vandermonde determinant in (\ref{matrix_def}) in the exponential form and considering the effective action for eigenvalues
\be
S = \frac{1}{\hbar}\sum_{a=1}^N V(z_a) - 2\sqrt{\beta}\sum_{1\leq a < b \leq N}\log(z_a - z_b).
\ee
From this perspective the equation of motion for the eigenvalue $z_a$ takes form $\frac{1}{\hbar}V'(z_a)=2\sqrt{\beta}\sum_{b\neq a}\frac{1}{z_a - z_b}$; multiplying both sides of this equation by $\frac{1}{x-z_a}$, setting $\beta=1$, and taking large $N$ limit, we again obtain (\ref{spcurve}). 

Note that even though the spectral curve is defined for $\beta=1$, the dependence of the partition function and other expectation values on $\beta$ can be reintroduced in the formalism of  the refined topological recursion. This formalism, as well as more details about the geometry of the spectral curve, will be presented in section \ref{sec-refined}. Alternatively, one may also try to keep a dependence on $\beta$ in the above derivation. For arbitrary $\beta$, instead of (\ref{spcurve-pre}) and (\ref{spcurve}) one then finds (also in terms of (\ref{yVomega}))
\be
\begin{split}
0 &= \hat{\hbar} \omega'(x) + \omega(x)^2 - \omega(x)V'(x) - \frac{1}{4} f_{cl}(x) =  \\
& = y(x)^2 - V'(x)^2 - f_{cl}(x) + 2\hat{\hbar}\big(V''(x) - y'(x) \big)
 \label{spcurve-Ric}
\end{split}
\ee
where we introduced
\be
\hat{\hbar} =  -\frac{\epsilon_1 + \epsilon_2}{2} = \frac{\mu}{N} (1-\beta^{-1}).   \label{hbar-Ric} 
\ee
The equation (\ref{spcurve-Ric}) is a nonlinear differential equation for $\omega(x)$ or $y(x)$, known as the Riccati equation, and upon a redefinition $\omega(x)=\hat{\hbar}\frac{u'(x)}{u(x)}$ it can be transformed into a Schr{\"o}dinger linear differential equation
 \be
 \hat{\hbar}^2u''(x) = \hat{\hbar}u'(x)V'(x) + \frac{1}{4} f_{cl}(x) u(x)
 \ee
for a new function $u(x)$. This equation was interpreted in \cite{Eynard:2008mz,Chekhov:2009mm,Chekhov:2010zg} as defining a quantum spectral curve, with $\hat{\hbar}$ identified as the quantization parameter (the Planck constant). Nonetheless, this interpretation is not directly related to our approach in this paper -- in particular, while we identify an infinite family of quantum curves corresponding naturally to different quantization parameters, none of these parameters takes form of $\hat{\hbar}$ given in (\ref{hbar-Ric}). 

Finally, the second significant aspect of the loop equation (\ref{ward_id}) is its equivalence to a set of constraint equations on the partition function, imposed by operators $\ell_n$
\be
\boxed{ \ \ell_n Z = 0 \qquad \textrm{for}\ n\geq -1.  \ }    \label{LnZ}
\ee
This shows in particular that the partition function $Z$ can be identified with the vacuum state $|0\rangle$ in conformal field theory interpretation. The operators $\ell_n$ arise from the expansion of the loop equation in powers of $x$ and satisfy the Virasoro algebra, hence (\ref{LnZ}) are referred to as Virasoro constraints. Operators $\ell_n$ can also be identified as the modes of the energy-momentum tensor associated to a chiral boson field \cite{Kostov:1999xi,ACDKV}
\begin{equation}
\phi(x)=-\frac{g_s}{\epsilon_2}N\log x + \frac{1}{g_s}\sum_{n=0}^{\infty}t_n x^n
-\frac{g_s}{2} \sum_{n=1}^{\infty} \frac{1}{nx^n}\partial_{t_n} 
=\frac{1}{g_s}V(x)-\frac{g_s}{\epsilon_2}\sum_{a=1}^N\log (x-z_a)   \label{phi-field}
\end{equation}
in an auxiliary conformal field theory, where we identified the action of $\partial_{t_n}$ on the partition function (\ref{matrix_def}) with the computation of the expectation value $\langle-\frac{2}{\epsilon_2}\sum_a z_a^n\rangle$. In what follows we review more details of this construction from a more general perspective and present an explicit form of operators $\ell_n$ in (\ref{Ln}).


\subsection{$\alpha/\beta$-deformed matrix integrals and wave-functions}

Consider now the insertion of the following expression into the matrix integral (\ref{matrix_def})
\begin{equation}
\psi^{\textrm{ins}}_{\alpha} (x)=\prod_{a=1}^N(x-z_a)^{-\frac{2\alpha}{\epsilon_{2}}}.
\label{fermion_alpha}
\end{equation}
For $\beta=1$ this is the eigenvalue representation of the determinant det$(x-M)$ raised to a power parameterized by the parameter $\alpha$, where $z_a$ denote eigenvalues of a matrix $M$ and the superscript ``ins'' is to stress that this expression is to be inserted under the matrix integral. We note that this type of integrals and corresponding loop equations were considered also in \cite{Chekhov:2010xj}, however in a different context. For a time being, let us treat an insertion of (\ref{fermion_alpha}) simply as an $x$-deformation of the matrix model partition function (\ref{matrix_def}), and derive Virasoro constraints analogous to (\ref{LnZ}). Such generalized constraints are expressed through generalized Virasoro operators $\ell^{\alpha}_n(x)$ that depend on both $x$ and $\alpha$ and can be defined for all values of these parameters. These are not yet the operators that will be used as building blocks of quantum curves, however the ingredients used in this derivation and the Virasoro constraints imposed in terms of $\ell^{\alpha}_n(x)$ are also useful in what follows.

Let us absorb the expression (\ref{fermion_alpha}) into the potential and -- instead of (\ref{matrix_def}) -- consider 
\begin{equation}
\psi_{\alpha}(x) \equiv  \left<\psi^{\textrm{ins}}_{\alpha}(x)\right>
=\frac{1}{(2\pi)^NN!}
\int_{\mathbb{R}^N} \Delta(z)^{2\beta} e^{-\frac{\sqrt{\beta}}{\hbar}\sum_{a=1}^N\widetilde{V}(z_a;x)}
\prod_{a=1}^Ndz_a,
\label{wave_Z2}
\end{equation}
where the modified potential takes form
\begin{equation}
\widetilde{V}(y;x)=V(y)+\alpha\log (x-y).
\label{tw_pot}
\end{equation}
We refer to (\ref{wave_Z2}) as the $\alpha/\beta$-deformed matrix integral, and also often call $\psi_{\alpha}(x)$ a wave-function, or a brane partition function in the context of topological string theory.

Note that via the bosonization formula, the operator $\psi^{\textrm{ins}}_{\alpha}(x)$ in (\ref{fermion_alpha}) can be interpreted as a fermionic operator with momentum $\alpha/g_s$ associated to the bosonic field (\ref{phi-field}), up to a tree level overall factor
\begin{equation}
e^{2\frac{\alpha}{g_s}\phi(x)}
=e^{-\frac{2\alpha}{\epsilon_1\epsilon_2}V(x)}\psi^{\textrm{ins}}_{\alpha}(x)
=e^{-\frac{2\alpha}{\epsilon_1\epsilon_2}V(x)}\prod_{a=1}^N(x-z_a)^{-\frac{2\alpha}{\epsilon_{2}}}.
\label{fermion_alpha_2}
\end{equation}
This is one reason why it is natural to include the tree level factor $\exp\big(-\frac{2\alpha}{\epsilon_1\epsilon_2}V(x)\big)$ in the definition of the wave-function and, instead of (\ref{wave_Z2}), also consider 
\be
\boxed{ \ \widehat{\psi}_{\alpha}(x) = e^{-\frac{2\alpha}{\epsilon_1\epsilon_2}V(x)} \, \psi_{\alpha}(x)   = e^{-\frac{2\alpha}{\epsilon_1\epsilon_2}V(x)} \Big\langle \prod_{a=1}^N(x-z_a)^{-\frac{2\alpha}{\epsilon_{2}}}   \Big\rangle   \ }
  \label{Psi-alpha-bis}
\ee
which was already introduced in (\ref{Psi-alpha}). Using $\epsilon_1$ and $\epsilon_2$, the conformal dimension of the primary operator (\ref{fermion_alpha_2}) with the momentum $\alpha$ can be expressed as
\begin{equation}
\Delta_{\alpha}=\frac{\alpha}{g_s}\Big(\frac{\alpha}{g_s} - Q \Big) = -\frac{\alpha(\epsilon_1+\epsilon_2-\alpha)}{g_s^2}.   \label{Delta}
\end{equation}
Furthermore, note that in terms of $\epsilon_1$ and $\epsilon_2$ the special values of momenta given in (\ref{alpha-rs-cft}) take simple form
\be
\alpha_{r,s} = -\frac{r-1}{2}\epsilon_1 - \frac{s-1}{2}\epsilon_2.     \label{alpha-rs-2}
\ee
These discrete values will play a crucial role in what follows. In addition, sometimes we also consider the wave-function normalized by the partition function $Z$, which we denote as
\be
\boxed{\ 
\Psi_{\alpha} = \frac{\widehat{\psi}_{\alpha}(x)}{Z}.   \label{PsiPsi-alpha}
\  }
\ee

Both (\ref{wave_Z2}) and (\ref{Psi-alpha-bis}) are invariant under the infinitesimal transformation $z_a \to z_a+\frac{\varepsilon}{y-z_a}$, so that we can derive the corresponding loop equation and its representation in terms of Virasoro constraints. The loop equation in this case takes form analogous to (\ref{ward_id}), however with the potential $V(z_a)$ replaced by $\widetilde{V}(z_a;x)$
\begin{align}
0=\int_{{\IR}^N} &
\Delta(z)^{2\beta}
e^{-\frac{\sqrt{\beta}}{\hbar}\sum_{a=1}^N\widetilde{V}(z_a,x)}  \prod_{a=1}^Ndz_a
\label{t_ward_id}  \\
&\times
\Big(\sum_{a=1}^N\frac{1}{(y-z_a)^2}-\frac{2\epsilon_1}{\epsilon_2}\sum_{1\le a<b\le N}\frac{1}{(y-z_a)(y-z_b)}-\frac{2}{\epsilon_2}\sum_{a=1}^N\frac{\partial_{z_a}\widetilde{V}(z_a,x)}{y-z_a}\Big). \nonumber 
\end{align}
We will identify Virasoro constraints imposed by operators $\ell_n^{\alpha}(x)$ associated to this loop equation after an interlude, where we introduce more notational details and discuss various operator expressions.


\subsection{Interlude -- notation and operator expressions}    \label{ssec-interlude}

In this section we introduce some additional notation and present various operator expressions used throughout the rest of the paper.

First, note that various expectation values involving a dependence on integration variables $z_a$ can be represented in terms of operators acting on integrated expressions, such as the partition function $Z$ in (\ref{matrix_def}),  or wave-functions $\psi_{\alpha}(x)$ and $\widehat{\psi}_{\alpha}(x)$. Indeed, the dependence on powers of $z_a$ can be represented in terms of derivatives with respect to times that appear in the potential  $V(z_a)=\sum_{n=0}^{\infty}t_n z_a^n$ (in the matrix model integrand). 
In particular, when considering partition functions $Z$ (understood as an expectation value of the unity) or $\psi_{\alpha}(x)=\langle \psi_{\alpha}^{\textrm{ins}}(x)\rangle$, the insertion of powers of $z_a$ can be expressed in terms of time derivatives
\begin{equation}
\Big\langle \sum_{a=1}^N z_a^n \, \cdots  \Big\rangle = 
-\frac{\epsilon_2}{2}\partial_{t_n}  \Big\langle \cdots \Big\rangle.
\end{equation}
In particular note that
\be
\partial_{t_0} Z=-\frac{2N}{\epsilon_2} Z =\frac{4\mu}{\epsilon_1\epsilon_2}Z,\qquad \partial_{t_0} \psi_{\alpha}(x)= \frac{4\mu}{\epsilon_1\epsilon_2}\psi_{\alpha}(x),    \qquad
\partial_{t_0} \widehat{\psi}_{\alpha}(x) = \frac{4\mu-2\alpha}{\epsilon_1\epsilon_2} \widehat{\psi}_{\alpha}(x).   \label{partialt0} 
\ee
Expectation values of the potential $V(z_a)$ or its derivatives in $Z$ or $\psi_{\alpha}(x)$ can also be represented in an analogous way; for example, an insertion of the first derivative of the potential can be encoded by
\be
\Big\langle \sum_{a=1}^N V'(z_a) \cdots \Big\rangle =
-\frac{\epsilon_2}{2}\sum_{k=1}^{\infty} k t_k \partial_{t_{k-1}}   \Big\langle \cdots \Big\rangle       \label{Vprim-operator}
\ee
Furthermore, an operator that plays a prominent role in the formalism of matrix models is the loop insertion operator defined as
\begin{equation}
\partial_{V(x)}\equiv  \partial^{(1)}_{V(x)}=\frac{N}{x}-\frac{\epsilon_2}{2}\sum_{n=1}^{\infty}\frac{1}{x^{n+1}}\partial_{t_n}.   \label{loop-insertion}
\end{equation}
We also introduce its higher order generalizations
\be
\partial_{V(x)}^{(k)}=\frac{(-1)^{k-1}}{(k-1)!}\underbrace{\big[\partial_x, \big[\partial_x, \cdots \big[\partial_x}_{k-1}, \partial_{V(x)}^{(1)}\big]\cdots\big]\big],  \label{loop-insertion-k}
\ee
where we use commutators to assert that the derivatives $\partial_x$ act only on the $x$-dependence in (\ref{loop-insertion}), and not on a putative object to the right of $\partial_{V(x)}$. These operators act on partition functions $Z$ or $\psi_{\alpha}(x)$ as 
\begin{equation}
\partial^{(k)}_{V(x)}  \Big\langle \cdots \Big\rangle = 
\Big\langle \sum_{a=1}^N\frac{1}{(x-z_a)^k} \cdots \Big\rangle ,    \label{loop-insertion-bis}
\end{equation}
while acting on the derivative of the potential we get
\begin{equation}
\left[\partial_{V(x)}^{(k)}, V'(z)\right]=-\frac{\epsilon_2}{2}\frac{k}{(x-z)^{k+1}}.   \label{loop-insertion-bis-V}
\end{equation}

It is also useful to represent derivatives with respect to $x$ of $\psi^{\textrm{ins}}_{\alpha}(x)$ (defined in (\ref{fermion_alpha})) in terms of products of $(x-z_a)^{-1}$. In particular
\begin{align}
\partial_x\psi^{\textrm{ins}}_{\alpha}(x)
& =
-\frac{2\alpha}{\epsilon_2}\sum_{a=1}^N\frac{\psi^{\textrm{ins}}_{\alpha}(x)}{x-z_a},
\label{diff_psi_1}
\\
\partial_x^2\psi^{\textrm{ins}}_{\alpha}(x)
& =
\frac{2\alpha}{\epsilon_2}\sum_{a=1}^N\frac{\psi^{\textrm{ins}}_{\alpha}(x)}{(x-z_a)^2}
+\frac{4\alpha^2}{\epsilon_2^2}\sum_{a,b=1}^N\frac{\psi^{\textrm{ins}}_{\alpha}(x)}{(x-z_a)(x-z_b)},
\label{diff_psi_2}
\\
\partial_x^3\psi^{\textrm{ins}}_{\alpha}(x)
& =
-\frac{4\alpha}{\epsilon_2}\sum_{a=1}^N\frac{\psi^{\textrm{ins}}_{\alpha}(x)}{(x-z_a)^3}
-\frac{12\alpha^2}{\epsilon_2^2}\sum_{a,b=1}^N\frac{\psi^{\textrm{ins}}_{\alpha}(x)}{(x-z_a)^2
(x-z_b)} +
\nonumber\\
&
\quad -\frac{8\alpha^3}{\epsilon_2^3}\sum_{a,b,c=1}^N\frac{\psi^{\textrm{ins}}_{\alpha}(x)}{(x-z_a)
(x-z_b)(x-z_c)},
\label{diff_psi_3}
\end{align}
and in general the $n$-th derivative of $\psi^{\textrm{ins}}_{\alpha}(x)$ takes form
\begin{equation}
\partial_x^n\psi^{\textrm{ins}}_{\alpha}(x)=\sum_{p=1}^n\frac{\alpha^p}{\epsilon_2^p}\sum_{
\begin{subarray}{c}Y_1\ge Y_2\ge \ldots \ge Y_p \ge 1\\Y_1+\ldots+Y_p=n\end{subarray}
}C_{Y_1,Y_2,\ldots,Y_p}
\sum_{a_1,\ldots,a_p=1}^N\frac{\psi^{\textrm{ins}}_{\alpha}(x)}{(x-z_{a_1})^{Y_1}\cdots (x-z_{a_p})^{Y_p}},
\label{diff_n_psi}
\end{equation}
where $C_{Y_1,Y_2,\ldots,Y_p}\in{\IZ}$ are some constants; the total number of such constants is given by  $\mathfrak{p}(n)$, where $\mathfrak{p}(n)$ denotes the number of partitions of $n$. 

Another expression we often come across, already introduced in (\ref{f-cl}), is
\be
f(x)=2\epsilon_1\sum_{a=1}^N\frac{V'(x)-V'(z_a)}{x-z_a},   \label{fx}
\ee
and it is also useful to consider its derivatives with respect to $x$. Note that expectation values involving $\psi^{\textrm{ins}}_{\alpha}(x)$ and derivatives of $f(x)$ can be expressed using derivatives of $\psi^{\textrm{ins}}_{\alpha}(x)$ given above; for example, taking advantage of (\ref{diff_psi_1}) and (\ref{diff_psi_2}), we get
\begin{equation}
\begin{split}
\big(\partial_x f(x)\big)\,\psi^{\textrm{ins}}_{\alpha}(x)&=\Big(-2\epsilon_1\sum_{a=1}^N\frac{V'(x)-V'(z_a)}{(x-z_a)^2}
-\frac{\epsilon_1\epsilon_2}{\alpha}V''(x)\partial_x\Big) \psi^{\textrm{ins}}_{\alpha}(x), \\
\big(\partial_x^2 f(x)\big)\,\psi^{\textrm{ins}}_{\alpha}(x) &=\Big(4\epsilon_1\sum_{a=1}^N\frac{V'(x)-V'(z_a)}{(x-z_a)^3}
-4\epsilon_1\sum_{a=1}^N\frac{V''(x)}{(x-z_a)^2}
-\frac{\epsilon_1\epsilon_2}{\alpha}V'''(x)\partial_x   \Big)    \psi^{\textrm{ins}}_{\alpha}(x).    \label{dxfx}
\end{split}
\end{equation}
The operator representation of $f(x)$, acting on $Z$ or $\psi_{\alpha}(x)$, takes form
\begin{equation}
\hat{f}(x)=-\epsilon_1\epsilon_2\sum_{n=0}^{\infty}x^n\partial_{(n)},\ \ \ \ 
\partial_{(n)}=\sum_{k=n+2}^{\infty}kt_k\frac{\partial}{\partial t_{k-n-2}}.
\label{sp_hat_f_op}
\end{equation}
We denote the operator representation of derivatives $\partial_x^n f(x)$ through commutators
\begin{equation}
\partial_x^n\widehat{f}(x) \equiv [\partial_x, \partial_x^{n-1}\widehat{f}(x)]
\label{sp_hat_fn_op}
\end{equation}
which again explicitly asserts that $x$-derivatives act here on $x$-dependence of $\hat{f}(x)$ (and not on a putative object to the right of $\hat{f}$). From this definition it follows that, acting on the partition function $Z$ or $\psi_{\alpha}(x)$, for any $n\geq 0$ we get 
\be
\partial_x^n\widehat{f}(x)\Big\langle \cdots \Big\rangle=\Big\langle\partial_x^nf(x)\cdots\Big\rangle.    \label{fhat-f}
\ee
Note that 
\begin{align}
\begin{split}
& \big[\widehat{f}(x), \partial_x^nV(x)\big]=
-\epsilon_1\epsilon_2\sum_{k=n+2}^{\infty}kt_kx^{k-n-2}
\sum_{p=n}^{k-2}\frac{p!}{(p-n)!} = -\frac{\epsilon_1\epsilon_2}{n+1}\partial_x^{n+2}V(x),  \\
& \big[\widehat{f}(x), \partial_x^n\widehat{f}(x)\big]=
\epsilon_1^2\epsilon_2^2\sum_{k=n+2}^{\infty}x^{k-n-2}\partial_{(k)}
\sum_{p=n}^{k-2}\frac{(2p+2-k)p!}{(p-n)!} = -\frac{n \epsilon_1\epsilon_2}{(n+1)(n+2)}\partial_x^{n+2}\widehat{f}(x),  \label{hatfVdk-hatfhatf-aux}
\end{split}
\end{align}
where in the second step of both lines we used  the formula
\begin{equation}
\sum_{p=1}^k\frac{(p+n-2)!}{(p-1)!}=
\frac{(k+n-1)!}{n(k-1)!},\quad k,n\in{\IN}.
\end{equation}
Differentiating (\ref{hatfVdk-hatfhatf-aux}) with respect to $x$, by induction we find commutation relations
\begin{align}
\big[\partial_x^m\widehat{f}(x), \partial_x^n\widehat{f}(x)\big]&=
\frac{(m-n)m!n!}{(m+n+2)!}\epsilon_1\epsilon_2\partial_x^{m+n+2}\widehat{f}(x),
\label{hatfhatf}
\\
\big[\partial_x^m\widehat{f}(x), \partial_x^n V(x)\big]&=
-\frac{m! n!}{(m+n+1)!}\epsilon_1\epsilon_2 \partial_x^{m+n+2}V(x),
\label{hatfVdk}
\end{align}
of which we take advantage in various calculations. 

Furthermore, we often need to translate various equations for $\psi_{\alpha}(x)$ into corresponding equations for $\widehat{\psi}_{\alpha}(x)$. The difference between these two wave-functions is given by the overall factor involving the potential in (\ref{Psi-alpha-bis}). Therefore, for an operator ${\cal O}(x)$ acting on $\psi_{\alpha}(x)$, the corresponding operator acting on $\widehat{\psi}_{\alpha}(x)$ takes form
\begin{equation}
{\cal O}(x)+\frac{2\alpha}{\epsilon_1\epsilon_2}\big[{\cal O}(x),V(x)\big].
\label{tree_op_trans}
\end{equation}
In particular, for ${\cal O}(x)=\partial_x^{n-2}\hat{f}(x)$ acting on $\psi_{\alpha}(x)$, using (\ref{hatfVdk}) we find that the corresponding operator acting on $\widehat{\psi}_{\alpha}(x)$ reads
\begin{equation}
\partial_x^{n-2}\hat{f}(x)-\frac{2\alpha}{n-1}\partial_x^{n}V(x).
\label{tree_op_trans-fhat}
\end{equation}


\subsection{Virasoro operators for $\alpha/\beta$-deformed matrix integrals}    \label{Virasoro-generators}

Having introduced relevant notation, we can now identify Virasoro operators $\ell_n^{\alpha}(x)$ associated to the loop equation (\ref{t_ward_id}) for the $\alpha/\beta$-deformed matrix integral, i.e. for the $\beta$-deformed matrix model with the insertion of (\ref{fermion_alpha}). To this end it is useful to introduce a more general chiral boson operator
\begin{equation}
\phi(y;x)=\frac{1}{g_s}\widetilde{V}(y;x)-\frac{g_s}{\epsilon_2}\sum_{a=1}^N\log (y-z_a).
\label{chi_boson_m}
\end{equation}
Virasoro operators $\ell_n^{\alpha}(x)$ can then be read off as the modes in the 
expansion of the corresponding energy-momentum tensor
\be
\begin{split}
T(y;x)  & =  : \partial_y\phi(y;x)\partial_y\phi(y;x):+\ Q\partial_y^2\phi(y;x)  =  
 \sum_{n=-\infty}^{\infty}\ell_n^{\alpha}(x) y^{-n-2}   = \label{Tyx} \\
 &= T_{-}(y;x) + T_{+}(y;x), 
\end{split}
\ee
where by $T_{-}(y;x)$ and $T_{+}(y;x)$ we denote respectively the pieces of the expansion (in powers of $y$) of $T(y;x)$ that encode the modes $\ell_n^{\alpha}(x)$ with $n\le -2$ and $n\ge -1$. We find
\begin{align}
T_{-}(y;x)&=\sum_{n=-\infty}^{-2}\ell_n^{\alpha}(x)y^{-n-2} = \label{T_minus_em} \\   
& = \frac{1}{g_s^2}\big(\partial_y\widetilde{V}(y;x)\big)^2+
\frac{\epsilon_1+\epsilon_2}{g_s^2}\partial_y^2\widetilde{V}(y;x)
-\frac{2}{\epsilon_2}\sum_{a=1}^N\frac{\partial_y\widetilde{V}(y;x)-\partial_{z_a}\widetilde{V}(z_a;x)}{y-z_a} =
\nonumber\\
&=\frac{\Delta_{\alpha}}{(y-x)^2} + \frac{1}{y-x}\partial_x - \frac{1}{\epsilon_1 \epsilon_2} \Big(  V'(y)^2 + (\epsilon_1 + \epsilon_2)V''(y)  + \frac{2\alpha}{y-x} V'(y) + \widehat{f}(y) \Big),  \nonumber
\\
T_{+}(y;x)&=\sum_{n=-1}^{\infty}\ell_n^{\alpha}(x)y^{-n-2} =    \label{T_plus_em}  \\   
& = \sum_{a=1}^N\frac{1}{(y-z_a)^2}-\frac{2\epsilon_1}{\epsilon_2}\sum_{1\le a<b\le N}\frac{1}{(y-z_a)(y-z_b)}-\frac{2}{\epsilon_2}\sum_{a=1}^N\frac{\partial_{z_a}\widetilde{V}(z_a;x)}{y-z_a} = 
\nonumber\\
&=\frac{2\alpha}{\epsilon_2}\sum_{a=1}^N\frac{1}{(x-z_a)(y-z_a)}
+\frac{\epsilon_1+\epsilon_2}{\epsilon_2}\sum_{a=1}^N\frac{1}{(y-z_a)^2}  +  \nonumber\\ 
&\qquad \qquad \qquad \qquad \qquad \qquad
-\frac{\epsilon_1}{\epsilon_2}\sum_{a,b=1}^N\frac{1}{(y-z_a)(y-z_b)}
-\frac{2}{\epsilon_2}\sum_{a=1}^N\frac{V'(z_a)}{y-z_a},   \nonumber
\end{align}
where in the third line of (\ref{T_minus_em}) we expressed $T_{-}(y;x)$ as an operator on $\psi_{\alpha}(x)$. Note that in terms of (\ref{Qdef}), from the OPE 
\be
T(y_1;x)T(y_2;x)=\frac{1-6Q^2}{2(y_1-y_2)^4}+\frac{2T(y_2;x)}{(y_1-y_2)^2}+\frac{\partial_{y_2}T(y_2;x)}{y_1-y_2}+\ldots
\ee
the central charge is determined as $c=1-6Q^2$.

Also note that the loop equation (\ref{t_ward_id}) can be written as
\begin{equation}
\left<T_{+}(y;x)\psi^{\textrm{ins}}_{\alpha}(x)\right>=
\int_{{\IR}^N} \Delta(z)^{2\beta}
e^{-\frac{\sqrt{\beta}}{\hbar}\sum_{a=1}^N\widetilde{V}(z_a;x)}T_{+}(y;x)
\prod_{a=1}^Ndz_a=0,
\label{t_ward_id_2}
\end{equation}
Upon the expansion in powers of $y$, this loop equation is equivalent to the set of Virasoro constraints on the wave-function (\ref{wave_Z2}) 
\begin{equation}
\boxed{
\quad \ell_n^{\alpha}(x) \psi_{\alpha}(x) =0,\ \ \ \ n\ge -1. \quad
}
\label{vir_const_wave_2}
\end{equation}

We can now explicitly write down Virasoro generators defined in (\ref{Tyx}). From the appropriate expansions of expressions in (\ref{T_minus_em}) and (\ref{T_plus_em}), and taking advantage of operator representation introduced in section \ref{ssec-interlude}, we find 
\begin{equation}
\boxed{\quad 
\ell_n^{\alpha}(x)=
\begin{cases}
\ell_n+\ell_n^{\textrm{Witt}}(x)
+\alpha\sum_{k=0}^nx^k\partial_{t_{n-k}},
&
\textrm{if}\ n\ge -1
\\
\ell_n+\ell_n^{\textrm{Witt}}(x)
+\frac{2\alpha}{\epsilon_1\epsilon_2}\sum_{k=0}^{-n-2}(k+1)t_{k+1}x^{k+n+1}-\Delta_{\alpha}(n+1)x^n,
&
\textrm{if}\ n\le -2
\end{cases}\quad
}
\label{twi_vira_g}
\end{equation}
where
\begin{equation}
\boxed{\qquad 
\ell_n=
\begin{cases}
-\frac{\epsilon_1\epsilon_2}{4}\sum_{k=0}^n\partial_{t_k}\partial_{t_{n-k}}
-\frac{\epsilon_1+\epsilon_2}{2}(n+1)\partial_{t_n}
+\sum_{k=1}^{\infty}kt_k\partial_{t_{n+k}},
&
\textrm{if}\ n\ge -1
\\
\frac{1}{\epsilon_1\epsilon_2}\sum_{k=0}^{-n-2}(k+1)(k+n+1)t_{k+1}t_{-n-k-1}-\frac{\epsilon_1+\epsilon_2}{\epsilon_1\epsilon_2}n(n+1)t_{-n}
\\
\hspace{15em}
+\sum_{k=0}^{\infty}(k-n)t_{k-n}\partial_{t_k},
&
\textrm{if}\ n\le -2
\end{cases}\qquad    \label{Ln}
}
\end{equation}
The modes $\ell_n^{\alpha}(x)$ in (\ref{twi_vira_g}) form an interesting representation of Virasoro algebra. These modes are sums of three pieces. The first piece, denoted by $\ell_n$, is $x$-independent and involves only times $t_k$. These $\ell_n$ are in fact the Virasoro generators for the original $\beta$-deformed ensemble (\ref{matrix_def}); they impose Virasoro constraints (\ref{LnZ}) and satisfy the Virasoro algebra
\begin{equation}
[\ell_{m}, \ell_{n}]=(m-n)\ell_{m+n}+\frac{c}{12}(m^3-m)\delta_{m+n,0}.
\end{equation}
The second, time-independent piece is given by the generators of the Witt algebra
\begin{equation}
\ell_n^{\textrm{Witt}}(x)=-x^{n+1}\partial_x    \label{LWitt}
\end{equation}
that satisfy commutation relations 
$
[\ell_{m}^{\textrm{Witt}}(x), \ell_{n}^{\textrm{Witt}}(x)]=(m-n)\ell_{m+n}^{\textrm{Witt}}(x)$. 
The third piece in $\ell_n^{\alpha}(x)$ is a mixed term, with dependence on both times and $x$. Altogether we see that (\ref{twi_vira_g}) also gives a representation of the Virasoro algebra
\begin{equation}
[\ell_{m}^{\alpha}(x), \ell_{n}^{\alpha}(x)]=(m-n)\ell_{m+n}^{\alpha}(x)+\frac{c}{12}(m^3-m)\delta_{m+n,0}.
\end{equation}

Note that Virasoro generators $\ell_{n}^{\alpha}(x)$ for $n=-1,0,1$, which generate $SL(2,\mathbb{C})$ algebra, can be written as 
\begin{align}
\begin{split}
\ell_{-1}^{\alpha}(x) &=\ell_{-1}^{\textrm{Witt}}(x)
-\frac{2}{\epsilon_2}\sum_{a=1}^NV'(z_a),
\label{l_m123_g}  \\ 
\ell_{0}^{\alpha}(x) & =\ell_{0}^{\textrm{Witt}}(x)
-\frac{2\mu}{\epsilon_1\epsilon_2}(2\mu+\epsilon_1+\epsilon_2
-2\alpha)
-\frac{2}{\epsilon_2}\sum_{a=1}^Nz_aV'(z_a), \\
\ell_{1}^{\alpha}(x) &=\ell_{1}^{\textrm{Witt}}(x)
+\frac{4\mu}{\epsilon_1\epsilon_2}\alpha x
+\frac{2}{\epsilon_2}(2\mu+\epsilon_1+\epsilon_2
-\alpha)\sum_{a=1}^N z_a
-\frac{2}{\epsilon_2}\sum_{a=1}^Nz_a^2V'(z_a).
\end{split}
\end{align}
The dependence on $V'(z_a)$ can be equivalently expressed in terms of time derivatives as given in (\ref{twi_vira_g}). 
These constraints relate derivatives of the wave-function with respect to $x$ to derivatives with respect to times. Taking advantage of such relations will enable us, at least for some specific matrix models, to derive time-independent quantum curves from time-dependent ones.


\subsection{Virasoro operators as building blocks of quantum curves}

In deriving Virasoro generators $\ell_n^{\alpha}(x)$ that impose constraints (\ref{vir_const_wave_2}) we treated the wave-function $\psi_{\alpha}(x)$ simply as an $x$-deformation of the matrix model partition function (\ref{matrix_def}). It is however also very interesting to interpret the expectation value of (\ref{fermion_alpha}), in the language of conformal field theory, as the expectation value of an insertion of a local operator at position $x$. From this perspective it is natural to consider another representation of Virasoro generators acting on the wave-function (\ref{wave_Z2}), introduced as
\begin{equation}
L_n\    \psi_{\alpha}(x) =\oint_{y=x}\frac{dy}{2\pi i}(y-x)^{n+1}T(y;x) \psi_{\alpha}(x) 
= \oint_{y=x}\frac{dy}{2\pi i}(y-x)^{n+1}T_-(y;x) \psi_{\alpha}(x).   \label{L-n}
\end{equation}
In the second equality we used the fact that $T_+(y;x)$ annihilates $\psi_{\alpha}(x)$, see (\ref{t_ward_id_2}). It follows that in order to find an explicit representation of generators $L_n$, we need to identify the coefficient of $(y-x)^{n-2}$ in (\ref{T_minus_em}). We immediately get
\be
L_0\,  =\Delta_{\alpha},   
\qquad \qquad L_{-1} =\partial_x - \frac{2\alpha}{\epsilon_1\epsilon_2} V'(x), \label{Lm1_fer_op-psi}
\ee
with $\Delta_{\alpha}$ given in (\ref{Delta}); furthermore, expanding (\ref{T_minus_em}) in powers of $(y-x)$, for $n\geq 2$ we find
\be
L_{-n} = - \frac{1}{\epsilon_1\epsilon_2 (n-2)!} \Big(  \partial_x^{n-2} \big(  V'(x)^2 \big) + (\epsilon_1+\epsilon_2 + \frac{2\alpha}{n-1}) \partial_x^n V(x)  + \partial_x ^{n-2}\hat{f}(x) \Big),
\ee
where $\hat{f}(x)$ and its derivatives are defined in (\ref{sp_hat_f_op}) and (\ref{sp_hat_fn_op}).

In an analogous way we define operators $\widehat{L}_n$ acting on $\widehat{\psi}_{\alpha}(x)$
\begin{equation}
\widehat{L}_n\    \widehat{\psi}_{\alpha}(x)=\oint_{y=x}\frac{dy}{2\pi i}(y-x)^{n+1}T_-(y;x)\widehat{\psi}_{\alpha}(x).   \label{Lhat-n}
\end{equation}
These operators are closely related to $L_n$ given above: the exponential prefactor in (\ref{Psi-alpha-bis}) simply removes the term proportional to $V'(x)$ in $L_{-1}$ and results in replacing $\partial_x ^{n-2}\hat{f}(x)$ by (\ref{tree_op_trans-fhat}). Ultimately operators $\widehat{L}_{-n}$ do not depend explicitly on $\alpha$ and take form
\begin{empheq}[box=\fbox]{equation}
\begin{split}
\widehat{L}_0\,  &=\Delta_{\alpha},  \qquad  \qquad   \widehat{L}_{-1} =\partial_x ,  \label{Lm_fer_op}  \\ 
\, \widehat{L}_{-n} = - \frac{1}{\epsilon_1\epsilon_2 (n-2)!} \Big(   \partial_x^{n-2} \big( & V'(x)^2  \big)  + (\epsilon_1+\epsilon_2) \partial_x^n V(x)  + \partial_x ^{n-2}\hat{f}(x) \Big),\  \textrm{for}\ n\geq 2 \,
\end{split}
\end{empheq}
Using (\ref{hatfhatf}) and (\ref{hatfVdk}) it is straightforward to check that $\widehat{L}_{-n}$ with positive $n$ satisfy Virasoro algebra $\big[\widehat{L}_{-m}, \widehat{L}_{-n}\big]=(n-m)\widehat{L}_{-m-n}$. For first several values of $n$ these operators take form
\begin{align}
\widehat{L}_{-2}&=
-\frac{1}{\epsilon_1\epsilon_2}\Big(V'(x)^2+(\epsilon_1+\epsilon_2)V''(x)
+\hat{f} (x)\Big),
\label{Lm2_fer_op}
\\
\widehat{L}_{-3} &=
-\frac{1}{\epsilon_1\epsilon_2}\Big(2V'(x)V''(x)+(\epsilon_1+\epsilon_2)
V'''(x)
+  \partial_x\hat{f} (x)\Big).
\label{Lm3_fer_op}
\end{align}
Note that in case $V(x)$ is a polynomial of degree $m$, in the representation (\ref{Lm_fer_op}) only a finite number of Virasoro generators is non-trivial and $\widehat{L}_{-n}=0$ for $n\geq 2m+1$. The operators $\widehat{L}_{-n}$ play a prominent role in this work -- as we show in the next section, they form building blocks of higher level quantum curves.

It is also useful to consider Virasoro operators that act on the normalized wave-function (\ref{PsiPsi-alpha}). By writing $Z^{-1}\widehat{L}_{-n}\widehat{\psi}_{\alpha}(x) =Z^{-1}\widehat{L}_{-n}Z\widehat{\Psi}_{\alpha}(x) \equiv \widehat{\mathcal L}_{-n}\widehat{\Psi}_{\alpha}(x)$ these operators take form
\begin{align}
\widehat{\mathcal L}_{-n}&=Z^{-1}\widehat{L}_{-n}Z =   \label{calLn} \\
&=
-\frac{1}{\epsilon_1\epsilon_2(n-2)!}\Big(\partial_x^{n-2}\big(V'(x)^2\big)+(\epsilon_1+\epsilon_2)\partial^n_x V(x)
+\partial_x^{n-2}\widehat{f}(x) + \big[\partial_x^{n-2}\widehat{f}(x), \log Z \big]\Big).    \nonumber
\end{align}


\newpage 

\section{Quantum curves as Virasoro singular vectors}    \label{sec-quantum}

It is well known that the expectation value of a determinant $\textrm{det}(x-M)$ in a matrix model (with matrices $M$ being integrated over) satisfies a differential equation, which can be interpreted as a quantum version of the spectral curve \cite{abmodel}. Such equations, including a dependence on times (i.e. parameters of the matrix model potential), can be derived also in the  $\beta$-deformed matrix model, and can be written in a form analogous to BPZ equations for degenerate fields at level 2, see  \cite{ACDKV}. This motivates us to ask whether more general expectation values, namely $\alpha/\beta$-deformed matrix integrals $\widehat{\psi}_{\alpha}(x)$ introduced in (\ref{Psi-alpha-bis})  
for an arbitrary parameter $\alpha$, also satisfy finite order differential equations. We find the following amusing answer to this question, which is essentially the main result of this paper: 
\begin{itemize}
\item $\widehat{\psi}_{\alpha}(x)$ satisfy higher order differential equations, which can be written in terms of differential operators $\widehat{A}^{\alpha}_n$ of order $n$ in $x$ 
\be
\widehat{A}^{\alpha}_n \widehat{\psi}_{\alpha}(x) = 0,  \label{Aalpha-psialpha}
\ee
only for specific, discrete values of the parameter $\alpha=\alpha_{r,s}$ given in (\ref{alpha-rs-2}) with $n=rs$; these values of $\alpha$ correspond to the momenta of Virasoro degenerate fields,
\item differential equations (\ref{Aalpha-psialpha}) take the same form as (higher level) BPZ equations in conformal fields theory (\ref{BPZ-cft}), with $\widehat{\psi}_{\alpha}(x)$ playing a role of the correlation function, and with Virasoro operators (that enter the operator $\widetilde{A}_{r,s}$ in (\ref{BPZ-cft})) represented now by $\widehat{L}_n$ derived in (\ref{Lm_fer_op}),
\item equivalently, operators $\widehat{A}^{\alpha_{r,s}}_n$ that annihilate $\widehat{\psi}_{\alpha_{r,s}}(x)$ are in one-to-one correspondence with Virasoro singular vectors and can be written as in (\ref{Lrs}), however with Virasoro generators represented as in (\ref{Lm_fer_op}).
\end{itemize}
The above statements essentially follow from the chiral boson interpretation of the $\alpha/\beta$-deformed matrix integral discussed in section \ref{sec-loop}. It is however also useful to prove them explicitly, and in this section we  conduct this task up to level 5. We show that such an explicit derivation leads to non-trivial results: for example, for a given positive integer $n$ we find universal, $\alpha$-dependent expressions for all level $n$ quantum curves 
\be
\widehat{A}^{\alpha}_n = \sum_{p_1+p_2+\ldots+p_k=n}  \widehat{c}_{p_1,p_2,\ldots,p_k}(\alpha)\,   \widehat{L}_{-p_1} \widehat{L}_{-p_2}\cdots \widehat{L}_{-p_k},       \label{chat-Lhat}
\ee
such that upon the substitution $\alpha=\alpha_{r,s}$, for all $r$ and $s$ satisfying $n=rs$, the coefficients $\widehat{c}_{p_1,p_2,\ldots,p_k}(\alpha)$ specialize to $c^{r,s}_{p_1,p_2,\ldots,p_k}(b)$ in (\ref{Lrs}). Moreover, upon the substitution $\alpha=\alpha_{r,s}$ with $rs<n$, the expression (\ref{chat-Lhat}) factorizes into a form that contains a factor representing the correct singular vector $|\phi_{r,s}\rangle$ at level $rs$. Therefore $\widehat{c}_{p_1,p_2,\ldots,p_k}(\alpha)$ determined at level $n$ encode information about coefficients $c^{r,s}_{p_1,p_2,\ldots,p_k}(b)$ in (\ref{Lrs}) for all singular vectors $|\phi_{r,s}\rangle$ with $rs\leq n$.

For brevity, and to stress the relation to singular vectors and degenerate fields in conformal field theory, we often call differential operators $\widehat{A}^{\alpha}_n$ that annihilate $\widehat{\psi}_{\alpha_{r,s}}(x)$ as quantum curves at higher levels, or simply higher level quantum curves.


\subsection{General construction}    \label{ssec-general-construction}

To derive differential equations satisfied by $\widehat{\psi}_{\alpha}(x)$ we show that its $n$-th derivative with respect to $x$ can be related to certain constraint equations that follow from (\ref{t_ward_id_2}). More precisely, it turns out that for special values of $\alpha$ a linear combination of such constraint equations can be written as $\partial^n_x \widehat{\psi}_{\alpha}(x)$ plus other terms that only involve derivatives of lower orders. In this way we obtain a constraint equation written as a differential equation that involves $\partial^n_x \widehat{\psi}_{\alpha}(x)$ and other derivatives of lower order, which is a quantum curve equation that we are after. Furthermore, the special values of $\alpha$ for which $\partial^n_x \widehat{\psi}_{\alpha}(x)$ can be rewritten in terms of constraint equations are precisely the momenta $\alpha_{r,s}$ of degenerate fields (\ref{alpha-rs-2}), and for these values the quantum curves can be written in the form of operators that encode Virasoro singular vectors (\ref{Lrs}), with Virasoro generators represented by  (\ref{Lm_fer_op}).

Let us discuss the above statements in more detail. For a time being it will be more convenient not to include the prefactor in (\ref{Psi-alpha-bis}), and to work with $\psi_{\alpha}(x)$ instead of $\widehat{\psi}_{\alpha}(x)$. On one hand, recall that $\partial_x^n\psi^{\textrm{ins}}_{\alpha}(x)$ can be expressed as in (\ref{diff_n_psi}), i.e. as a linear combination of $\mathfrak{p}(n)$ expressions (where $\mathfrak{p}(n)$ is the number of partitions of $n$) of the form
\be
\sum_{a_1,\ldots,a_p=1}^N\frac{\psi^{\textrm{ins}}_{\alpha}(x)}{(x-z_{a_1})^{Y_1}\cdots (x-z_{a_p})^{Y_p}}, \label{psi-Y1Yp}
\ee
where coefficients of this linear combination are given by integers $C_{Y_1,Y_2,\ldots,Y_p}$ (labeled by partitions $(Y_1,\ldots,Y_p)$ of $n$) multiplied by $\alpha^p\epsilon_2^{-p}$. In what follows, by the rank of an expression of the form (\ref{psi-Y1Yp}) we understand the sum of exponents of singular terms $\frac{1}{x-z_{a_i}}$. In particular, the rank of (\ref{psi-Y1Yp}), with $\sum_i Y_i=n$, is equal to $n$. We also define a rank of a linear combination of the terms of the form (\ref{psi-Y1Yp}) as the largest rank of all terms involved in this combination; the terms with the largest rank in this combination are also referred to as the most singular terms.

On the other hand we consider the loop equation for the $\alpha/\beta$-deformed matrix integral written in the form (\ref{t_ward_id_2}), where $T_+(y;x)$ is given in (\ref{T_plus_em}). This loop equation depends on a parameter $y$, therefore expanding it in powers of $y$ we obtain an infinite set of constraint equations. Let us introduce
\begin{equation}
T_{+}^{(k)}(x)=\frac{(-1)^k}{(k-2)!}\frac{\partial^{k-2}}{\partial y^{k-2}}
T_{+}(y;x)\Big|_{y=x},\ \ \ \
k\ge 2,    \label{T+derivatives}
\end{equation}
in terms of which an infinite number of constraints can be written in the form
\begin{equation}
\big<T_{+}^{(k)}(x)\psi^{\textrm{ins}}_{\alpha}(x)\big>=0.
\label{t_ward_id_3}
\end{equation}
Furthermore, we consider additional constraints that arise from acting with higher order loop insertion operators $\partial_{V(x)}^{(k)}$ (introduced in (\ref{loop-insertion-k})) on (\ref{t_ward_id_3})
\begin{equation}
\left(\partial_{V(x)}^{(1)}\right)^{k_1}\left(\partial_{V(x)}^{(2)}\right)^{k_2}
\cdots \left(\partial_{V(x)}^{(p)}\right)^{k_p}\big<T_{+}^{(k)}(x)\psi^{\textrm{ins}}_{\alpha}(x)\big>=0.
\label{t_ward_id_lv}
\end{equation}
From the representation of $T_+(y;x)$ given in the last line of (\ref{T_plus_em}), as well as the action of the higher order loop insertion operators given in (\ref{loop-insertion-bis}) and (\ref{loop-insertion-bis-V}), it follows that (\ref{t_ward_id_lv}) can be expressed as the expectation value of expressions that involve terms of the form (\ref{psi-Y1Yp}) with $\sum_i Y_i \leq k+\sum_{i=1}^p ik_i$, and other terms that involve derivatives of the matrix model potential. Let us now consider a linear combination of the form 
\begin{align}
\sum_{p=0}^{n-2}\sum_{
\begin{subarray}{c}\mu_1\ge \mu_2\ge \ldots \ge \mu_p \ge 0\\\mu_1+\ldots+\mu_p=p\end{subarray}
}c_{\mu_1,\mu_2,\ldots,\mu_p}
\partial_{V(x)}^{(\mu_1)}\partial_{V(x)}^{(\mu_2)}
\cdots \partial_{V(x)}^{(\mu_p)}T_{+}^{(n-p)}(x)\psi^{\textrm{ins}}_{\alpha}(x),
\label{ward_comb}
\end{align}
with coefficients $c_{\mu_1,\mu_2,\ldots,\mu_p}$. The rank of this expression is also $n$. We claim that one can adjust the coefficients $c_{\mu_1,\mu_2,\ldots,\mu_p}$ (and also $\alpha$) in such a way, that the most singular terms in (\ref{ward_comb}) altogether reproduce (\ref{diff_n_psi}). Therefore, with such a choice of $c_{\mu_1,\mu_2,\ldots,\mu_p}$ and $\alpha$, the expression (\ref{ward_comb}) can be written as $\partial_x^n\psi^{\textrm{ins}}_{\alpha}(x)$ plus terms of lower rank that are also uniquely specified. Altogether, after recalling that expectation values of all terms in the combination (\ref{ward_comb}) vanish (i.e. all those terms give rise to constraint equations), we obtain a differential equation in $x$ of order $n$  for various expectation values of $\psi^{\textrm{ins}}_{\alpha}(x)$, that also include derivatives of the matrix model potential. Moreover, by modifying (\ref{ward_comb}) by further constraints of lower rank that involve the action of $\partial^k_x V(x)$ or $\partial^k_x\widehat{f}(x)$ on (\ref{T+derivatives}), and adjusting appropriately coefficients of those additional constraints, all derivatives of the potential can be represented in terms of time derivatives. Therefore ultimately we obtain a time-dependent, order $n$ differential equation in $x$ for $\langle\psi^{\textrm{ins}}_{\alpha}(x)\rangle = \psi_{\alpha}(x)$, which can be simply rewritten as a time-dependent quantum curve equation for $\widehat{\psi}_{\alpha}(x)$. 

It is however not obvious that it is possible to adjust the coefficients $c_{\mu_1,\mu_2,\ldots,\mu_p}$ and $\alpha$ in (\ref{ward_comb}) to reproduce (\ref{diff_n_psi}). First of all, note that the number of coefficients $c_{\mu_1,\mu_2,\ldots,\mu_p}$ is equal to $\sum_{p=0}^{n-2}\mathfrak{p}(p)$, and together with an indeterminate momentum $\alpha$ the combination (\ref{ward_comb}) depends on $1+ \sum_{p=0}^{n-2}\mathfrak{p}(p)$ parameters. On the other hand, there are $\mathfrak{p}(n)$ coefficients $C_{Y_1,Y_2,\ldots,Y_p}$ in (\ref{diff_n_psi}). We show below that 
\be
1+ \sum_{p=0}^{n-2}\mathfrak{p}(p) \geq \mathfrak{p}(n),  \qquad n\geq 2,    \label{equations-cC}
\ee
so the number of constraint equations is sufficient to determine $C_{Y_1,Y_2,\ldots,Y_p}$ in terms of $c_{\mu_1,\mu_2,\ldots,\mu_p}$. Moreover, even though (\ref{equations-cC}) is in general an inequality, in fact -- except for the terms with smaller rank -- not all constraint equations are independent, and we claim that the number of independent constraints is precisely sufficient to adjust indeterminates $c_{\mu_1,\mu_2,\ldots,\mu_p}$ and $\alpha$ in a way that reproduces (\ref{diff_n_psi}). Moreover, to adjust these indeterminates we need to solve a system of equations which has a unique solution for $c_{\mu_1,\mu_2,\ldots,\mu_p}$, and several solutions for $\alpha$ which take form of the degenerate momenta $\alpha_{r,s}$ with $rs=n$, given in (\ref{alpha-rs-2}). This is how all degenerate momenta at a given level $n$ arise naturally in this construction. In the rest of this section we illustrate the above statements in explicit examples and  construct higher level quantum curves up to level $n=5$.

Before we discuss explicit examples of higher level quantum curves, let us prove that the inequality (\ref{equations-cC}) indeed holds (note that for $n=2,3,4$ this is the equality). To this end consider a generating function
\begin{equation}
g(x)=\sum_{n=2}^{\infty} \Big(\sum_{p=0}^{n-2}\mathfrak{p}(p)+1-\mathfrak{p}(n)\Big) x^n
=\frac{1}{1-x}\sum_{n=2}^{\infty}
\big(\mathfrak{p}(n-2)+\mathfrak{p}(n-1)-\mathfrak{p}(n)\big)x^n.
\end{equation}
By the recurrence formula
\begin{align}
\begin{split}
\mathfrak{p}(n)&=\sum_{k\in{\IZ}\backslash\{0\}}
(-1)^{k-1}\mathfrak{p}\big(n-\frac{k}{2}(3k-1)\big) =
\\
&=
\mathfrak{p}(n-1)+\mathfrak{p}(n-2)-\mathfrak{p}(n-5)-\mathfrak{p}(n-7)+\mathfrak{p}(n-12)+\ldots
\end{split}
\end{align}
where $\mathfrak{p}(n)=0$ for $n<0$, we see that $\mathfrak{p}(n-2)+\mathfrak{p}(n-1)-\mathfrak{p}(n)\ge 0$. Therefore the coefficients of the series expansion of $g(x)$ around $x=0$ form an increasing sequence of positive integers, and the inequality (\ref{equations-cC}) follows.


\subsection{Quantum curve at level 1}    \label{ssec-level1}

In contrast to higher levels, a quantum curve at level 1 takes a very simple form. At level 1 the value of the degenerate momentum is zero, which corresponds to $r=s=1$ in the expression (\ref{alpha-rs-2}), and in this case the wave-function (\ref{Psi-alpha-bis}) reduces to the partition function (\ref{matrix_def}) , $\widehat{\psi}_{\alpha=0}(x)=Z$. In Virasoro algebra a singular vector (\ref{singular-vect}) at level 1 takes form $L_{-1}|0\rangle$. In matrix model representation the partition function $Z$ is identified with the vacuum state $|0\rangle$ and $\widehat{L}_{-1}=\partial_x$, therefore the quantum curve equation at level 1 takes form
\be
\widehat{A}^{\alpha=0}_{1}  \widehat{\psi}_{\alpha=0}(x) = \widehat{L}_{-1}  \widehat{\psi}_{\alpha=0}(x) = \partial_x Z = 0,
\ee
which is a statement that the matrix model partition function $Z$ does not depend on $x$.


\subsection{Quantum curves at level 2}    \label{ssec-level2}

Let us consider quantum curves at level 2. At this level we expect to find second order differential equations for the wave-function, therefore we consider the second derivative of $\psi^{\textrm{ins}}_{\alpha}(x)$ which takes form (\ref{diff_psi_2}).
Following the construction presented in section \ref{ssec-general-construction} we would like to relate this expression to one (in this case) constraint equation 
\begin{equation}
\big<T_{+}^{(2)}(x)\psi^{\textrm{ins}}_{\alpha}(x)\big>=0,
\label{T_p_0}
\end{equation}
where
$$
T_{+}^{(2)}(x) = T_+(x;x) = 
\frac{2\alpha+\epsilon_1+\epsilon_2}{\epsilon_2}\sum_{a=1}^N\frac{1}{(x-z_a)^2}
-\frac{\epsilon_1}{\epsilon_2}\sum_{a,b=1}^N\frac{1}{(x-z_a)(x-z_b)}
-\frac{2}{\epsilon_2}\sum_{a=1}^N\frac{V'(z_a)}{x-z_a}.
$$
Note that the first and the second term in  $c_1T_{+}^{(2)}(x)\psi^{\textrm{ins}}_{\alpha}(x)$ combine into $\partial_x^2\psi^{\textrm{ins}}_{\alpha}(x)$ in (\ref{diff_psi_2}) if and only if
\begin{equation}
\frac{2\alpha+\epsilon_1+\epsilon_2}{\epsilon_2}c_1=\frac{2\alpha}{\epsilon_2},\qquad
-\frac{\epsilon_1}{\epsilon_2}c_1=\frac{4\alpha^2}{\epsilon_2^2}.
\end{equation}
These equations have a solution only for  $c_1=-\frac{4\alpha^2}{\epsilon_1\epsilon_2}$ and either of the two values of $\alpha$
\be
\boxed{\quad \alpha=-\frac{\epsilon_1}{2}, -\frac{\epsilon_2}{2}, \quad }
\ee
and these values of $\alpha$ are precisely the degenerate momenta (\ref{alpha-rs-2}) at level 2. With the above special values of $c_1$ and $\alpha$, the equation  $c_1\big<T_{+}^{(2)}(x)\psi^{\textrm{ins}}_{\alpha}(x)\big>=0$ can be rewritten as
\begin{equation}
\Big(\partial_x^2-\frac{4\alpha}{\epsilon_1\epsilon_2}V'(x)\partial_x
-\frac{4\alpha^2}{\epsilon_1^2\epsilon_2^2}\widehat{f}(x)\Big)\left<\psi^{\textrm{ins}}_{\alpha}(x)\right>=0,\ \ \
\textrm{for}\ \alpha=-\frac{\epsilon_1}{2}, -\frac{\epsilon_2}{2},   \label{BPZ-level-2-psi}
\end{equation}
where $\widehat{f}(x)$ is the differential operator defined in (\ref{sp_hat_f_op}). 
By including the tree term introduced in (\ref{fermion_alpha_2}) we obtain differential equations for $\widehat{\psi}_{\alpha}(x)$
\begin{equation}
\boxed{ \
\Big(\partial_x^2-\frac{4\alpha^2}{\epsilon_1^2\epsilon_2^2}V'(x)^2-\frac{4\alpha^2}{\epsilon_1^2\epsilon_2^2}(\epsilon_1+\epsilon_2)V''(x)
-\frac{4\alpha^2}{\epsilon_1^2\epsilon_2^2}\widehat{f}(x)\Big)
\widehat{\psi}_{\alpha}(x)=0,\ \ \
\textrm{for}\ \alpha=-\frac{\epsilon_1}{2}, -\frac{\epsilon_2}{2}. \ }   \label{BPZ-level-2bis}
\end{equation}
Furthermore, using $\widehat{L}_{-1}=\partial_x$ and the representation of $\widehat{L}_{-2}$ given in  (\ref{Lm2_fer_op}), these differential equations can be rewritten in the form of BPZ equations
\begin{equation}
\boxed{\ 
\widehat{A}_{2}^{\alpha}\widehat{\psi}_{\alpha}(x)  = 0, \qquad 
\widehat{A}_{2}^{\alpha} = \widehat{L}_{-1}^2+\frac{4\alpha^2}{\epsilon_1\epsilon_2}\widehat{L}_{-2},\ \ \
\textrm{for}\ \alpha=-\frac{\epsilon_1}{2}, -\frac{\epsilon_2}{2}, \ }
\label{BPZ-level-2}
\end{equation}
and the operators $\widehat{A}_{2}^{\alpha}$ for the above choice of $\alpha$ are identified as the quantum curves we have been after. At this level, a variant of the above calculation was originally conducted in \cite{ACDKV}. Note that substituting the values $\alpha=-\frac{\epsilon_1}{2}, -\frac{\epsilon_2}{2}$,  the operator $\widehat{A}_{2}^{\alpha}$ specializes respectively to $(\widehat{L}_{-1}^2+b^{\pm 2}\widehat{L}_{-2})$ with $b^2=\frac{\epsilon_1}{\epsilon_2}$, which have form of $A_{2,1}$ in (\ref{null-level234}) or $A_{1,2}$ (with analogous form, however with $b$ replaced by $b^{-1}$), and encode Virasoro singular vectors at level 2. Equivalently, for these values of $\alpha$ the equations (\ref{BPZ-level-2}) take the same form as BPZ equations at level 2
$$
(\widehat{L}_{-1}^2+b^{\pm 2}\widehat{L}_{-2})\widehat{\psi}_{\alpha}(x)=0.
$$


\subsection{Quantum curves at level 3}   \label{ssec-level3}

We derive now quantum curves at level 3. Recall that the third derivative of $\psi^{\textrm{ins}}_{\alpha}(x)$ takes form (\ref{diff_psi_3}).
We should relate this expression to two constraint equations (at this level) of the form (\ref{t_ward_id_lv})
 \be
\big<T_{+}^{(3)}(x)\psi^{\textrm{ins}}_{\alpha}(x)\big> = 0,
\qquad
\partial_{V(x)}^{(1)}\big<T_{+}^{(2)}(x)\psi^{\textrm{ins}}_{\alpha}(x)\big> = 0,
\label{T_p_12}
\ee
where
\begin{align}
T_{+}^{(3)}(x)&=
\frac{2(\alpha+\epsilon_1+\epsilon_2)}{\epsilon_2}\sum_{a=1}^N\frac{1}{(x-z_a)^3}
-\frac{2\epsilon_1}{\epsilon_2}\sum_{a,b=1}^N\frac{1}{(x-z_a)^2(x-z_b)}
-\frac{2}{\epsilon_2}\sum_{a=1}^N\frac{V'(z_a)}{(x-z_a)^2}, \nonumber \\
\partial_{V(x)}^{(1)}T_{+}^{(2)}(x)&=
\sum_{a=1}^N\frac{1}{(x-z_a)^3}
+\frac{2\alpha+\epsilon_1+\epsilon_2}{\epsilon_2}\sum_{a,b=1}^N\frac{1}{(x-z_a)^2(x-z_b)} +
\nonumber\\
&\ \
-\frac{\epsilon_1}{\epsilon_2}\sum_{a,b,c=1}^N\frac{1}{(x-z_a)(x-z_b)(x-z_c)}
-\frac{2}{\epsilon_2}\sum_{a,b=1}^N\frac{V'(z_a)}{(x-z_a)(x-z_b)}.  \nonumber
\end{align}
It turns out that the most singular terms in the linear combination of these constraints
\begin{equation}
T_{+}^{(3,1)}(x)\psi^{\textrm{ins}}_{\alpha}(x) =
\left(c_1T_{+}^{(3)}(x)+c_2\partial_{V(x)}^{(1)}T_{+}^{(2)}(x)\right)\psi^{\textrm{ins}}_{\alpha}(x)
\end{equation}
reproduce the right hand side of (\ref{diff_psi_3}) if and only if 
$$
c_1=\frac{2\alpha^2\big(2\alpha(2\alpha+\epsilon_1+\epsilon_2)+3\epsilon_1\epsilon_2\big)}{\epsilon_1^2\epsilon_2^2},\ \ \ \
c_2=\frac{8\alpha^3}{\epsilon_1\epsilon_2^2},
$$
and for $\alpha$ taking one of the following values
\be
\boxed{\ \alpha=-\frac{\epsilon_1}{2}, -\frac{\epsilon_2}{2}, -\epsilon_1, -\epsilon_2, \ }
\label{c1_c2_alpha_3}
\ee
which are precisely the values of degenerate momenta (\ref{alpha-rs-2}) at levels 2 and 3. Specializing to these values of $c_1$, $c_2$ and $\alpha$, the operator $T_{+}^{(3,1)}(x)$ acting on $\psi^{\textrm{ins}}_{\alpha}(x)$ can be written as
\be
\begin{split}
T_{+}^{(3,1)}(x)&=
\partial_x^3-\frac{2}{\epsilon_2}V'(x)\left(c_1\sum_{a=1}^N\frac{1}{(x-z_a)^2}+c_2\sum_{a,b=1}^N\frac{1}{(x-z_a)(x-z_b)}\right)   +  \\
&\ \
+\frac{2}{\epsilon_2}\left(c_1\sum_{a=1}^N\frac{V'(x)-V'(z_a)}{(x-z_a)^2}+c_2\sum_{a,b=1}^N\frac{V'(x)-V'(z_a)}{(x-z_a)(x-z_b)}\right).
\label{Tp_c12_a}
\end{split}
\ee
However, this expression does not yet provide the equation that we are after -- our aim is to identify a differential equation that can be written entirely in terms of operators represented by time derivatives. To this end we consider 
\begin{equation}
T_{+}^{(3,2)}(x)=T_{+}^{(3,1)}(x)-\frac{2}{\epsilon_2}c_3V'(x)T_{+}^{(2)}(x),
\end{equation}
and realize that for $c_3=-4\alpha^3/(\epsilon_1^2\epsilon_2)$ and values of $\alpha$ given in (\ref{c1_c2_alpha_3}) it is possible to bring all derivatives of the potential that arise in this expression into a form which appears in (\ref{f-cl}) and (\ref{dxfx}). In consequence we can represent derivatives of the potential in $T_{+}^{(3,2)}(x)$ by means of operators $\widehat{f}(x)$ and $\partial_x\widehat{f}(x)$ defined in (\ref{sp_hat_f_op}) and (\ref{sp_hat_fn_op}), and we obtain a third order partial differential equation
\begin{equation}
T_{+}^{(3,2)}(x)\left<\psi^{\textrm{ins}}_{\alpha}(x)\right>=0,\ \ \
\textrm{for}\ \alpha=-\frac{\epsilon_1}{2}, -\frac{\epsilon_2}{2}, -\epsilon_1, -\epsilon_2,
\end{equation}
where
\be
\begin{split}
T_{+}^{(3,2)}(x)&=
\partial_x^3-\frac{6\alpha}{\epsilon_1\epsilon_2}V'(x)\partial_x^2
+\frac{8\alpha^2}{\epsilon_1^2\epsilon_2^2}V'(x)^2\partial_x
+\frac{8\alpha^3}{\epsilon_1^3\epsilon_2^3}V'(x)\widehat{f}(x) +   \\
&\ \
-\frac{c_1}{\epsilon_1\epsilon_2} \partial_x\widehat{f}(x) 
-\frac{c_1}{\alpha}V''(x)\partial_x
-\frac{c_2}{2\epsilon_1\alpha}\widehat{f}(x)\partial_x.
\end{split}
\ee 
Finally, by including the tree term introduced in (\ref{fermion_alpha_2}), we obtain a differential equation for $\widehat{\psi}_{\alpha}(x)$, which can be written as
\begin{equation}
\boxed{\ \widehat{A}_3^{\alpha}\widehat{\psi}_{\alpha}(x)=0,\qquad
\textrm{for}\ \alpha=-\frac{\epsilon_1}{2}, -\frac{\epsilon_2}{2}, -\epsilon_1, -\epsilon_2 \ } \label{BPZ_level_3bis}
\end{equation}
where
\begin{align}
\widehat{A}_3^{\alpha}&
=
\partial_x^3-\frac{4\alpha^2}{\epsilon_1^2\epsilon_2^2}\Big(V'(x)^2+(\epsilon_1+\epsilon_2)V''(x)
+\widehat{f}(x)\Big)\partial_x +
\nonumber\\
&\ \ \ 
-\frac{2\alpha^2}{\epsilon_1^3\epsilon_2^3}\big(2\alpha(2\alpha+\epsilon_1+\epsilon_2)+3\epsilon_1\epsilon_2\big)
\big(2V'(x)V''(x)+(\epsilon_1+\epsilon_2)V'''(x)\big) +
\nonumber\\
&\ \ \ 
-\frac{2\alpha^2}{\epsilon_1^3\epsilon_2^3}\big(2\alpha(2\alpha+\epsilon_1+\epsilon_2)+3\epsilon_1\epsilon_2\big)
\partial_x\widehat{f}(x).
\end{align} 
Furthermore, using the representation (\ref{Lm_fer_op}), for values of $\alpha$ given in (\ref{c1_c2_alpha_3}) the operator $\widehat{A}_3^{\alpha}$ can be written as
\begin{empheq}[box=\fbox]{equation}
\begin{split}
\ \widehat{A}_3^{\alpha} &=
\widehat{L}_{-1}^3+\frac{4\alpha^2}{\epsilon_1\epsilon_2}\widehat{L}_{-2}\widehat{L}_{-1}
+\frac{2\alpha^2}{\epsilon_1^2\epsilon_2^2}\big(2\alpha(2\alpha+\epsilon_1+\epsilon_2)+3\epsilon_1\epsilon_2\big)\widehat{L}_{-3} = \  \label{BPZ-level-3} \\ 
&  = \widehat{L}_{-1}\widehat{A}_2^{\alpha}
+\frac{2\alpha^2}{\epsilon_1^2\epsilon_2^2}(2\alpha+\epsilon_1)(2\alpha+\epsilon_2)\widehat{L}_{-3}
\end{split}
\end{empheq}
where $\widehat{A}_2^{\alpha}$ is defined in (\ref{BPZ-level-2}), and to write the expression in the second line we used the relation $\big[\widehat{L}_{-1}, \widehat{L}_{-2}\big]=\widehat{L}_{-3}$. Amusingly, this identification of $\widehat{A}_3^{\alpha}$ captures simultaneously all singular vectors up to level 3; or equivalently, (\ref{BPZ_level_3bis}) takes form of BPZ equations for all degenerate fields up to level 3. Indeed, singular vectors at level 3 correspond to values $\alpha=-\epsilon_1,\epsilon_2$, for which (\ref{BPZ-level-3})  takes form of either $A_{3,1}$ given in (\ref{null-level234}), or $A_{1,3}$ of analogous form, however with $b$ replaced by $b^{-1}$. On the other hand, for $\alpha=-\epsilon_1/2,\epsilon_2/2$ that correspond to singular vectors at level 2, the second term in the second line of (\ref{BPZ-level-3})  vanishes, so that $\widehat{A}_3^{\alpha}$ factorizes and essentially reduces to the action of the operator $\widehat{A}_2^{\alpha}$, that indeed encodes singular vectors at level 2, see (\ref{BPZ-level-2}). 

We stress that the identification of the operator $\widehat{A}_3^{\alpha}$ and its $\alpha$-dependent coefficients is not restricted to our matrix model representation -- replacing $\widehat{L}_{-n}$ by abstract Virasoro operators (\ref{L-Virasoro}), we can interpret (\ref{BPZ-level-3}) as an (abstract) operator that encodes all Virasoro singular vectors up to level 3, upon the substitution of relevant values of $\alpha$ given in (\ref{c1_c2_alpha_3}).


\subsection{Quantum curves at level 4}   \label{ssec-level4}

To derive quantum curves at level 4 we relate the fourth derivative $\partial^4_x \psi^{\textrm{ins}}_{\alpha}(x)$, which takes form (\ref{diff_n_psi}), to the constraint equations at level 4
\begin{align}
\begin{split}
\big<T_{+}^{(4)}(x)\psi^{\textrm{ins}}_{\alpha}(x)\big>&=0,
\qquad
\partial_{V(x)}^{(1)}\big<T_{+}^{(3)}(x)\psi^{\textrm{ins}}_{\alpha}(x)\big>=0,
\\
\partial_{V(x)}^{(2)}\big<T_{+}^{(2)}(x)\psi^{\textrm{ins}}_{\alpha}(x)\big>&=0,
\qquad
\left(\partial_{V(x)}^{(1)}\right)^2\big<T_{+}^{(2)}(x)\psi^{\textrm{ins}}_{\alpha}(x)\big>=0,
\label{T_p_level4}
\end{split}
\end{align}
where
\begin{align}
T_{+}^{(4)}(x)&=
\frac{2\alpha+3(\epsilon_1+\epsilon_2)}{\epsilon_2}\sum_{a=1}^N\frac{1}{(x-z_a)^4}
-\frac{2\epsilon_1}{\epsilon_2}\sum_{a,b=1}^N\frac{1}{(x-z_a)^3(x-z_b)} +  \\
&\ \
-\frac{\epsilon_1}{\epsilon_2}\sum_{a,b=1}^N\frac{1}{(x-z_a)^2(x-z_b)^2}
-\frac{2}{\epsilon_2}\sum_{a=1}^N\frac{V'(z_a)}{(x-z_a)^3},
\nonumber \\
\partial_{V(x)}^{(1)}T_{+}^{(3)}(x)&=
\sum_{a=1}^N\frac{1}{(x-z_a)^4}
+\frac{2(\alpha+\epsilon_1+\epsilon_2)}{\epsilon_2}\sum_{a,b=1}^N\frac{1}{(x-z_a)^3(x-z_b)} +
 \\
&\ \
-\frac{2\epsilon_1}{\epsilon_2}\sum_{a,b,c=1}^N\frac{1}{(x-z_a)^2(x-z_b)(x-z_c)}
-\frac{2}{\epsilon_2}\sum_{a,b=1}^N\frac{V'(z_a)}{(x-z_a)^2(x-z_b)},
\nonumber \\
\partial_{V(x)}^{(2)}T_{+}^{(2)}(x)&=
\sum_{a=1}^N\frac{2}{(x-z_a)^4}
+\frac{2\alpha+\epsilon_1+\epsilon_2}{\epsilon_2}\sum_{a,b=1}^N\frac{1}{(x-z_a)^2(x-z_b)^2}  + \\
&\ \
-\frac{\epsilon_1}{\epsilon_2}\sum_{a,b,c=1}^N\frac{1}{(x-z_a)^2(x-z_b)(x-z_c)}
-\frac{2}{\epsilon_2}\sum_{a,b=1}^N\frac{V'(z_a)}{(x-z_a)(x-z_b)^2}, \nonumber
\end{align}
and
\begin{align}
\left(\partial_{V(x)}^{(1)}\right)^2T_{+}^{(2)}(x)&=
\sum_{a,b=1}^N\frac{2}{(x-z_a)^3(x-z_b)}
+\frac{2\alpha+\epsilon_1+\epsilon_2}{\epsilon_2}\sum_{a,b,c=1}^N\frac{1}{(x-z_a)^2(x-z_b)(x-z_c)}  +  \nonumber  \\
&\hspace{-4em}
-\frac{\epsilon_1}{\epsilon_2}\sum_{a,b,c,d=1}^N\frac{1}{(x-z_a)(x-z_b)(x-z_c)(x-z_d)}
-\frac{2}{\epsilon_2}\sum_{a,b,c=1}^N\frac{V'(z_a)}{(x-z_a)(x-z_b)(x-z_c)}.  
\end{align}
Now we find that $\partial^4_x \psi^{\textrm{ins}}_{\alpha}(x)$ arises as the most singular term in a linear combination
\begin{equation}
T_{+}^{(4,1)}(x)\psi^{\textrm{ins}}_{\alpha}(x)=
\Big(c_1T_{+}^{(4)}(x)+c_2\partial_{V(x)}^{(1)}T_{+}^{(3)}(x)+c_3\partial_{V(x)}^{(2)}T_{+}^{(2)}(x)+c_4\left(\partial_{V(x)}^{(1)}\right)^2T_{+}^{(2)}(x)\Big)\psi^{\textrm{ins}}_{\alpha}(x)
\end{equation}
only for a specific choice of $c_1,c_2,c_3$ and $c_4$ and, amusingly, only for specific values of $\alpha$
\begin{equation}
\boxed{ \ \alpha=-\frac{\epsilon_1}{2}, -\frac{\epsilon_2}{2}, -\epsilon_1, -\epsilon_2, 
-\frac{3\epsilon_1}{2}, -\frac{3\epsilon_2}{2}, -\frac{\epsilon_1+\epsilon_2}{2} = \alpha_{r,s},\qquad 2\leq rs\leq 4.\  }
\label{level-4-s-mom}
\end{equation}
Again, these are values of all degenerate momenta (\ref{alpha-rs-2}) up to level 4; in particular the value $\alpha_{2,2}=-(\epsilon_1+\epsilon_2)/2$ corresponds to the choice $r=s=2$. To obtain an equation for $\psi_{\alpha}(x)$ written in terms of time-derivatives, we additionally consider the following combination of constraint equations
\begin{align}
T_{+}^{(4,2)}(x)&=T_{+}^{(4,1)}(x)-\frac{2}{\epsilon_2}V'(x)
\Big(c_5T_{+}^{(3)}(x)+c_6\partial_{V(x)}^{(1)}T_{+}^{(2)}(x)\Big)+\frac{4c_5}{\epsilon_2^2}c_7V'(x)^2T_{+}^{(2)}(x) +
\nonumber\\
&\ \
+\frac{2}{\epsilon_2}c_8V''(x)T_{+}^{(2)}(x)+\frac{c_9}{\epsilon_1\epsilon_2}\widehat{f}(x)T_{+}^{(2)}(x),
\end{align}
and adjust constants $c_5,\ldots,c_9$ in a way that brings terms of the form (\ref{psi-Y1Yp}) into combinations given in  (\ref{diff_psi_1}), (\ref{diff_psi_2}) and (\ref{diff_psi_3}), and simultaneously combines derivatives of the potential into the form (\ref{f-cl}) and (\ref{dxfx}). It turns out that it can be achieved for a unique choice of $c_5,\ldots,c_9$, and again for all values of $\alpha$ (\ref{level-4-s-mom}) corresponding to degenerate momenta up to level 4; with this choice of parameters we obtain the fourth order partial differential equation
\begin{equation}
T_{+}^{(4,2)}(x)\left<\psi^{\textrm{ins}}_{\alpha}(x)\right>=0,\qquad
\textrm{for $\alpha$ given in (\ref{level-4-s-mom}). }
\end{equation}
Including the tree term introduced in (\ref{fermion_alpha_2}), after some algebra, this equation can be written as a time-dependent quantum curve equation at level 4
\begin{equation}
\boxed{\   \widehat{A}_4^{\alpha}\widehat{\psi}_{\alpha}(x)=0,
\qquad
\textrm{for $\alpha$ given in (\ref{level-4-s-mom}) } } 
 \label{BPZ-level-4}
\end{equation}
and in terms of $\widehat{A}_2^{\alpha}$ and $\widehat{A}_3^{\alpha}$ given respectively in (\ref{BPZ-level-2}) and (\ref{BPZ-level-3}), $\widehat{A}_4^{\alpha}$ takes form
\begin{empheq}[box=\fbox]{equation}
\begin{split}
\ \widehat{A}_4^{\alpha}&=
\widehat{L}_{-1}\widehat{A}_3^{\alpha}
+\frac{4\alpha(\alpha+\epsilon_1)(\alpha+\epsilon_2)}{\epsilon_1\epsilon_2(5\alpha+3\epsilon_1+3\epsilon_2)}\widehat{L}_{-2}\widehat{A}_2^{\alpha}\, + \label{BPZ-level-4c}  \\  
&\ \
-\frac{2\alpha(2\alpha+\epsilon_1)(2\alpha+\epsilon_2)(\alpha+\epsilon_1)(\alpha+\epsilon_2)}
{\epsilon_1^3\epsilon_2^3(5\alpha+3\epsilon_1+3\epsilon_2)}
\Big(\epsilon_1\epsilon_2\widehat{L}_{-1}\widehat{L}_{-3}
-2(2\alpha+\epsilon_1)(2\alpha+\epsilon_2)\widehat{L}_{-4}\Big).  \ 
\end{split}
\end{empheq}
This equation indeed specializes to the form analogous to the structure of Virasoro singular vectors upon the substitution of values of $\alpha$ given in (\ref{level-4-s-mom}). Setting $b^2=\epsilon_1/\epsilon_2$, for $\alpha=-\frac{3\epsilon_1}{2}$  the operator $\widehat{A}_4^{\alpha}$ takes form of the operator $A_{4,1}$ given  in (\ref{null-level234}), and for $\alpha=-\frac{3\epsilon_2}{2}$ we obtain an analogous expression (with $b$ replaced by $b^{-1}$), of the form of $A_{1,4}$. Amusingly, for $\alpha=\alpha_{2,2}=-\frac{\epsilon_1+\epsilon_2}{2}$, the operator $\widehat{A}_4^{\alpha}$ has the structure of the additional singular vector at level 4 given by (\ref{null-level4-mixed}). Furthermore, for values $\alpha=-\epsilon_1,-\epsilon_2$ the expression (\ref{BPZ-level-4c}) reduces simply to $\widehat{L}_{-1}\widehat{A}_3^{\alpha}$, whose non-trivial part is given by an operator (\ref{BPZ-level-3}) that encodes singular vectors up to level 3. Finally, for $\alpha=-\frac{\epsilon_1}{2},-\frac{\epsilon_2}{2}$ the second line of (\ref{BPZ-level-4c}) drops out, and from the second line of (\ref{BPZ-level-3}) it follows that altogether (\ref{BPZ-level-4c}) factorizes to a form with a non-trivial factor being simply $\widehat{A}_2^{\alpha}$, so it indeed encodes singular vectors at level 2, see (\ref{BPZ-level-2}).


\subsection{Quantum curves at level 5}

To find quantum curves at level 5 we consider a linear combination of $\sum_{p=0}^{3}\mathfrak{p}(p)=7$ constraint equations
\begin{align}
0&=
\big<T_{+}^{(5)}(x)\psi^{\textrm{ins}}_{\alpha}(x)\big>=\partial_{V(x)}^{(1)}\big<T_{+}^{(4)}(x)\psi^{\textrm{ins}}_{\alpha}(x)\big>
=\partial_{V(x)}^{(2)}\big<T_{+}^{(3)}(x)\psi^{\textrm{ins}}_{\alpha}(x)\big> =
\nonumber\\
&
=\left(\partial_{V(x)}^{(1)}\right)^2\big<T_{+}^{(3)}(x)\psi^{\textrm{ins}}_{\alpha}(x)\big>
=\partial_{V(x)}^{(3)}\big<T_{+}^{(2)}(x)\psi^{\textrm{ins}}_{\alpha}(x)\big>
=\partial_{V(x)}^{(1)}\partial_{V(x)}^{(2)}\big<T_{+}^{(2)}(x)\psi^{\textrm{ins}}_{\alpha}(x)\big> =
\nonumber\\
&
=\left(\partial_{V(x)}^{(1)}\right)^3\big<T_{+}^{(2)}(x)\psi^{\textrm{ins}}_{\alpha}(x)\big>.
\end{align}
It turns out that -- except for the terms with smaller rank -- only 6 of these constraints are independent, therefore there are 7 parameters that we can adjust (i.e. 6 coefficients in a linear combination of independent constraints and the value of $\alpha$), in order to match $\mathfrak{p}(5)=7$ coefficients $C_{Y_1,Y_2,\ldots,Y_p}$ in a derivative $\partial_x^5 \psi^{\textrm{ins}}_{\alpha}(x)$, which takes form (\ref{diff_n_psi}). In this way we obtain a system of equations which, as usual, has a unique solution for the coefficients in the linear combination of constraints, and several possible solutions for $\alpha$ that now take form of degenerate momenta (\ref{alpha-rs-2}) up to level 5
\begin{equation}
\boxed{ \ \alpha=-\frac{\epsilon_1}{2}, -\frac{\epsilon_2}{2}, -\epsilon_1, -\epsilon_2, 
-\frac{3\epsilon_1}{2}, -\frac{3\epsilon_2}{2}, -\frac{\epsilon_1+\epsilon_2}{2}, 
-2\epsilon_1, -2\epsilon_2=\alpha_{r,s}, \qquad 2\leq r s \leq 5. \  }
\label{level_5_s_mom}
\end{equation}
After some algebra, we find that the quantum curve equations at level 5 take form
\begin{empheq}[box=\fbox]{equation}
\begin{split}
\ \widehat{A}_5^{\alpha}\widehat{\psi}_{\alpha}(x) &= 0,\qquad
\textrm{for $\alpha$ given in (\ref{level_5_s_mom})} \\
\label{BPZ_level_5c}
\widehat{A}_5^{\alpha} &= \widehat{L}_{-1}\widehat{A}_4^{\alpha}
+2\delta_1\delta_2\gamma_4\big(2\widehat{L}_{-2}\widehat{A}_3^{\alpha} + \gamma_3\widehat{L}_{-3}\widehat{A}_2^{\alpha}\big) \\
&\quad -4\delta_2\gamma_2\gamma_3\gamma_4
\big(\delta_1\widehat{L}_{-1}\widehat{L}_{-4} - (\gamma_1+3\delta_1)\widehat{L}_{-5}\big)\ 
\end{split}
\end{empheq}
where we denote
\begin{align}
\begin{split}
&
\gamma_1=\frac{\alpha^2}{\epsilon_1\epsilon_2},\qquad
\gamma_2=\frac{(2\alpha+\epsilon_1)(2\alpha+\epsilon_2)}{\epsilon_1\epsilon_2},\qquad
\gamma_3=\frac{(\alpha+\epsilon_1)(\alpha+\epsilon_2)}{\epsilon_1\epsilon_2},\\
&
\gamma_4=\frac{(2\alpha+3\epsilon_1)(2\alpha+3\epsilon_2)(2\alpha+\epsilon_1+\epsilon_2)}{\epsilon_1\epsilon_2\alpha},
\\
&
\delta_1=\frac{\alpha}{5\alpha+3\epsilon_1+3\epsilon_2},\qquad
\delta_2=\frac{\alpha}{7\alpha+6\epsilon_1+6\epsilon_2},
\end{split}
\end{align}
while $\widehat{A}_2^{\alpha}$, $\widehat{A}_3^{\alpha}$, and $\widehat{A}_4^{\alpha}$ are given by (\ref{BPZ-level-2}), (\ref{BPZ-level-3}), and (\ref{BPZ-level-4c}), and using the above constants they can also be written as
\begin{align}
\begin{split}
\widehat{A}_2^{\alpha}&=\widehat{L}_{-1}^2+4\gamma_1\widehat{L}_{-2},
\\
\widehat{A}_3^{\alpha}&=\widehat{L}_{-1}\widehat{A}_2^{\alpha}
+2\gamma_1\gamma_2\widehat{L}_{-3},
\\
\widehat{A}_4^{\alpha}&=
\widehat{L}_{-1}\widehat{A}_3^{\alpha}
+4\delta_1\gamma_3\widehat{L}_{-2}\widehat{A}_2^{\alpha}
-2\delta_1\gamma_2\gamma_3
\big(\widehat{L}_{-1}\widehat{L}_{-3}
-2\gamma_2\widehat{L}_{-4}\big).  \label{BPZ-levels-234}
\end{split}
\end{align}
Analogously as we observed in other examples, substituting $\alpha=\alpha_{r,s}$ with $rs=5$ the operator $\widehat{A}_5^{\alpha}$ in (\ref{BPZ_level_5c}) takes form of an operator that encodes singular vectors at level 5, while for $rs<5$ it factorizes into operators that encode singular vectors at lower levels.





\newpage 

\section{Double quantum structure and various limits}   \label{sec-double}

So far we have shown how to assign an infinite number of (time-dependent) quantum curves to a given matrix model.  In this section we discuss an interesting feature of these quantum curves, namely the fact that they are quantum in a double sense, and analyze corresponding classical limits and various perturbative expansions. As we explain in what follows, in order to consider classical limits one should analyze normalized partition functions $\Psi_{\alpha}(x)$ defined in (\ref{PsiPsi-alpha}), as usual for special values of momenta $\alpha=\alpha_{r,s}$ given in (\ref{alpha-rs-2}), and the relevant representation of Virasoro operators (\ref{calLn}).

The double quantum character of quantum curves has to do with the presence of two parameters,  $g_s$ and $b^2=-\beta$. 
First, as usual in the context of matrix models, quantum curves can be interpreted as arising from quantization of a classical spectral curve. In particular, such an interpretation is well known \cite{abmodel} for quantum curves that we identify at level 2. In this case the quantum parameter (the Planck constant) is identified with $g_s$, or equivalently with $1/N$ in the large $N$ ('t Hooft) limit. Quantum curves can then be written as $\widehat{A}=\widehat{A}(\hat{x},\hat{y})$ in terms of operators $\hat{x} = x$ and $\hat{y}\sim g_s\partial_x$ that satisfy the relation $[\hat{y},\hat{x}]\sim g_s$. In the large $N$ limit, when $g_s\to 0$, the operators $\hat{x}$ and $\hat{y}$ become commuting variables $x$ and $y$; if in addition we set $\beta=1$, the (classical) spectral curve (\ref{spcurve}) is written as $A(x,y)=0$. Below we show that, in an analogous way, quantum curves at higher levels can be also interpreted as arising from quantization of the spectral curve $A(x,y)=0$, or more precisely of its multiple copy. In what follows we refer to the classical limit in the above sense as the classical 't Hooft limit.

There is however the second quantum structure encoded in quantum curves, related to their interpretation in conformal field theory. In this context, e.g. as in Liouville theory \cite{Zamolodchikov:2003yb,Hadasz:2003he,Hadasz:2005gk}, the classical limit is the limit of an infinite central charge, which corresponds to an infinite or zero value of the parameter $b$ (or equivalently $\beta$). In this limit, in the context of Liouville theory, singular vector equations reduce to equations of motion in classical Liouville theory, which have form of certain differential equations written in terms of $\partial_x$.
More precisely, to relate our framework to the classical Liouville theory one needs to take a double scaling limit, so that apart from the limit of the parameter $b$, also $g_s$ is taken to zero in such a way, that the product $\beta^{1/2}g_s=-\epsilon_1$ or $g_s \beta^{-1/2}=\epsilon_2$ is fixed, while the second parameter (respectively $\epsilon_2$ or $\epsilon_1$) vanishes. In terms of $\epsilon_1$ and $\epsilon_2$ this limit is precisely the Nekrasov-Shatashvili limit \cite{NS}.

The two quantum structures mentioned above have nice interpretation both in the language of matrix models, as well as in the language of Virasoro algebra. In the language of matrix models, the first limit (leading to the classical spectral curve) is the usual 't Hooft large $N$ limit, while the second limit corresponds to very particular $\beta$ ensemble, with vanishing or infinite value of the parameter $\beta$. On the other hand, from the viewpoint of Virasoro algebra, the second limit is the standard classical limit considered in Liouville theory, while the first limit (leading to the spectral curve) is equivalent to the limit in which all Virasoro operators $\widehat{L}_{-n}$ are set to zero for $n\geq 3$, while $\widehat{L}_{-1}$ and $\widehat{L}_{-2}$ are set to be commuting. Such a limit has been introduced by Feigin and Fuchs and analyzed in \cite{Feigin:1988se,Kent:1991wm}.

In this section we discuss the two quantum structures and corresponding classical limits mentioned above. We also discuss how to reconstruct wave-functions and quantum curves in a perturbative expansion in $g_s$. Furthermore, we analyze an expansion of the unnormalized wave-functions $\psi_{\alpha}(x)$ in the limit of large values of $x$, and show that from this expansion one can extract time-dependent contributions to the matrix model partition function $Z$ in (\ref{matrix_def}).


\subsection{Classical ('t Hooft) limit}    \label{ssec-tHooft}

In the classical 't Hooft limit quantum curves $\widehat{A}(\hat{x},\hat{y})$ are expected to reduce to classical algebraic curves. In this section we show that in this limit quantum curves at level 2 reduce to the matrix model spectral curve, while quantum curves at higher levels reduce to classical expressions that factorize into products of several factors, which all represent the underlying spectral curve. We note that in order to take the classical ('t Hooft) limit  one has to subtract matrix model partition function $Z$ from the wave-function, as otherwise it would result in a divergence, as follows e.g. from the asymptotics (\ref{asymptotics-psi}). Therefore to analyze the classical limit we need first to rewrite the quantum curve as the equation for the normalized wave-function $\Psi_{\alpha}(x)$ defined in (\ref{PsiPsi-alpha}).

Let us consider first the quantum curves at level 2, for $\beta=1$. From (\ref{BPZ-level-2bis}), for either value $\alpha=-\epsilon_1/2$ or $-\epsilon_2/2$, we immediately get
\be
\left( g_s^2 \partial_x^2 - V'(x)^2 -\widehat{f}(x)\right)\widehat{\psi}_{\alpha}(x) = 0.
\ee
To rewrite this equation as a differential equation for $\Psi_{\alpha}(x)$ we divide it by the partition function $Z$ and rewrite the action of $\widehat{f}(x)$ as an additional term $f(x)$ under the expectation value defining $\widehat{\psi}_{\alpha}(x)$, as follows from (\ref{fhat-f}). Furthermore, using the factorization of expectation values in the large $N$ limit and the definition (\ref{f-cl}), the above equation reduces to 
the classical spectral curve $y^2 - V'(x)^2 - f_{cl}(x) = 0$ in (\ref{spcurve})
by identifying $y$ with the classical limit of $g_s\partial_x$. 
Note that from the representation (\ref{sp_hat_f_op}) we can also write
\be
f_{cl}(x) = - \lim_{\epsilon_1,\epsilon_2\to 0}  \epsilon_1 \epsilon_2\, \sum_{n=0}^{\infty} x^n \partial_{(n)} \log Z,  \label{f-clas}
\ee
and introducing the classical limit of the $\widehat{\mathcal L}_{-2}$ operator in (\ref{calLn})
\begin{equation}
\widehat{\mathcal L}_{-2}^{cl} = \lim_{\epsilon_1,\epsilon_2 \to 0} \epsilon_1 \epsilon_2\,  \widehat{\mathcal L}_{-2} = -V'(x)^2-f_{cl}(x)
\end{equation}
the classical curve (\ref{spcurve}) can be written as $y^2 + \widehat{\mathcal L}_{-2}^{cl} = 0$.

More generally, from the explicit form of quantum curves at various levels determined in section \ref{sec-quantum}, as well as from the asymptotics (\ref{det_op_exp}), it follows that for a quantum curve corresponding to the momentum $\alpha_{r,s}$ it is natural to identify the quantum parameter as (the absolute value of)  
\be
\hbar_{r,s} = -\frac{\epsilon_1\epsilon_2}{2\alpha_{r,s}}.   \label{hbar-rs}
\ee
In particular, for quantum curves corresponding to the values $\alpha_{r,1}$ or $\alpha_{1,s}$, the quantum parameter is identified respectively as $\hbar_{r,1}=\frac{\epsilon_2}{r-1}$ and $\hbar_{1,s}=\frac{\epsilon_1}{s-1}$. It follows that for arbitrary value of $\beta$ quantum curves at level 2 also reduce in the classical limit to the equation of the form (\ref{spcurve}), however with $y$ identified with a classical limit of $\epsilon_2\partial_x$ or $\epsilon_1\partial_x$, respectively for the quantum curves corresponding to $\alpha_{2,1}$ and $\alpha_{1,2}$. 

On a side track, note that none of the values (\ref{hbar-rs}) reproduces $\hat{\hbar}$ introduced in (\ref{hbar-Ric}), which was interpreted as the Planck constant associated to another construction of a quantum spectral curve, proposed in \cite{Eynard:2008mz,Chekhov:2009mm,Chekhov:2010zg}. One may also notice that keeping the dependence on $\epsilon_1$ and $\epsilon_2$ to the first order, the classical limit of the level 2 quantum curve (\ref{BPZ-level-2bis}) could be interpreted as $y^2 - V'(x)^2 - f_{cl}(x) + 2\hat{\hbar}V''(x)=0$. This is similar to the Riccati equation (\ref{spcurve-Ric}),  
however without the term involving $y'(x)$. This is another manifestation that quantum curves discussed in this paper are not directly related to those introduced in \cite{Eynard:2008mz,Chekhov:2009mm,Chekhov:2010zg}.

Let us consider now the classical limit for quantum curves at higher levels. Recall that quantum curves at level $n$ have the same structure as singular vectors (\ref{Lrs}) and are represented as sums of terms of the form $\widehat{L}_{-p_1}\widehat{L}_{-p_2}\cdots \widehat{L}_{-p_k}$, with $p_1+\ldots+p_k=n$, and with the representation of $\widehat{L}_{-p}$ given in (\ref{Lm_fer_op}). Among those terms there is always one of the form $\widehat{L}_{-1}^n$ (that gives rise to the differential equation in $x$ of order $n$), and therefore, to obtain a classical limit, the quantum curve equation needs to be multiplied by $\hbar_{r,s}^n$, so that $(\hbar_{r,s}\partial_x)^n$ can be identified in the limit with $y^n$. The homogeneity of the operator $\widehat{A}^{\alpha}_n$ implies then that each $\widehat{L}_{-p}$ in the expression for the quantum curve gets multiplied by $\hbar_{r,s}^p$. Now note that all $\widehat{L}_{-p}$ for $p\geq 2$ are proportional to $(\epsilon_1\epsilon_2)^{-1}$. Therefore multiplying $\widehat{L}_{-2}$ by $\hbar_{r,s}^2$ and taking the limit $\hbar_{r,s}\to 0$ gives some finite expression, which moreover does not include any time derivatives. On the other hand, $\widehat{L}_{-p}$ for $p\geq 3$ multiplied by $\hbar_{r,s}^p$ vanishes in the limit $\hbar_{r,s}\to 0$, and in consequence all summands that include at least one $\widehat{L}_{-p}$ with $p\geq 3$ in the expression for the quantum curve vanish in the classical limit. Therefore we conclude that the classical limit is simply the limit where $\widehat{L}_{-p}$ with $p\geq 3$ are set to zero, while $\widehat{L}_{-1}$ and $\widehat{L}_{-2}$ are set to be commuting. This is the limit analyzed in \cite{Feigin:1988se,Kent:1991wm}. In particular in this limit the quantum curve equations factorize -- for example, quantum curves corresponding to momenta $\alpha_{r,1}$ reduce to
\be
\begin{split}
0 &= \prod_{k=1}^{r/2}  \Big( y^2 - \frac{(2k-1)^2}{(r-1)^2} \big(V'(x)^2 + f_{cl}(x) \big)  \Big)  ,\ \quad \qquad \textrm{for $r$ even}     \\
0 &= y \prod_{k=1}^{(r-1)/2}  \Big( y^2 - \frac{4k^2}{(r-1)^2} \big(V'(x)^2 + f_{cl}(x) \big)  \Big),\ \, \qquad \textrm{for $r$ odd}    \label{class-curves-higher-level}
\end{split}
\ee
with $y$ identified with the limit of $\frac{\epsilon_2}{r-1}\partial_x$. Note that each factor in those expressions essentially represents the spectral curve (\ref{spcurve}), with $(V'(x)^2 + f_{cl}(x))$ term rescaled by a simple factor. In this sense higher level quantum curves can be interpreted as arising from quantization of a multiple copy of the original spectral curve.

Note that classical curves at higher levels, corresponding to momenta $\alpha_{r,1}$, can be also obtained from the recursion relation
\begin{equation}
a_{0}^{r+1}=1,\ \ \
a_{1}^{r+1}=ry,\ \ \
a_{q+1}^{r+1}=ry a_{q}^{r+1}
+q(r-q+1)\widehat{\mathcal L}^{\textrm{cl}}_{-2}a_{q-1}^{r+1},
\label{class_B_op}
\end{equation}
which is $\epsilon_2\to 0$ limit of the recursion (\ref{NS_B_op}) that we discuss in the next section, and where we defined $a_n^{r+1}=\lim_{\epsilon_2\to 0}\widehat{a}_n^{r+1}$. Solving this recursion leads to the expression for the classical curve of the form $A(x,y)\equiv a^{r+1}_{r+1}=0$, which reproduces the result (\ref{class-curves-higher-level}).


\subsection{Nekrasov-Shatashvili -- classical Liouville limit}  \label{ssec-NSlimit}

The second interesting limit to consider is the classical limit in Liouville theory. It turns out to be equivalent to the Nekrasov-Shatashvili limit, whereupon one of $\epsilon_1,\epsilon_2$ parameters is set to zero, and the other one is kept constant. For definiteness let us choose the case $\epsilon_1\to 0$, which in view of $b^2=\frac{\epsilon_1}{\epsilon_2}$ is the limit of vanishing $b$, which is indeed a classical limit in Liouville theory. In terms of parameters $b$ and $g_s$, in order to keep $\epsilon_2$ constant one needs to take a double scaling limit with both of these parameters vanishing with a constant ratio.

In the limit $\epsilon_1\to 0$ it is natural to consider behavior of quantum curves and wave-functions labeled by the momenta
\begin{equation}
\alpha=\alpha_{r+1,1}=-\frac{r}{2}\epsilon_1.
\label{p_ep_2_br}
\end{equation}
As explained earlier, we should consider wave-functions normalized by the partition function (\ref{PsiPsi-alpha}), which in the $\epsilon_1\to 0$ limit we denote by $\Psi^{\textrm{NS}}_{\alpha}(x)$. As follows e.g. from the representation (\ref{det_op_exp}), these wave-functions factorize
\be
\Psi^{\textrm{NS}}_{-\frac{r}{2}\epsilon_1}(x) 
\equiv \lim_{\epsilon_1\to 0}\Psi_{-\frac{r}{2}\epsilon_1}(x)
=\left(\Psi^{\textrm{NS}}_{-\frac{1}{2}\epsilon_1}(x)\right)^r.
\ee
We write equations satisfied by these wave-functions as
\be
\widehat{\mathcal{A}}_{r+1}^{\textrm{NS}}  \Psi^{\textrm{NS}}_{-\frac{r}{2}\epsilon_1}(x) = 0, 
\ee
where the quantum curve $\widehat{\mathcal{A}}_{r+1}^{\textrm{NS}}$ arises from the limit of (\ref{chat-Lhat})  
\begin{equation}
\widehat{\mathcal{A}}_{r+1}^{\textrm{NS}} = \lim_{\epsilon_1\to 0} \epsilon_2^{r+1}Z^{-1}\widehat{A}_{r+1}^{-\frac{r}{2}\epsilon_1}Z.
\end{equation}
As usual this quantum curve has a structure (\ref{Lrs}), however this time with Virasoro operators taking form of the $\epsilon_1\to 0$ limit of operators (\ref{calLn}), which we denote as 
\begin{equation}
\widehat{\mathcal L}_{-n}^{\textrm{NS}}=\lim_{\epsilon_1\to 0} \epsilon_1\epsilon_2 \widehat{\mathcal L}_{-n}
= -\frac{1}{(n-2)!}\Big(\partial_x^{n-2}\big(V'(x)^2\big)+\epsilon_2
\partial_x^n V(x)+F^{(0)}_{n-2}(x,\epsilon_2)\Big),
\end{equation}
where for the deformed prepotential
\begin{equation}
F^{(0)}(\epsilon_2)=-\lim_{\epsilon_1\to 0}\epsilon_1\epsilon_2\log Z
\end{equation}
we have defined
\begin{equation}
F^{(0)}_{k}(x,\epsilon_2)=\sum_{n=k}^{\infty}\frac{n!}{(n-k)!}x^{n-k}\partial_{(n)}F^{(0)}(\epsilon_2).
\end{equation}
Note that $F^{(0)}_{0}(x,0)=f_{cl}(x)$ given in (\ref{f-clas}).

In particular in the $\epsilon_1\to 0$ limit the quantum curve equation (\ref{BPZ-level-2}) at level 2 takes form
\begin{equation}
\widehat{\mathcal{A}}_2^{\textrm{NS}}\Psi^{\textrm{NS}}_{-\frac{1}{2}\epsilon_1}(x)
= \Big(\epsilon_2^2\partial_x^2+\widehat{\mathcal L}^{\textrm{NS}}_{-2}\Big)\Psi^{\textrm{NS}}_{-\frac{1}{2}\epsilon_1}(x)=0.
\end{equation}
To present a quantum curve equation for $\Psi^{\textrm{NS}}_{-\frac{r}{2}\epsilon_1}(x)$, we define inductively differential operators $\widehat{a}_{q}^{r+1}$ for $q=0,1,\ldots,r+1$ 
\begin{equation}
\widehat{a}_{0}^{r+1}=1,\ \ \
\widehat{a}_{1}^{r+1}=\epsilon_2\partial_x,\ \ \
\widehat{a}_{q+1}^{r+1}=\epsilon_2\partial_x\widehat{a}_{q}^{r+1}
+q(r-q+1)\widehat{\mathcal L}^{\textrm{NS}}_{-2}\widehat{a}_{q-1}^{r+1},
\label{NS_B_op}
\end{equation}
and by induction we find
\begin{equation}
\widehat{a}_{q+1}^{r+1}\Psi^{\textrm{NS}}_{-\frac{r}{2}\epsilon_1}(x)=
r(r-1)(r-2)\cdots (r-q)\left(\Psi^{\textrm{NS}}_{-\frac{1}{2}\epsilon_1}(x)\right)^{r-q-1}
\left(\epsilon_2\partial_x \Psi^{\textrm{NS}}_{-\frac{1}{2}\epsilon_1}(x)\right)^{q+1}.
\end{equation}
It follows that the wave-function $\Psi^{\textrm{NS}}_{-\frac{r}{2}\epsilon_1}(x)$ satisfies 
an ordinary differential equation  of order $(r+1)$, and the corresponding quantum curve is identified as
\begin{equation}
\widehat{\mathcal{A}}_{r+1}^{\textrm{NS}} = \widehat{a}_{r+1}^{r+1}.
\label{NS_diff_p}
\end{equation}
For example, in this way we obtain
\be
\begin{split}
\widehat{\mathcal{A}}_{2}^{\textrm{NS}}&= \epsilon_2^2\partial_x^2+\widehat{\mathcal L}^{\textrm{NS}}_{-2},  \\
\widehat{\mathcal{A}}_{3}^{\textrm{NS}}&=
\epsilon_2^3\partial_x^3+4\epsilon_2\widehat{\mathcal L}^{\textrm{NS}}_{-2}\partial_x+2\epsilon_2\widehat{\mathcal L}^{\textrm{NS}}_{-3},
\\
\widehat{\mathcal{A}}_{4}^{\textrm{NS}}&=
\epsilon_2^4\partial_x^4
+10\epsilon_2^2\widehat{\mathcal L}^{\textrm{NS}}_{-2}\partial_x^2
+10\epsilon_2^2\widehat{\mathcal L}^{\textrm{NS}}_{-3}\partial_x
+9\big(\widehat{\mathcal L}^{\textrm{NS}}_{-2}\big)^2
+6\epsilon_2^2\widehat{\mathcal L}^{\textrm{NS}}_{-4},
\\
\widehat{\mathcal{A}}_{5}^{\textrm{NS}}&=
\epsilon_2^5\partial_x^5
+20\epsilon_2^3\widehat{\mathcal L}^{\textrm{NS}}_{-2}\partial_x^3
+30\epsilon_2^3\widehat{\mathcal L}^{\textrm{NS}}_{-3}\partial_x^2
+64\epsilon_2\big(\widehat{\mathcal L}^{\textrm{NS}}_{-2}\big)^2\partial_x
+36\epsilon_2^3\widehat{\mathcal L}^{\textrm{NS}}_{-4}\partial_x +  \\
&\ \ \
+64\epsilon_2\widehat{\mathcal L}^{\textrm{NS}}_{-2}\widehat{\mathcal L}^{\textrm{NS}}_{-3}
+24\epsilon_2^3\widehat{\mathcal L}^{\textrm{NS}}_{-5}.
\end{split}
\ee
Note that using the Virasoro algebra
$\partial_x^n \widehat{\mathcal L}^{\textrm{NS}}_{-2}  = n! \widehat{\mathcal L}^{\textrm{NS}}_{-n-2}$, each $\widehat{\mathcal{A}}_{r+1}^{\textrm{NS}}$ can be expressed in terms of $\partial_x$ and (derivatives of) $\widehat{\mathcal L}^{\textrm{NS}}_{-2}$ only. If we further identify the energy-momentum tensor in classical Liouville theory as $T^{(c)}\equiv \widehat{\mathcal L}^{\textrm{NS}}_{-2}$, then the operators $\widehat{\mathcal{A}}_{r+1}^{\textrm{NS}}$ take the same form as operators imposing differential equations for the fields $e^{-r\varphi/2}$ in the classical Liouville theory, as discussed in \cite{Zamolodchikov:2003yb}. 

One can also consider a further limit $\epsilon_2\to 0$, whereupon all results in this section reduce to those discussed in section \ref{ssec-tHooft}. In particular, in such a limit the relations (\ref{NS_B_op}) reduce to recursion relations (\ref{class_B_op}) that encode classical curves (\ref{class-curves-higher-level}).


\subsection{$g_s$-expansion: quantum curves from wave-functions}    \label{ssec-reconstruct}

In section \ref{ssec-tHooft} we discussed classical 't Hooft limit of vanishing $g_s$, whereupon quantum curves reduce to classical algebraic curves. Let us discuss now how to reconstruct quantum curves in the form of a series of perturbative corrections in $g_s$ to the classical curve. First, note that in general, for small values of $g_s$, the wave-function $\widehat{\psi}_{\alpha}(x)$ normalized by the partition function $Z$ has the following asymptotic expansion
\be
\Psi_{\alpha} = \frac{\widehat{\psi}_{\alpha}(x)}{Z} = \exp\Big(\sum_{m=0}^{\infty} g_s^{m-1} S_m\Big),   \label{psi-alpha-Sm}
\ee 
where $S_m$ depend on both $x$ and times $t_k$. As we explain in section \ref{ssec-det}, the coefficients $S_m$ can be reconstructed for example by means of the topological recursion. The precise form of this expansion is given in  (\ref{det_op_exp}); at this moment however let us only assume, that there is a way to reconstruct the expansion of the form (\ref{psi-alpha-Sm}). If the above expansion is substituted to the (possibly time-dependent) quantum curve equation it is supposed to satisfy, one can expand this equation in powers of $g_s$ and analyze order by order.

Such perturbative analysis can be presented in a general, explicit form, if we assume that (\ref{psi-alpha-Sm}) satisfies a time-independent quantum curve equation. Such circumstances are not unexpected -- in various cases, in particular for the Gaussian and Penner models discussed in section \ref{sec-examples}, one can indeed get rid of the time dependence, and turn quantum curve equations into ordinary differential equations in variable $x$. Expanding such equations into a series in $g_s$ one obtains a hierarchy of differential equations, which we summarize below following \cite{abmodel}. In particular, as stressed in \cite{abmodel}, from this hierarchy one can reconstruct the form of the operator $\widehat{A}(\hat x, \hat y)$ that is supposed to annihilate the wave-function constructed as in (\ref{det_op_exp}) or, say, (\ref{wave_involution}). 

More precisely, assume that a wave-function $\Psi_{\alpha}(x)$ satisfies an equation of the form
\begin{equation}
\widehat{A}(\hat x,\hat y)\, \Psi_{\alpha}(x) = 0,  \label{AZ}
\end{equation}
where $\hat y = g_s \partial_x$. Also assume that the wave-function has an asymptotic expansion of the form
\be
\Psi_{\alpha}(x) = \exp\Big(\sum_{m=0}^{\infty} g_s^{m-1} S_m\Big),   \label{psi-Sm}
\ee 
where $S_m=S_m(x)$. Furthermore, as $\widehat{A}(\hat x, \hat y)$ is an operator expression, choose the ordering such that $g_s \partial_x$ are given to the right of $x$, and write 
\begin{equation}
\widehat{A}(\hat x, \hat y) = \widehat{A}_0 + g_s \widehat{A}_1 + g_s^2 \widehat{A}_2 + \ldots 
\label{Ahatpert}
\end{equation}
where $\widehat{A}_0$ is identified with the classical curve, $A_0=\widehat{A}_0=A=A(x,y)$. Substituting (\ref{psi-Sm}) and (\ref{Ahatpert}) into (\ref{AZ}) we get a hierarchy of equations \cite{abmodel}
\begin{equation}
\sum_{r=0}^n \D_{r} A_{n-r} =  0,
\label{hierarchy-SA}    
\end{equation}
where $A_{n-r}$ are symbols of the operators $\widehat{A}_{n-r}$, and $\D_r$ are differential operators in $\partial_y$ of degree $2r$, whose coefficients are polynomial expressions in derivatives of $S_m$. The operators $\D_r$ are defined via the generating function
\begin{equation}
\sum_{r=0}^{\infty} g_s^r \D_r =
\exp \left( \sum_{n=1}^{\infty} g_s^n \frak{d}_n \right), \qquad \textrm{where} \quad  \frak{d}_n = \sum_{r=1}^{n+1} \frac{S_{n+1-r}^{(r)}}{r!} (\partial_y)^r.
\label{Sigma-def}
\end{equation}
For example, for small values of $n$ we find
\bea
\frak{d}_1 & = & \frac{1}{2} S''_0 \partial_y^2 + S'_1 \partial_y \,, \nonumber \\
\frak{d}_2 & = & \frac{1}{6}S'''_0 \partial_y^3 + \frac{1}{2} S''_1 \partial_y^2 + S'_2 \partial_y \,, \nonumber \\
\frak{d}_3 & = & \frac{1}{4!}S^{(4)}_0 \partial_y^4 +\frac{1}{3!}S'''_1 \partial_y^3 + \frac{1}{2} S''_2 \partial_y^2 + S'_3 \partial_y \,, \nonumber
\eea
where we denote $\partial_x$ derivatives by a prime. It follows that
\be
\begin{split}
\D_0 & =  1 \,, \nonumber \\
\D_1 & =  \frac{S''_0}{2}  \partial_y^2 + S'_1 \partial_y \,,  \nonumber \\
\D_2 & =  \frac{(S''_0)^2}{8} \partial_y^4 +  \frac{1}{6}\big(S'''_0 + 3S''_0 S'_1 \big) \partial_y^3
+ \frac{1}{2}\big(S''_1 + (S'_1)^2 \big) \partial_y^2 + S'_2 \partial_y \, . \nonumber 
\end{split}
\ee
In consequence, the first equation (at order $g_s^0$) in the hierarchy is simply the classical curve equation $A=A(x,y)=0$, at the order $g_s^1$ we find an equation 
\be
\Big(\frac{S''_0}{2}\partial_y^2 + S'_1 \partial_y \Big) A + A_1 = 0,
\ee
and equations at higher orders of $g_s$ take form \eqref{hierarchy-SA}. 

To sum up, if the quantum curve $\widehat{A}(\hat x, \hat y)$ is known, the hierarchy (\ref{hierarchy-SA}) can be used to determine (\ref{psi-Sm}) order by order; vice versa, if $S_m$ are known -- for example from the topological recursion, as in (\ref{det_op_exp}) or (\ref{wave_involution}) -- one can perturbatively reconstruct the operator $\widehat{A}(\hat x, \hat y)$.


\subsection{$x$-expansion: partition functions from quantum curves}     \label{ssec-x-expansion}

So far in this section we subtracted the matrix model partition function  $Z$ from the wave-function and considered the normalized expressions $\Psi_{\alpha}(x)$, and corresponding quantum curves, in various limits. It is however also useful to consider the unnormalized wave-function $\psi_{\alpha}(x)$ in yet another limit, namely the limit of large $x$ -- as we discuss now, in this way one can reconstruct the form of time-dependent contributions to the partition function $Z$.

As follows from the definition (\ref{wave_Z2}), the leading dependence on $x$ in the large $x$ limit takes form
\be
\psi_{\alpha}(x)  \sim x^{-\frac{2\alpha N}{\epsilon_2}} Z= x^{\frac{4\mu \alpha}{\epsilon_1\epsilon_2}} Z.
\ee
Taking into account subleading corrections in negative powers of $x$, we can write
\be
\psi_{\alpha}(x)  = \exp\Big(  \frac{4\mu \alpha}{\epsilon_1\epsilon_2}\log x + \sum_{k=0}^{\infty}S_k x^{-k}    \Big),    \label{psi-x-expansion}
\ee
where $S_k$ depends now on $g_s$ and times $t_k$ (and possibly $\beta$). Assuming that this wave-function is annihilated by a certain time-dependent quantum curve, one can analyze such an equation order by order in $x$. As $\psi_{\alpha}(x)$ is not normalized by $Z$, this partition function must be entirely encoded in the term $S_0$, and from the knowledge of the time-dependent equation satisfied by $\psi_{\alpha}(x)$ one can extract time-dependent information contained in $Z$. We illustrate how to reconstruct the partition function in this way in the example of the Penner model in section \ref{ssec-Penner}. For detailed discussion of this point see also \cite{Andersen:2016ida}.


\newpage 

\section{Quantum curves and the (refined) topological recursion}    \label{sec-refined}

In section \ref{sec-quantum} we derived a general form of quantum curves (\ref{chat-Lhat}), i.e. differential operators that annihilate wave-functions $\widehat{\psi}_{\alpha}(x)$, from the analysis of loop equations for the $\alpha/\beta$-deformed matrix integral. On the other hand, it is known that for $\alpha=\alpha_{1,2}$ or $\alpha=\alpha_{2,1}$, the $g_s$ expansion of the expectation value $\widehat{\psi}_{\alpha}(x)$ can be reconstructed by means of the topological recursion, both in the unrefined \cite{abmodel} and refined \cite{Kozcaz:2010af} case. In this section we explain how to reconstruct the wave-functions $\psi_{\alpha}(x)$ in (\ref{wave_Z2}) (or equivalently $\widehat{\psi}_{\alpha}(x)$) in the $g_s$ expansion for all values $\alpha=\alpha_{r,s}$ in (\ref{alpha-rs-2}). The topological recursion is also the main tool in this process. 

Before discussing how wave-functions can be reconstructed, in this section (together with relevant appendices) we first summarize and present a few new results within the formalism of refined topological recursion. In particular we present a detailed analysis of the one-cut case, which in the presence of the $\beta$-deformation is already quite non-trivial. We believe that this summary will be useful for all readers interested in various applications of the topological recursion -- while some results presented here can be also found elsewhere in literature \cite{Chekhov:2006rq,Eynard:2008mz,Chekhov:2009mm,Chekhov:2010zg,Chekhov:2010xj,Brini:2010fc,Marchal:2011iu,Manabe:2012bq}, we find it useful to assemble them in one place.

After summarizing the formalism of refined topological recursion we explain how it can be used to reconstruct wave-functions $\psi_{\alpha}(x)$ at arbitrary levels, in principle for arbitrary algebraic curves. In addition, we realize that one can define wave-functions in various ways, by making different choices of reference points in their defining integrals. In consequence, for a given model one can introduce different quantum curves corresponding to different choices of the reference point. We discuss two choices of such reference points -- at infinity, which leads to the results we presented earlier, and as a conjugate point, which leads to the results discussed in mathematical literature. It is desirable to study this issue, and perhaps even larger families of quantum curves parameterized by various reference points, in more detail.

The general formalism presented in this section will be employed in the analysis of a few matrix models with specific potentials in section \ref{sec-examples}.


\subsection{Topological recursion -- the idea and main ingredients}

The topological recursion, also referred to as the Eynard-Orantin recursion, is a formalism that assigns the so called symplectic invariants and multi-resolvents to a given algebraic curve \cite{eyn-or}. In case such an algebraic curve arises as a spectral curve of some matrix model, the symplectic invariants in question are identified with the coefficients of the large $N$ expansion of the free energy, i.e. the logarithm of the partition function of this matrix model, and multi-resolvents have an explicit definition as certain matrix model expectation values. Indeed, originally the topological recursion was found in the analysis of matrix models \cite{Eynard:2004mh,Chekhov:2005rr}, and only subsequently was it reinterpreted and generalized by Eynard and Orantin to the realm of arbitrary algebraic curves \cite{eyn-or}. In addition also the refined version of the topological recursion was introduced, by reformulating and generalizing the analysis of loop equations for $\beta$-deformed matrix models \cite{Chekhov:2006rq,Eynard:2008mz,Chekhov:2009mm,Chekhov:2010zg}. 

In this brief section we present the main idea and general features of this formalism; more details are discussed in the following sections. As in this paper we primarily work with matrix models, the summary below is also presented from this viewpoint, whereupon the topological recursion arises as a relation between certain expectation values. The reader should however bear in mind that, more generally, the recursion can be regarded as a procedure that to a given curve assigns free energies and multi-resolvents, which have many nice properties.

To start with, for the $\beta$-deformed matrix model (\ref{matrix_def}) we define connected $h$-point differentials
\begin{equation}
W_h(x_1,\ldots,x_h)=\beta^{h/2}\bigg<\prod_{i=1}^h\sum_{a=1}^N\frac{dx_i}{x_i-z_a}\bigg>^{(\mathrm{c})},
\label{h_conn_diff}
\end{equation}
where $\left<\mathcal{O}\right>^{(\mathrm{c})}$ denotes the connected part of the normalized expectation value $\left<\mathcal{O}\right>/Z$.
In the large $N$ limit (\ref{tHooft}),
the $h$-point differential has an asymptotic expansion
\begin{equation}
W_h(x_1,\ldots,x_h)=\sum_{g,\ell=0}^{\infty}\hbar^{2g-2+h+\ell}\gamma^{\ell}W^{(g,h)}_{\ell}(x_1,\ldots,x_h),
\label{W_diff_asy}
\end{equation}
where
\begin{equation}
\gamma=\beta^{1/2}-\beta^{-1/2}=-Q.   \label{gamma-beta}
\end{equation}

A crucial role in the formalism of the topological recursion is played by $W^{(g,h)}_{\ell}$ introduced in (\ref{W_diff_asy}), also referred to as multi-resolvents. First, we can write the leading differential (disk contribution) as $W^{(0,1)}_1=\omega(x)dx$, where the resolvent $\omega(x)$ was introduced in (\ref{resolvent}); via (\ref{yVomega}) this differential encodes the spectral curve $y=y(x)$ of a matrix model, which can be written as in (\ref{spcurve}), or equivalently (\ref{spectral_gen}) below. Other multi-differentials $W^{(g,h)}_{\ell}$ satisfy (for an appropriate range of $(g,h,\ell)$) the recursion relations (\ref{ref_loop_eq}) that give name to the topological recursion formalism. These recursion relations arise as a non-trivial reformulation of loop equations generalizing the loop equation discussed in section \ref{ssec-loopeqs}, see \cite{Eynard:2008mz,Chekhov:2009mm,Chekhov:2010zg}. Moreover, $W^{(g,h)}_{\ell}$ determine (stable) free energies of the matrix model. Indeed, in the limit (\ref{tHooft}) the free energy (i.e. the logarithm of the partition function $Z$) has the asymptotic expansion
\begin{equation}
F = \log Z = \sum_{g,\ell=0}^{\infty}\hbar^{2g-2+\ell}\gamma^{\ell}F_{g,\ell},
\label{free_energy_beta}
\end{equation}
and stable free energies, i.e. coefficients $F_{g,\ell}$ for $2-2g-\ell<0$, can be determined as
\begin{equation}
\boxed{\ F_{g,\ell}=\frac{1}{2-2g-\ell}\oint_{\mathcal{A}}\frac{1}{2\pi i}\Phi(z)W^{(g,1)}_{\ell}(z),\ }
\label{free_ene_rec}
\end{equation}
where the integration contour $\mathcal{A}$ is defined below (\ref{spectral_gen}) and $\Phi(z)$ is defined by
\begin{equation}
\Phi'(z)=
-\frac12 y(z).
\end{equation}
Unstable free energies, i.e. $F_{0,0}, F_{0,1}, F_{1,0}$ and $F_{0,2}$, need to be determined separately, and we present their detailed expressions in appendix \ref{sec-Funstable}.

The topological recursion was formulated first in unrefined limit $\beta=1$ \cite{eyn-or}, and in this case various simplifications arise. In particular, in this limit we get $\gamma=0$ in (\ref{gamma-beta}), and in consequence multi-differentials and free energies in (\ref{W_diff_asy}) and (\ref{free_energy_beta}) get contributions only from $\ell=0$ sector
\be
W_h(x_1,\ldots,x_h)=\sum_{g=0}^{\infty}\hbar^{2g-2+h} W^g_h(x_1,\ldots,x_h),\qquad 
F = \sum_{g=0}^{\infty}\hbar^{2g-2} F_{g},\qquad \textrm{for}\ \beta=1,    \label{WF-unrefined}
\ee
where $W^g_h=W^{(g,h)}_{0}$ and $F_g=F_{g,0}$. For $\beta=1$ also the recursion relations (\ref{ref_loop_eq}) simplify -- in particular integrals they contain are replaced by the evaluation of certain residues, and the derivative term in the third line of (\ref{ref_loop_eq}) drops out. 


\subsection{Wave-functions and quantum curves for various reference points}    \label{ssec-det}

The main object of interest in this paper is the wave-function $\widehat{\psi}_{\alpha}(x)$ satisfying  the quantum curve equation. This wave-function can be thought of as a generalization of a determinant expectation value in $\beta=1$ matrix model
\be
\big\langle \textrm{det}(x-M) \big\rangle = \frac{1}{(2\pi)^N N!} \int DM\, \textrm{det}(x-M) e^{-\frac{1}{\hbar}\textrm{Tr}\,V(M)}   \label{det-xM}
\ee
where the integral is performed over an ensemble of hermitian matrices $M$. This determinant expectation value plays an important role in matrix models and related topics: for example, in integrable systems it represents the Baker-Akhiezer function \cite{Dijkgraaf:1991qh}, and in topological string theory it encodes partition function of a topological brane \cite{ADKMV,abmodel}. 

The wave-functions $\widehat{\psi}_{\alpha}(x)$ that we consider in this paper are two-parameter deformations of (\ref{det-xM}). One of these deformations is the $\beta$-deformation of a matrix model. As the second deformation we introduce the parameter $\alpha$, which appears in the exponent of the determinant insertion. In the undeformed case and for $\alpha=\alpha_{1,2}$, the $g_s$ expansion of the wave-function was presented e.g. in \cite{abmodel}. It is straightforward to generalize it to arbitrary values of $\alpha$ and $\beta$. Including the classical piece given by the potential $V(x)$, the asymptotic expansion of the wave-function given in (\ref{Psi-alpha-bis}), normalized by the partition function $Z$ in (\ref{matrix_def}), takes form 
\begin{empheq}[box=\fbox]{equation}
\begin{split}
\quad \log \frac{\widehat{\psi}_{\alpha}(x)}{Z}&
=
-\frac{2\alpha}{\epsilon_1 \epsilon_2}V(x)+
\sum_{h=1}^{\infty}\frac{1}{h!}\Big(-\frac{2\alpha}{g_s}\Big)^h \int^{x}_{\infty}\cdots\int^{x}_{\infty}W_h(x_1',\ldots,x_h') \\
&=
\sum_{g,\ell=0, h=1}^{\infty}  \frac{(-1)^{g+\ell+h-1}}{h!} 2^{2-2g-\ell} \alpha^h (\epsilon_1\epsilon_2)^{g-1}
(\epsilon_1+\epsilon_2)^{\ell}F^{(g,h)}_{\ell}(x,\ldots,x) \quad
\label{det_op_exp}
\end{split}
\end{empheq}
where we have defined
\begin{align}
F^{(g,h)}_{\ell}(x_1,\ldots,x_h)=\int^{x_1}_{\infty}\cdots\int^{x_h}_{\infty}
W^{(g,h)}_{\ell}(x_1',\ldots,x_h')-\frac{1}{2}V(x)\delta_{g,0}\delta_{h,1}\delta_{\ell,0}.
\label{int_W_d}
\end{align}
From our perspective the fact that the wave-function $\widehat{\psi}_{\alpha}(x)$ is determined in this way by multi-differentials $W^{(g,h)}_{\ell}$ is the main reason to consider the latter ones, and so unavoidably the recursion relations they satisfy. Furthermore, several comments are in order.

First, note that from the expansions (\ref{free_energy_beta}) and (\ref{det_op_exp}) we get the following asymptotic behavior
\begin{equation}
\log \widehat{\psi}_{\alpha}(x)=-\frac{4}{\epsilon_1\epsilon_2}F_{0,0}+\mathcal{O}(g_s^{-1}),\qquad 
\log \frac{\widehat{\psi}_{\alpha}(x)}{Z}=\frac{4\alpha}{\epsilon_1\epsilon_2}F^{(0,1)}_{0}(x)+\mathcal{O}(g_s^{0}).     \label{asymptotics-psi}
\end{equation}

Second, note that setting $\alpha=\alpha_{1,n+1}=-n\epsilon_2/2$, taking the unrefined limit $\beta=1$, and ignoring the potential factor, the expansion (\ref{det_op_exp}) reduces to
\be
\psi_{\beta=1,n}(x) = Z_{\beta=1}\exp\left(
\sum_{g=0,h=1}^{\infty}\frac{n^h}{h!} \hbar^{2g-2+h} \int^{x}_{\infty}\cdots\int^{x}_{\infty}W^g_h(x_1',\ldots,x_h') \right),  \label{psi-n}
\ee
where $Z_{\beta=1}$ is $\beta=1$ limit of the partition function $Z$, and $W^g_h$ are introduced in (\ref{WF-unrefined}). For $n=1$ this expression indeed reproduces (at least for the genus zero spectral curve) the expansion of the determinant expectation value (\ref{det-xM}), for a detailed derivation  see e.g. \cite{abmodel}. Moreover, as discussed in \cite{abmodel}, the expansion (\ref{psi-n}) for $n=1$ is supposed to be annihilated by the quantum curve that in our formalism arises at level 2. This is consistent with the statement that for the value of $\alpha=-\epsilon_2/2$ we get the quantum curve at level 2, as shown in section \ref{ssec-level2}. More generally, for arbitrary positive $n$, the expression (\ref{psi-n}) reproduces $\beta=1$ limit of the wave-function corresponding to a singular vector at level $(n+1)$. Also note, that for the choice $\alpha=-\epsilon_1/2$ in (\ref{det_op_exp}) (corresponding to the second singular vector at level 2), in the unrefined limit the wave-function can be represented as the expectation value of the inverse of the determinant $\langle\textrm{det}(x-M)^{-1}\rangle$, which in the context of topological string theory represents an anti-brane.

Third, a crucial subtlety in the expression (\ref{det_op_exp}) is the choice of the reference point of the integration (i.e. the lower limit of the integrals of $W_h(x_1',\ldots,x_h')$).  In (\ref{det_op_exp}) we chose it to be a point at infinity, which has two important features: first of all this definition makes sense in the $\beta$-deformed case, and as we show in what follows, for such a choice we get wave-functions that are indeed annihilated by quantum curves derived in section \ref{sec-quantum}.

It has been proposed that also another choice of the reference point in the definition of the wave-function, i.e. the conjugate point $\overline{x}$ (for its precise definition see section \ref{ssec-beta-details}), may be suitable from some viewpoints \cite{Norbury-quantum,Dumitrescu:2015mpa}. More precisely, for such a choice of the reference point, in the unrefined limit, let us introduce the following wave-function 
\begin{equation}
\overline{\psi}_{\beta=1,n}(x) = Z_{\beta=1} \exp\Big(\sum_{g=0, h=1}^{\infty}  \frac{n^h}{h!} \hbar^{2g-2+h} \overline{F}^{(g,h)}(x,\ldots,x)  \Big),
\label{wave_involution}
\end{equation}
where, specializing to the one-cut spectral curve (\ref{spectral_one}), we define
\begin{align}
&
\overline{F}^{(0,2)}(x_1,x_2)=\log\frac{2}{x_1-x_2+\sqrt{\sigma(x_1)}+\sqrt{\sigma(x_2)}},\\
&
\overline{F}^{(g,h)}(x_1,\ldots,x_h)=\frac{1}{2^h}\int^{x_1}_{\overline{x}_1}\cdots\int^{x_h}_{\overline{x}_h}
W^g_h (x_1',\ldots,x_h')\ \ \textrm{for}\ (g,h)\neq (0,2).
\end{align}
The main feature of these $\overline{F}^{(g,h)}$ is that their derivatives reproduce $W^g_h$
\begin{equation}
d_{x_1}\cdots d_{x_h}\overline{F}^{(g,h)}(x_1,\ldots,x_h)=W^g_h(x_1,\ldots,x_h),  \label{dF-W}
\end{equation}
and they satisfy the involution condition
\begin{equation}
\overline{F}^{(g,h)}(x_1,\ldots,\overline{x}_i,\ldots,x_h)=-\overline{F}^{(g,h)}(x_1,\ldots,x_i,\ldots,x_h),\ \ \ i=1,\ldots,h, \label{convolution}
\end{equation}
for $(g,h)\neq (0,2)$. More abstractly, reversing the logic, these two conditions may be chosen as consistent defining conditions that determine $\overline{F}^{(g,h)}$. Nonetheless, the definition of the wave-function (\ref{wave_involution}) based on these conditions has two important drawbacks. First, it cannot be generalized to the $\beta$-deformed case, i.e. in the $\beta$-deformed model the condition (\ref{dF-W}) does not hold. Second, (\ref{wave_involution}) is inconsistent with the form of (i.e. it is not annihilated by) the quantum curves derived in section \ref{sec-quantum}. However, in genus zero examples we will show that there exists another quantum curve that annihilates (\ref{wave_involution}). Moreover, we also show that (at least in genus zero examples that we consider), the wave-functions (\ref{psi-n}) and (\ref{wave_involution}) at level $(n+1)$ are related by a simple shift of the 't Hooft coupling $\mu$
\be
\psi_{\beta=1,n}(x) \vert_{\mu\to\mu - \frac{n\hbar}{2}} = \overline{\psi}_{\beta=1,n}(x).   \label{psi-refpts-relation}
\ee

To sum up, the following picture arises. From the refined topological recursion one can reconstruct the asymptotic expansion of the wave-function $\widehat{\psi}_{\alpha}(x)$ defined in (\ref{Psi-alpha-bis}), choosing the infinity as the reference point in integrals in (\ref{det_op_exp}). This wave-function, for specific values of $\alpha$ corresponding to singular vectors (\ref{alpha-rs-2}), is annihilated by quantum curves determined in section \ref{sec-quantum}. Moreover, in the unrefined case $\beta=1$, one can introduce another quantum curve, that annihilates wave-functions defined via integrals with different (i.e. the conjugate) reference points and satisfying conditions (\ref{dF-W}) and (\ref{convolution}). 

In section \ref{sec-examples} we will illustrate various examples of wave-functions and quantum curves, reconstructed as explained above. However, prior to that, in the rest of this section we provide more details and some new results concerning the refined version of the topological recursion.


\subsection{More (refined) details}  \label{ssec-beta-details}

We present now more details about the refined version of the topological recursion. Its crucial ingredient is the spectral curve introduced in (\ref{spcurve}), i.e. an algebraic curve that in the unrefined case encodes distribution of eigenvalues of a matrix model. In the following, we assume that the spectral curve has $s$ cuts and takes form
\be
\begin{split}
& y(x)=M(x)\sqrt{\sigma(x)},    \\
&
\sigma(x)=\prod_{i=1}^{2s}(x-q_i),\ \ \ \ M(x)=c\prod_{i=1}^f(x-\alpha_i)^{m_i},
\label{spectral_gen}
\end{split}
\ee
where $M(x)$ is called the moment function. The spectral curve is a Riemann surface that consists of two sheets, so that two values $y(x)$ are assigned to a given $x$. The first sheet, also called the physical sheet, is defined by the condition
\be
 \sqrt{\sigma(x)} \underset{x\to\infty}{\simeq}  x^{s}, 
\ee
and the second sheet is characterized by
\be
\sqrt{\sigma(x)} \underset{x\to\infty}{\simeq} - x^{s}.
\ee
For a point $x$ in the physical sheet, the corresponding point in the second sheet is called a conjugate point and it is denoted by $\overline{x}$. It follows that
\be
y(\overline{x})=-y(x),\qquad  \sqrt{\sigma({\overline{x}})} = -\sqrt{\sigma(x)},\qquad M(\overline{x})=M(x).
\ee
The two sheets meet at $2s$ points $q_{i}$ called branch points, characterized by the condition $\overline{q}_{i}=q_{i}$. Pairs of branch points form components of the branch cut $D=\bigcup_{i=1}^{s}D_i$, $D_i=[q_{2i-1}, q_{2i}]$, of the spectral curve. Let $\mathcal{A}=\bigcup_{i=1}^{s}\mathcal{A}_{i}$ denote the counterclockwise contour surrounding the branch cut. The branch points $q_i$ can be then determined by $2s$ conditions
\begin{align}
\label{bp_cond_1}
&
\bullet\ \textrm{asymptotic condition}:\ \
\oint_{\mathcal{A}}\frac{dz}{2\pi i}\frac{z^kV'(z)}{\sqrt{\sigma(z)}}=2\mu\delta_{k,s},\ \ \ k=0,1,\ldots,s,\\
&
\bullet\ \textrm{filling fraction}:\ \ 
\oint_{\mathcal{A}_{i}}y(z)dz=-4\pi i\beta^{1/2}\hbar N_i,\ \ \ i=1,\ldots,s-1,
\label{fill_frac}
\end{align}
where $N_i$ denotes the number of eigenvalues in the branch cut $D_i$; note that $\sum_{i=1}^{s}N_i=N$.  In what follows we analyze in detail, in particular, the one-cut ($s=1$) case, where $\sigma(x)$ takes form 
\begin{equation}
\sigma(x)=(x-a)(x-b). \label{one_cut_ab}
\end{equation}

Let us provide now detailed expressions for the multi-differentials $W^{(g,h)}_{\ell}$ introduced in (\ref{W_diff_asy}). First, the disk differential is given by \cite{Migdal:1984gj, Ambjorn:1992gw, Akemann:1996zr, Eynard:2004mh} (see also a review \cite{Marino:2004eq})
\begin{equation}
W^{(0,1)}_{0}(x)=\frac12\oint_{\mathcal{A}}\frac{d\lambda}{2\pi i}\frac{V'(\lambda)}{x-z}\sqrt{\frac{\sigma(x)}{\sigma(z)}}dx,
\label{W010_s}
\end{equation}
and it encodes the spectral curve $y=y(x)$ in a way already presented in (\ref{yVomega}), i.e.
\begin{equation}
W^{(0,1)}_{0}(x)=\frac12\big(V'(x)-y(x)\big)dx.
\label{W010_ss}
\end{equation}

The leading differential (annulus contribution) in the asymptotic expansion (\ref{W_diff_asy}) for $h=2$ is given by \cite{Eynard:2004mh}
\begin{equation}
W^{(0,2)}_{0}(x_1,x_2)=B(x_1,x_2)-\frac{dx_1dx_2}{(x_1-x_2)^2},
\end{equation}
where the Bergman kernel $B(x_1,x_2)$ is a bilinear differential with no pole except $x_1=x_2$, and defined by the conditions:
\be
\begin{split}
&
\bullet\ B(x_1,x_2)\mathop{\sim}_{x_1 \to x_2} \frac{dx_1dx_2}{(x_1-x_2)^2}+\mathrm{(regular\  terms)}   \\
&
\bullet\ \oint_{x_2\in{\mathcal{A}}_{i}}B(x_1,x_2)=0,\ \ \ i=1,\ldots,s-1.
\end{split}
\ee
In the one-cut case (\ref{one_cut_ab}) the Bergman kernel is given by
\begin{equation}
B(x_1,x_2)=\frac{dx_1dx_2}{2(x_1-x_2)^2}\bigg(1+\frac{x_1x_2-\frac12(a+b)(x_1+x_2)+ab}{\sqrt{\sigma(x_1)\sigma(x_2)}}\bigg).
\label{Berg_one}
\end{equation}

The subleading differential (M\"obius strip contribution) in the $\hbar$ expansion (\ref{W_diff_asy}), for $h=1$, is given by \cite{Chekhov:2005rr, Chekhov:2006rq, Brini:2010fc}
\begin{equation}
W^{(0,1)}_1(x)=\oint_{\mathcal{A}}\frac{1}{2\pi i}\frac{dS(x,z)}{y(z)}
\frac{\partial}{\partial z}W^{(0,1)}_0(z).
\end{equation}
Here $dS(x_1,x_2)$ is the third type differential which is a 1-form in $x_1$ and a multivalued function of $x_2$, defined by the conditions:
\be
\begin{split}
&
\bullet\ dS(x_1,x_2)\mathop{\sim}_{x_1 \to x_2} \frac{dx_1}{x_1-x_2}+\mathrm{reg.},\ \ \ \
\bullet\ dS(x_1,x_2)\mathop{\sim}_{x_1 \to \overline{x}_2} -\frac{dx_1}{x_1-x_2}+\mathrm{reg.},  \\
&
\bullet\ \oint_{x_2\in{\mathcal{A}}_{i}}dS(x_1,x_2)=0,\ \ \ i=1,\ldots,s-1,
\label{third_type_def}
\end{split}
\ee
where ``reg.'' denotes regular (non-singular) terms, and $\overline{x}_2$ is the conjugate point. Assuming the analyticity of $V''(x)$ inside $\mathcal{A}$, from (\ref{W010_ss}) one obtains
\begin{equation}
W^{(0,1)}_1(x)=-\frac12\oint_{\mathcal{A}}\frac{dz}{2\pi i}\frac{y'(z)}{y(z)}dS(x,z).
\label{W011_assume}
\end{equation}
In the one-cut case (\ref{one_cut_ab}), the third type differential (\ref{third_type_def}) is given by
\begin{equation}
dS(x_1,x_2)=\frac{\sqrt{\sigma(x_2)}}{\sqrt{\sigma(x_1)}}\frac{dx_1}{x_1-x_2},
\label{third_type_one}
\end{equation}
and an explicit formula for the M\"obius strip differential (\ref{W011_assume}) takes form \cite{Brini:2010fc}
\begin{equation}
W^{(0,1)}_1(x)=-\frac{dy(x)}{2y(x)}+\frac{dx}{2\sqrt{\sigma(x)}}\bigg[1+\sum_{i=1}^fm_i\bigg(1+\frac{\sqrt{\sigma(\alpha_i)}}{x-\alpha_i}\bigg)\bigg].
\label{W011_one}
\end{equation}

The multi-differentials $W^{(0,1)}_{0}(x)$, $W^{(0,2)}_{0}(x_1,x_2)$ and $W^{(0,1)}_1(x)$ presented above can be thought of as initial conditions for the refined recursion relations. To present these relations it is convenient to define
\be
\begin{split}
&
\mathcal{W}^{(0,1)}_{0}(x)=0,\ \ \ \mathcal{W}^{(0,2)}_{0}(x_1,x_2)=W^{(0,2)}_{0}(x_1,x_2)+\frac{dx_1dx_2}{2(x_1-x_2)^2},   \\
&
\mathcal{W}^{(g,h)}_{\ell}(x_H)=W^{(g,h)}_{\ell}(x_H)\ \textrm{for}\ (g,h,\ell)\neq(0,1,0),\ (0,2,0),
\end{split}
\ee
and for a multilinear differential $f(x,x_1,\ldots,x_h)dxdx_1\cdots dx_h$ to denote
\begin{equation}
\frac{f(x,x_1,\ldots,x_h)dxdx_1\cdots dx_h}{dx}=f(x,x_1,\ldots,x_h)dx_1\cdots dx_h.
\end{equation}
In terms of the data above, the statement of the refined topological recursion \cite{Chekhov:2006rq, Chekhov:2010xj, Brini:2010fc} is that the differentials $W^{(g,h)}_{\ell}(x_H)$ for $(g,h,\ell)\neq(0,1,0)$, $(0,2,0)$, $(0,1,1)$ satisfy the relations:
\begin{empheq}[box=\fbox]{equation}
\begin{split}
&\quad W^{(g,h+1)}_{\ell}(x,x_H)=\oint_{\mathcal{A}}\frac{1}{2\pi i}\frac{dS(x,z)}{y(z)dz}\bigg(W^{(g-1,h+2)}_{\ell}(z,z,x_H)  + \\
&
\hspace{8em}+\sum_{k=0}^{g}\sum_{n=0}^{\ell}\sum_{\emptyset=J\subseteq H}\mathcal{W}^{(g-k,|J|+1)}_{\ell-n}(z,x_J)\mathcal{W}^{(k,|H|-|J|+1)}_{n}(z,x_{H \backslash J}) + \quad  \label{ref_loop_eq}
\\
& \hspace{8em}+dz^2\frac{\partial}{\partial z}\frac{W^{(g,h+1)}_{\ell-1}(z,x_H)}{dz}\bigg) 
\end{split}
\end{empheq}
where $H=\{1,2,\ldots,h\}\supset J=\{i_1,i_2,\ldots,i_j\}$, and $H\backslash J=\{i_{j+1},i_{j+2},\ldots,i_h\}$. 

In case $W^{(g,h)}_{\ell}$ come from a matrix model as in (\ref{W_diff_asy}), the relations (\ref{ref_loop_eq}) should be understood as a theorem that relates various expectation values. On the other hand, for a large class of algebraic curves -- not necessarily spectral curves (\ref{spectral_gen}) of some matrix model -- (\ref{ref_loop_eq}) can be regarded as defining relations for multi-differentials $W^{(g,h)}_{\ell}$.


\subsection{One-cut solution in the Zhukovsky variable}    \label{ssec-zhukovsky}

Let us present now the formalism of the refined topological recursion in the one-cut case (\ref{one_cut_ab}) with $a<b$. First, we introduce the Zhukovsky variable ${\mathrm{z}}$ 
\begin{equation}
x({\mathrm{z}})=\frac{a+b}{2}-\frac{a-b}{4}({\mathrm{z}}+{\mathrm{z}}^{-1}).
\label{Zhukovsky_v}
\end{equation}
In terms of this variable the spectral curve  (\ref{spectral_gen}) takes form
\be
y(x)=M(x)\sqrt{\sigma(x)}, \qquad  \sqrt{\sigma(x)}=\frac{b-a}{4}({\mathrm{z}}-{\mathrm{z}}^{-1}),
\label{spectral_one}
\ee
and its first and the second sheet are mapped respectively to the outside and inside of the unit disk $|{\mathrm{z}}|\le 1$, while the branch points $x=a, b$ are mapped respectively to ${\mathrm{z}}=-1, +1$. For completeness, also note that
\be
\frac{dx}{\sqrt{\sigma(x)}}=\frac{d{\mathrm{z}}}{{\mathrm{z}}},\qquad
\frac{\sqrt{\sigma(x_2)}}{x_1-x_2}=\frac{{\mathrm{z}}_1}{{\mathrm{z}}_1-{\mathrm{z}}_2}-\frac{{\mathrm{z}}_1}{{\mathrm{z}}_1-{\mathrm{z}}_2^{-1}},
\ee
and the third type differential (\ref{third_type_one}) takes form
\be
d\widehat{S}({\mathrm{z}}_1,{\mathrm{z}}_2)\equiv dS(x_1({\mathrm{z}}_1),x_2({\mathrm{z}}_2))=\frac{d{\mathrm{z}}_1}{{\mathrm{z}}_1-{\mathrm{z}}_2}-\frac{d{\mathrm{z}}_1}{{\mathrm{z}}_1-{\mathrm{z}}_2^{-1}}.    \label{third_type_z}
\ee
Furthermore, under the map (\ref{Zhukovsky_v}) the zeros or poles $\alpha_i$ of the moment function $M(x)$ in the spectral curve (\ref{spectral_gen}) are mapped to $2f$ points ${s}_i^{\pm 1}$ (so that we can assume $|{s}_i|>1$), $i=1,\ldots,f$,
\begin{equation}
\alpha_i({s}_i)=\frac{a+b}{2}-\frac{a-b}{4}({s}_i+{s}_i^{-1}), \qquad |{s}_i|>1
\label{alpha_s}
\end{equation}
and therefore
\begin{equation}
M(x)=c\prod_{i=1}^f(x-\alpha_i)^{m_i}=c\prod_{i=1}^f\left(\frac{b-a}{4}\frac{({\mathrm{z}}-s_i)({\mathrm{z}}-s_i^{-1})}{{\mathrm{z}}}\right)^{m_i}.
\end{equation}
We also define
\be
\begin{split}
&
\widehat{y}({\mathrm{z}})d{\mathrm{z}}=y(x({\mathrm{z}}))dx,
\\
& \widehat{W}^{(g,h)}_{\ell}({\mathrm{z}}_1,\ldots,{\mathrm{z}}_h)=W^{(g,h)}_{\ell}(x_1({\mathrm{z}}_1),\ldots,x_h({\mathrm{z}}_h)),
\label{loop_var_ch_zhu}
\end{split}
\ee
so that in particular the annulus differential (Bergman kernel) (\ref{Berg_one}) and the M\"obius strip differential (\ref{W011_one}) take form
\begin{align}
&
\widehat{W}^{(0,2)}_{0}({\mathrm{z}}_1,{\mathrm{z}}_2)=\frac{d{\mathrm{z}}_1d{\mathrm{z}}_2}{({\mathrm{z}}_1{\mathrm{z}}_2-1)^2},\\
&
\widehat{W}^{(0,1)}_1({\mathrm{z}})=\bigg(\frac{1}{{\mathrm{z}}}-\frac{1}{2({\mathrm{z}}-1)}-\frac{1}{2({\mathrm{z}}+1)}+\sum_{i=1}^fm_i\Big(\frac{1}{{\mathrm{z}}}-\frac{1}{{\mathrm{z}}-s_i^{-1}}\Big)\bigg)d{\mathrm{z}}.
\end{align}
Furthermore, we introduce
\be
\begin{split}
&
\widehat{\mathcal{W}}^{(0,1)}_{0}({\mathrm{z}})=0,\ \ \ \widehat{\mathcal{W}}^{(0,2)}_{0}({\mathrm{z}}_1,{\mathrm{z}}_2)=\widehat{W}^{(0,2)}_{0}({\mathrm{z}}_1,{\mathrm{z}}_2)+\frac{({\mathrm{z}}_1^2-1)({\mathrm{z}}_2^2-1)d{\mathrm{z}}_1d{\mathrm{z}}_2}{2({\mathrm{z}}_1-{\mathrm{z}}_2)^2({\mathrm{z}}_1{\mathrm{z}}_2-1)^2},  \\
&
\widehat{\mathcal{W}}^{(g,h)}_{\ell}({\mathrm{z}}_H)=\widehat{W}^{(g,h)}_{\ell}({\mathrm{z}}_H)\ \textrm{for}\ (g,h,\ell)\neq(0,1,0),\ (0,2,0).
\end{split}
\ee

Using these definitions and relations, the refined topological recursion (\ref{ref_loop_eq}) in the Zhukovsky variable is expressed as the following relation  \cite{Marchal:2011iu}:
\begin{equation}
\widehat{W}^{(g,h+1)}_{\ell}({\mathrm{z}},{\mathrm{z}}_H)=\oint_{\widetilde{\mathcal{A}}}\frac{1}{2\pi i}\frac{d\widehat{S}({\mathrm{z}},\zeta)}{\widehat{y}(\zeta)d\zeta}\mathrm{Rec}^{(g,h+1)}_{\ell}(\zeta,{\mathrm{z}}_H),
\label{ref_loop_eq_z}
\end{equation}
for differentials $\widehat{W}^{(g,h)}_{\ell}(\mathrm{z}_H)$ with $(g,h,\ell)\neq(0,1,0)$, $(0,2,0)$, $(0,1,1)$, where $\widetilde{\mathcal{A}}$ denotes the contour surrounding the unit disk $|\zeta|=1$, and
\be
\begin{split}
\mathrm{Rec}^{(g,h+1)}_{\ell}(\zeta,{\mathrm{z}}_H)&=
\widehat{W}^{(g-1,h+2)}_{\ell}(\zeta,\zeta,{\mathrm{z}}_H)  +  \\
&\ \ \
+\sum_{k=0}^{g}\sum_{n=0}^{\ell}\sum_{\emptyset=J\subseteq H}\widehat{\mathcal{W}}^{(g-k,|J|+1)}_{\ell-n}(\zeta,{\mathrm{z}}_J)\widehat{\mathcal{W}}^{(k,|H|-|J|+1)}_{n}(\zeta,{\mathrm{z}}_{H \backslash J}) +  \\
&\ \ \ 
+d\zeta^2\left(\frac{\partial}{\partial\zeta}
+\frac{\partial^2\zeta}{\partial{w}^2}\left(\frac{\partial{w}}{\partial\zeta}\right)^2\right)\frac{\widehat{W}^{(g,h+1)}_{\ell-1}(\zeta,{\mathrm{z}}_H)}{d\zeta},
\label{ref_loop_rec}
\end{split}
\ee
with ${w}=\frac{a+b}{2}-\frac{a-b}{4}(\zeta+\zeta^{-1})$ and $|\zeta|>1$. 

In the Zhukovsky variable the integrand of (\ref{ref_loop_eq_z}) has no branch cut, so the integration can be expressed as the summation of residues inside the unit disk $|\zeta|=1$. In appendix \ref{app:integrands_rec} we write down explicitly several expressions for $\mathrm{Rec}^{(g,h)}_{\ell}$ that are used in subsequent calculations.


\newpage 

\section{Various specific models}    \label{sec-examples}

In this section we show that various results that we derived so far for a general matrix model (with generic, independent times in the potential) can be specialized to matrix models with specific, polynomial or logarithmic potentials. In such cases we present explicit form of quantum curves and wave-functions at various levels, and illustrate the use of the refined version of the topological recursion, etc. In particular, in the last example we show how various ingredients familiar in the study of Liouville theory or minimal models, such as relevant representations of the Virasoro algebra or BPZ equations, arise from the specialization of our formalism to the multi-Penner matrix model. Also, we show that quantum curves at level 2 take form of differential equations that define orthogonal polynomials for a given model (in particular Hermite polynomials for the Gaussian model, and Laguerre polynomials for the Penner model); it would be interesting to find a similar interpretation of solutions to higher level quantum curve equations, and its interpretation in conformal field theory. Furthermore, following the discussion in section \ref{ssec-det}, we explicitly construct different quantum curves within the same model, for different choices of reference points in integrals defining wave-functions. Note that even though we use the refined version of the topological recursion only in examples with spectral curves of genus zero, the ensuing analysis is already non-trivial. We leave the analysis of examples with higher genus curves for future work.

Let us also stress that one needs to be careful in specializing our general formalism to specific matrix models with fixed potentials: only after taking derivatives with respect to times $t_n$ can these times be specialized to particular values. Therefore, in general, in order to work with quantum curves one may not be able to fix all times in the potential. However, amusingly, in various important cases, in particular in Penner and multi-Penner models, various combinations of time-derivatives acting on wave-functions can be expressed as derivatives with respect to $x$, and in consequence time-dependent quantum curves can be turned into time-independent ones. In such cases one can fix all times in the potential to requisite values, and the resulting time-independent curves are often more advantageous to work with than time-dependent ones.


\subsection{Gaussian model}

As the first example we consider a $\beta$-deformed Gaussian matrix model, i.e. the integral (\ref{matrix_def}) with the quadratic potential
\be
V(x) = \frac{1}{2} x^2.   \label{V-gaussian}
\ee
While this is the simplest matrix model that has been analyzed thoroughly from many perspectives, it appears that a representation of the  Virasoro algebra and the higher level quantum curves that we find below have not been identified before. As the Gaussian model encodes a lot of information relevant in the context of moduli spaces of Riemann surfaces, combinatorics of graphs, string theory models, etc., it is desirable to interpret the meaning of higher level quantum curves associated to this model in all those cases.


\subsubsection{Virasoro algebra and quantum curves}

The potential (\ref{V-gaussian}) is a specialization of (\ref{V}) with only one non-zero time $t_2=\frac{1}{2}$. Upon this specialization the operator (\ref{sp_hat_f_op}) simplifies and its derivatives vanish
\be
\hat{f}(x)  = -\epsilon_1 \epsilon_2 \partial_{t_0}  , \qquad \qquad
 \partial_x^k \hat{f}(x)=0 \qquad\textrm{for} \ k\geq 1,
\ee
where the action of $\partial_{t_0}$ is given in (\ref{partialt0}). It follows that operators (\ref{Lm_fer_op}) (when acting on $\widehat{\psi}_{\alpha}(x)$) take form  
\be
\widehat{L}_{-1} =\partial_x ,\qquad 
\widehat{L}_{-2} = -\frac{x^2 +\epsilon_1+\epsilon_2 -4\mu + 2\alpha}{\epsilon_1\epsilon_2}, \qquad
\widehat{L}_{-3} = -\frac{2x}{\epsilon_1\epsilon_2}, \qquad
\widehat{L}_{-4} = -\frac{1}{\epsilon_1\epsilon_2},
\label{Lm_fer_op-gaussian}
\ee
where $\mu$ is the 't Hooft parameter defined in (\ref{tHooft}), and $\widehat{L}_{-n}=0$ for $n\geq 5$. Note that in this way we obtain an interesting realization of a subalgebra of the Virasoro algebra (truncated at $\widehat{L}_{-5}$) in terms of differential operators in one variable $x$; as discussed in the next section, for polynomial potentials of higher degree in an analogous way we obtain realization of larger Virasoro subalgebras, in terms of differential operators in several variables.

The operators (\ref{Lm_fer_op-gaussian}) can be used as building blocks of higher level quantum curves acting on $\widehat{\psi}_{\alpha}(x)$. For example, at level 2, equations (\ref{BPZ-level-2bis}) or (\ref{BPZ-level-2}) take form
\be
\big(\widehat{L}_{-1}^2 + b^{\pm 2} \widehat{L}_{-2} \big) \widehat{\psi}_{\alpha}(x) = 
\Big( \partial_x^2-\frac{4\alpha^2}{\epsilon_1^2\epsilon_2^2}
(x^2+\epsilon_1+\epsilon_2-4\mu+2\alpha)
 \Big)  \widehat{\psi}_{\alpha}(x) = 0 ,\ \ \
\textrm{for}\ \alpha=-\frac{\epsilon_1}{2}, -\frac{\epsilon_2}{2}    \label{BPZ2-gaussian-pre}
\ee
As usual we use the notation $b^2=\frac{\epsilon_1}{\epsilon_2}=-\beta$, and a choice of a sign in the exponent of $b^{\pm 2}$ corresponds respectively to the choice of $\alpha=-\frac{\epsilon_1}{2}$ or $\alpha= -\frac{\epsilon_2}{2}$. Note that only for these values of $\alpha$ the above equalities hold and the above equation makes sense. Specializing to $\alpha=-\frac{\epsilon_1}{2}$ or $\alpha= -\frac{\epsilon_2}{2}$, and (for simplicity) taking unrefined limit $\beta=1$, from (\ref{BPZ2-gaussian-pre}) we obtain unrefined quantum curves
\be
\big(g_s^2\partial_x^2 -x^2 + 4\mu  \pm g_s \big) \widehat{\psi}_{\alpha}(x) =0, \label{BPZ2-gaussian}
\ee
and then (a unique) algebraic curve in the classical limit 
\be
y^2 -x^2 + 4\mu = 0,       \label{gaussian-class-Psi}
\ee
with $y$ identified with the classical limit of $g_s \partial_x$.

At level 3, quantum curves (\ref{BPZ-level-3}) (again, for simplicity) in the unrefined limit, take form
\be
\Big( g_s^3\partial_x^3 - 4(x^2 -4\mu \pm 2g_s)g_s\partial_x + 4g_s x \Big) \widehat{\psi}_{\alpha}(x) =0,
\label{BPZ3-gaussian}
\ee
where $\pm$ corresponds respectively to the choice of $\alpha=-\epsilon_1, -\epsilon_2$. In the classical limit we find an algebraic curve which factorizes as
\be
2y\big((2y)^2-4x^2+16\mu\big) = 0,  \label{gaussian-class-Psi-level3}
\ee
which is in agreement with (\ref{class-curves-higher-level}), and where (identifying the Planck constant as in (\ref{hbar-rs})) this time  $y$ represents the classical limit of $\frac{1}{2}g_s\partial_x$, while the second factor represents nothing but the curve (\ref{gaussian-class-Psi}) found at level 2.
This illustrates the statement that the quantum curve at level 3 can be regarded as a quantization of the classical curve.

Note that quantum curve equations at level 2 given above are essentially equations defining Hermite polynomials, which are orthogonal polynomials for the Gaussian model. It would be interesting to find analogous interpretation of solutions of higher level quantum curve equations and its meaning in conformal field theory.

Also note, that to obtain classical curves (\ref{gaussian-class-Psi}) and (\ref{gaussian-class-Psi-level3}), we first need to write (\ref{BPZ2-gaussian}) and (\ref{BPZ3-gaussian}) as equations for $\Psi_{\alpha}(x)=\widehat{\psi}_{\alpha}(x)/Z$; for the Gaussian model, which has no background dependence, this is achieved simply by dividing these equations by $Z$. Using standard expressions for singular vectors and the representation (\ref{Lm_fer_op-gaussian}), it is straightforward to write 
explicit form of quantum and classical curves at higher levels for the Gaussian model.




\subsubsection{Refined free energies from the topological recursion}   \label{sssec-gaussian}

For the Gaussian model with the potential $V(x)=x^2/2$ the spectral curve (\ref{spectral_gen}) takes form
\be
y^2 = x^2 - 4\mu,   \label{sp-curve-gaussian}
\ee
which of course agrees with the classical limit of the unrefined quantum curve (\ref{gaussian-class-Psi}).

For completeness, let us compute refined free energies following the approach presented in section \ref{ssec-zhukovsky}. Stable free energies, i.e. those for $\chi=2-2g-\ell<0$, are computed as (\ref{free_ene_rec}) via the refined topological recursion (\ref{ref_loop_eq_z}). To this end we need to evaluate $\mathrm{Rec}^{(g,h)}_{\ell}(\mathrm{z}_1,\ldots,\mathrm{z}_h)$, which are summarized in appendix \ref{app:integrands_rec}. From this computation, at various orders of $\chi$,  we get 
\begin{align}
&\chi=-1:\qquad F^{\mathrm{G}}_{1,1}=-\frac{1}{24{\mu}},\qquad   F^{\mathrm{G}}_{0,3}=0, \nonumber \\
&\chi=-2:\qquad F^{\mathrm{G}}_{2,0}=-\frac{1}{240{\mu}^2},\qquad 
F^{\mathrm{G}}_{1,2}=\frac{1}{180{\mu}^2}, \qquad  
F^{\mathrm{G}}_{0,4}=\frac{1}{720{\mu}^2}, \nonumber \\
&\chi=-3: \qquad F^{\mathrm{G}}_{2,1}=\frac{1}{240{\mu}^3},\qquad
F^{\mathrm{G}}_{1,3}=\frac{1}{720{\mu}^3},\qquad
F^{\mathrm{G}}_{0,5}=0,\nonumber \\
&\chi=-4:\qquad F^{\mathrm{G}}_{3,0}=\frac{1}{1008{\mu}^4},\quad
F^{\mathrm{G}}_{2,2}=\frac{11}{10080{\mu}^4},\quad
F^{\mathrm{G}}_{1,4}=-\frac{1}{840{\mu}^4},\quad
F^{\mathrm{G}}_{0,6}=-\frac{1}{5040{\mu}^4}.\nonumber
\end{align}
It follows that
\begin{align}
& F^{\mathrm{G}}_{\chi=-1} =
\gamma F^{\mathrm{G}}_{1,1}+\gamma^3 F^{\mathrm{G}}_{0,3}
=-\frac{1}{24{\mu}}\big(\beta^{1/2}-\beta^{-1/2}\big) = \widehat{F}_1^{\mathrm{odd}}(\beta,\mu),\nonumber\\
&F^{\mathrm{G}}_{\chi=-2} =
F^{\mathrm{G}}_{2,0}+\gamma^2 F^{\mathrm{G}}_{1,2}+\gamma^4 F^{\mathrm{G}}_{0,4}
=\frac{1}{720{\mu}^2}(\beta^2-5+\beta^{-2}) = \widehat{F}_2^{\mathrm{even}}(\beta,\mu),\nonumber\\
&F^{\mathrm{G}}_{\chi=-3}=
\gamma F^{\mathrm{G}}_{2,1}+\gamma^3 F^{\mathrm{G}}_{1,3}+\gamma^5 F^{\mathrm{G}}_{0,5}
=\frac{1}{720{\mu}^3}\big(\beta^{3/2}-\beta^{-3/2}\big) = \widehat{F}_2^{\mathrm{odd}}(\beta,\mu),\nonumber\\
&F^{\mathrm{G}}_{\chi=-4}=
F^{\mathrm{G}}_{3,0}+\gamma^2 F^{\mathrm{G}}_{2,2}+\gamma^4 F^{\mathrm{G}}_{1,4}+\gamma^6 F^{\mathrm{G}}_{0,6} = \nonumber \\
& \qquad \quad=-\frac{1}{10080{\mu}^4}(\beta+\beta^{-1})(2\beta^2-9+2\beta^{-2}) = \widehat{F}_3^{\mathrm{even}}(\beta,\mu),   \nonumber
\end{align}
correctly reproducing free energies of the Gaussian model presented in appendix \ref{ssec-gaussian}.

We can find unstable free energies following the discussion in appendix \ref{sec-Funstable} (however due to different normalization these unstable free energies take a little different form than those given in appendix \ref{ssec-gaussian}). The spectral curve (\ref{sp-curve-gaussian}) has one cut $D=[a,b]$ with $a=-b=-2\sqrt{\mu}$, so that evaluating the integral in (\ref{F_00_xi}) and (\ref{F_00_xi_f1}) we find respectively
\begin{equation}
-\frac{1}{4\pi i}\int_a^b dz y(z)V(z)=-\frac{1}{4}\mu^2, \qquad  \xi=-\mu+\mu \log \mu, 
\end{equation}
and it follows that
\begin{equation}
F_{0,0}^{\mathrm{G}}=\frac{1}{2}\mu^2\log \mu-\frac{3}{4}\mu^2.   \label{F00-gaussian}
\end{equation}
Furthermore, using (\ref{RP2_one_cut}), (\ref{torus_formula}) and (\ref{Klein_one_cut})  we find
\begin{equation}
\partial_{{\mu}}F^{\mathrm{G}}_{0,1}=1+\frac12\log {\mu},\qquad
F^{\mathrm{G}}_{1,0}=-\frac{1}{12}\log(16{\mu}),\qquad
F^{\mathrm{G}}_{0,2}=\frac{1}{12}\log(16{\mu}).
\end{equation}


\subsubsection{Quantum curves from the topological recursion}

We also verified that quantum curves (\ref{BPZ2-gaussian}), (\ref{BPZ3-gaussian}) and those level 4 (constructed following section \ref{ssec-level4}) agree with quantum curves and wave-functions obtained perturbatively from the topological recursion, as explained in section \ref{sec-refined}. To this end we constructed wave-functions $\widehat{\psi}_{\alpha}(x)$ in the form (\ref{det_op_exp}) from $W^{(g,h)}_\ell$ to several orders in $g_s$ and up to level 4, following details presented in section \ref{ssec-zhukovsky}, and checked that they satisfy relevant quantum curve equations. Having constructed $\widehat{\psi}_{\alpha}(x)$ in this way, we can also reconstruct quantum curves in the way presented in section \ref{ssec-reconstruct}. 

Similarly we reconstructed wave-functions $\overline{\psi}_{\beta=1,n}(x)$ for a choice of the conjugate point as a reference point (\ref{wave_involution}), verified that they are related to the original wave-functions via (\ref{psi-refpts-relation}), and reconstructed perturbatively corresponding quantum curves. In particular, the quantum curve at level 2 takes form 
\begin{equation}
\big(g_s^2\partial_x^2-x^2+4{\mu}\big) \overline{\psi}_{\beta=1,n=1}(x)=0,  \label{gaussian-limit-penner-qcurve}
\end{equation}
which agrees with the form of the Schr{\"o}dinger equation considered in \cite{Mulase:2012tm}, derived therein also based on the definition (\ref{wave_involution}).


\subsection{Cubic model and higher degree matrix models}

In a similar way we can analyze matrix models a potential $V(x)=\sum_{n=0}^k t_n x^n$, which is a polynomial of a fixed degree $k$, that depends on a finite set of times $(t_0,t_1,t_2,\ldots,t_k)$. For definiteness, let us consider the cubic matrix model with the potential
\be
V_3(x) =t_0 + t_1 x + t_2 x^2 + t_3 x^3.
\ee
We find that the operator (\ref{sp_hat_f_op}) and its derivatives, when acting on $\widehat{\psi}_{\alpha}(x)$, take form
\begin{align}
\begin{split}
\hat{f}(x) &= -8t_2\mu - 12t_3\mu x - 3t_3\epsilon_1\epsilon_2 \partial_{t_1} + 4 \alpha t_2 + 6\alpha t_3 x,   \\
\partial_x \hat{f}(x) &= 6t_3(\alpha-2\mu), \qquad \partial_x^k \hat{f}(x)=0 
\ \textrm{for} \ k\geq 2,
\end{split}
\end{align}
where $\mu$ is the 't Hooft parameter defined in (\ref{tHooft}). It follows that operators (\ref{Lm_fer_op}) take form  
\begin{gather}
\widehat{L}_{-2} = -\frac{V_3'^2  + V_3''(\epsilon_1+\epsilon_2)+ \hat{f}(x)}{\epsilon_1\epsilon_2}, \nonumber \\
\widehat{L}_{-3} = -\frac{2V_3' V_3'' + V_3'''(\epsilon_1+\epsilon_2) +6t_3(\alpha-2\mu)}{\epsilon_1\epsilon_2}, \label{Lm_fer_op-cubic}\\
\widehat{L}_{-4} = -\frac{V_3''^2 + V_3'V_3'''}{\epsilon_1\epsilon_2},\qquad 
\widehat{L}_{-5} = -\frac{V_3''V_3'''}{\epsilon_1\epsilon_2},\qquad 
\widehat{L}_{-6} = -\frac{V_3'''^2}{4\epsilon_1\epsilon_2},   \nonumber
\end{gather}
and as usual $\widehat{L}_0=\Delta_{\alpha}$ and $\widehat{L}_{-1} =\partial_x$, while $\widehat{L}_{-k}=0$ for $k\geq 7$. In the above expressions we can also consistently set $t_0, t_2$ and $t_3$ to some particular value; for example, for $t_0=t_2=0$ and $t_3=1$ we get
\begin{gather}
\widehat{L}_{-2} =- \frac{ 6(\epsilon_1+\epsilon_2-2\mu+\alpha) x + (t_1+3x^2)^2  }{\epsilon_1\epsilon_2} + 3\partial_{t_1},\nonumber \\
\widehat{L}_{-3} = -\frac{6(\epsilon_1+\epsilon_2-2\mu+\alpha+2t_1 x + 6x^3)}{\epsilon_1\epsilon_2}, \\
\widehat{L}_{-4} = -\frac{6(t_1+9x^2)}{\epsilon_1\epsilon_2},\qquad 
\widehat{L}_{-5} = -\frac{36x}{\epsilon_1\epsilon_2},\qquad 
\widehat{L}_{-6} = -\frac{9}{\epsilon_1\epsilon_2}.    \nonumber
 \end{gather}
In this way we obtain a realization of a subalgebra of Virasoro algebra in the space of functions in two variables $(x,t_1)$, which is truncated at $\widehat{L}_{-7}$. Using the above representations one can also easily write down quantum curves at arbitrary levels, using standard formulas for singular vectors and following the construction presented in section \ref{sec-quantum}. 

For more general models, restricting the potential to be a polynomial of degree $m$, we obtain a realization of a subalgebra of Virasoro algebra on a space of functions of several variables $(x,t_1,t_2,\ldots)$ that truncates at level $2m+1$, analogously to (\ref{Lm_fer_op-cubic}).


\subsection{Penner model}   \label{ssec-Penner}

As the next example we consider the Penner model, characterized by the potential
\be
V(x) = -x - \log(1-x).   \label{V-Penner}
\ee
Actually, we find it useful to introduce a one-parameter deformation of this potential
\be
V_t(x) = -x - \log(1-t x),    \label{Vt-Penner}
\ee
which corresponds to the following specialization of times in (\ref{V})
\be
t_0=0,\qquad t_1 = t-1,\qquad t_n = \frac{t^n}{n} \quad \textrm{for}\ n\geq 2.
\ee
We will refer to the model with such a potential as the $t$-deformed Penner model. 
Note that for the model with $t=1$, after rescaling
\begin{equation}
x \to T x,\qquad {\mu} \to T^2 {\mu},\qquad g_s \to T^2 g_s,    \label{Penner_Gaussian}
\end{equation}
and taking the limit $T\to 0$, all results presented in this section reduce to the results for the Gaussian model.


\subsubsection{Virasoro algebra and quantum curves}   \label{sssec-Penner-quantum}

The $t$-deformed Penner model has several interesting features. First of all, by (\ref{l_m123_g}) the Virasoro constraint $\ell_{-1}^{\alpha}(x) \psi_{\alpha}(x)=0$ can be rewritten as follows
\begin{align}
0  = \ell_{-1}^{\alpha}(x)     \psi_{\alpha}(x) =
\Big(-\partial_x + \frac{4\mu}{\epsilon_1\epsilon_2} (t-1) + t^2 \partial_t\Big)  \psi_{\alpha}(x),  \label{L-1alpha-Penner}
\end{align}
where we have used $\partial_xV_t(x)=t-1+t^2\partial_tV_t(x)$. This constraint equation (\ref{L-1alpha-Penner})
is equivalent to
\be
\left( \epsilon_1\epsilon_2(\partial_x-t^2 \partial_t)
- (t-1)(4\mu-2\alpha)\right)\widehat{\psi}_{\alpha}(x) = 0. \label{L-1alpha-Penner-bis}
\ee
This relation means that a derivative with respect to time deformation $t$ can be expressed in terms of $x$-derivative, which is crucial for the subsequent analysis. 
Furthermore, by (\ref{fhat-f}) an analogous computation leads to the following representation of the operator (\ref{sp_hat_f_op})
\begin{align}
\begin{split}
\hat{f}(x) \psi_{\alpha}(x)&=
\frac{2\epsilon_1 t^2}{1-tx}
\left\langle\sum_{a=1}^N\frac{\psi^{\textrm{ins}}_{\alpha}(x)}{1-tz_a}\right\rangle=
-\frac{\epsilon_1\epsilon_2 t^2}{1-tx}\left(t \partial_t + \partial_{t_0} \right) \psi_{\alpha}(x),    \label{fhat-Penner}
\end{split}
\end{align}
and similarly we find, on $\psi_{\alpha}(x)$ and $\widehat{\psi}_{\alpha}(x)$,
\be
\partial_x^k\hat{f}(x) =  -\frac{k! \epsilon_1\epsilon_2 t^{k+2}}{(1-tx)^{k+1}} \left(t\partial_t + \partial_{t_0} \right),
\ee
where the action of $\partial_{t_0}$ is given in (\ref{partialt0}).
Therefore $\hat{f}(x)$ and all its derivatives $\big(\partial_x^k\hat{f}\big)(x)$ can also be written in terms of a single derivative with respect to $t$, and we can take advantage of (\ref{L-1alpha-Penner}) and (\ref{L-1alpha-Penner-bis}) to write the action of these operators on $\psi_{\alpha}(x)$ and $\widehat{\psi}_{\alpha}(x)$ in terms of $\partial_x$; in particular
\begin{align}
\begin{split}
\hat{f}(x) \psi_{\alpha}(x) &=  -\frac{t}{1-tx}\left(\epsilon_1\epsilon_2 \partial_x+4\mu\right)  \psi_{\alpha}(x),
\\
\hat{f}(x) \widehat{\psi}_{\alpha}(x) &=  -\frac{t}{1-tx}
\left( \epsilon_1\epsilon_2 \partial_x+4\mu-2\alpha\right)  \widehat{\psi}_{\alpha}(x).  \label{fhat-Penner-x}
\end{split}
\end{align}
It follows that all $\widehat{L}_{-n}$ operators in (\ref{Lm_fer_op}) become simply ordinary differential operators in $x$. Therefore all higher level quantum curves in this model, which are built out of $\widehat{L}_{-n}$ operators, can be also expressed as ordinary differential operators in $x$, and become time-independent quantum curves. 





With the above representation of Virasoro operators, the differential equation (\ref{BPZ-level-2-psi}) for $\psi_{\alpha}(x)$ at level 2 in the $t$-deformed Penner model takes form
\be
\Big( \partial_x^2 + \frac{4\alpha}{\epsilon_1 \epsilon_2} \frac{ 1- t(1+x)}{1-tx} \partial_x + \frac{4\alpha^2}{\epsilon_1^2\epsilon_2^2} \frac{t^2}{1-tx}\left(\epsilon_1\epsilon_2 t \partial_t +4\mu\right)  \Big) \psi_{\alpha}(x) = 0.    \label{BPZ2-t-Penner-ddt}
\ee
Using (\ref{L-1alpha-Penner}) this can be written solely using derivatives $\partial_x$
\be
\Big( \partial_x^2 + \frac{4\alpha}{\epsilon_1 \epsilon_2} \frac{1+\alpha t - t(1+x)}{1-tx} \partial_x + \frac{16\mu\alpha^2}{\epsilon_1^2\epsilon_2^2}\frac{t}{1-tx} \Big) \psi_{\alpha}(x) = 0.    \label{BPZ2-t-Penner}
\ee
We also obtain the differential equation (\ref{BPZ-level-2bis}) for $\widehat{\psi}_{\alpha}(x)$ at level 2 in the $t$-deformed Penner model.
For $\alpha=-\epsilon_1 / 2$ this equation specializes to
\begin{align}
\Big( \epsilon_2^2 \partial_x^2  + & \frac{\epsilon_1\epsilon_2}{t^{-1}-x} \partial_x  +    \label{BPZ2-t-Penner-Psi-epsilon1}  \\ 
& - \frac{ x^2 +\left(4 \mu+\epsilon_1+2(1-t^{-1})\right)x
-4 \mu t^{-1}+\left(1-t^{-1}\right)^2
-\epsilon_1t^{-1}+\epsilon_1+\epsilon_2}{(t^{-1}-x)^2}  
\Big)  \widehat{\psi}_{-\frac{\epsilon_1 }{2}}(x) = 0    ,    \nonumber
\end{align}
and for $\alpha=-\epsilon_2/2$ we get the same equation, however with all $\epsilon_1$ and $\epsilon_2$ exchanged. To obtain the corresponding classical curves we need to rewrite the above equation as an equation for $\Psi_{\alpha}(x)=\widehat{\psi}_{\alpha}(x)/Z$. At this stage this is achieved simply by dividing the whole equation by $Z$, as we have already got rid of all time derivatives. After setting $\alpha=-\epsilon_1/2$  or $\alpha=-\epsilon_2/2$, in both cases the classical unrefined limit of (\ref{BPZ2-t-Penner-Psi-epsilon1}) can be written as
\be
(x - t^{-1})^2 y^2 - (x+1-t^{-1})^2 - 4\mu x+4\mu t^{-1}= 0,    \label{class-BPZ2-Pennert}
\ee
with $y$ identified with the classical limit of $g_s\partial_x$. For $t=1$ we get a classical curve for the original Penner model
\be
(1-x)^2 y^2 - x^2 - 4\mu x + 4\mu = 0.  \label{class-BPZ2-Penner}
\ee
Similarly one can analyze higher level quantum curves.

We also note that quantum curve equations at level 2 given above are essentially equations defining Laguerre polynomials, which are orthogonal polynomials for the Penner model. As already mentioned, it would be interesting to find analogous interpretation of solutions of higher level quantum curve equations.


\subsubsection{Refined free energies from the topological recursion}

We present now the use of the refined topological recursion for the Penner model. Once we determine its main ingredients, we present the computation of refined free energies in this model. In the next subsection we show that the wave-functions and quantum curves for the Penner model can also be reconstructed from the topological recursion. 

Let us consider the $t$-deformed Penner model with the potential (\ref{Vt-Penner}). Following the presentation in section \ref{ssec-beta-details}, we find that the spectral curve (\ref{spectral_gen}) of this model takes form
\begin{align}
&
y_{\mathrm{P}_t}(x)=M_{\mathrm{P}_t}(x)\sqrt{\sigma_{\mathrm{P}_t}(x)},\nonumber \\
& \sigma_{\mathrm{P}_t}(x)=(x+1-t^{-1})^2+4{\mu}x - 4{\mu}t^{-1},\qquad 
M_{\mathrm{P}_t}(x)=\frac{1}{t^{-1}-x},
\label{def_Penner_curve}
\end{align}
in agreement with (\ref{class-BPZ2-Pennert}). Amusingly, the spectral curve for the undeformed model
\begin{equation}
y_{\mathrm{P}}(x)=M_{\mathrm{P}}(x)\sqrt{\sigma_{\mathrm{P}}(x)},\qquad
\sigma_{\mathrm{P}}(x)=x^2+4{\mu}x-4{\mu}, \qquad
M_{\mathrm{P}}(x)=\frac{1}{1-x},
\label{Penner_curve}
\end{equation}
(that agrees with (\ref{class-BPZ2-Penner})) can be obtained from (\ref{def_Penner_curve}) not only by setting $t=1$, but also by a symplectic transformation 
\be
x\mapsto x-1+t^{-1}.   \label{x-symplectic-Penner}
\ee
This means that, apart from $F_{0,0}$, all free energies $F_{g,\ell}$ introduced in (\ref{free_energy_beta}) in this model are $t$-independent and agree with free energies associated to the spectral curve of the original Penner model (\ref{Penner_curve}). It is therefore sufficient to determine refined free energies associated to the curve (\ref{Penner_curve}), which we denote by $F^{\mathrm{P}}_{g,\ell}$. It is convenient to normalize them,  similarly as in (\ref{Gauss_Pen_norm}), by the free energies in the Gaussian model $F^{\mathrm{G}}_{g,\ell}$ found (from the topological recursion) in section \ref{sssec-gaussian}. To this end we define
\begin{equation}
F^{\mathrm{P}/\mathrm{G}}_{g,\ell}=F^{\mathrm{P}}_{g,\ell}-F^{\mathrm{G}}_{g,\ell},   \label{Penner-FPG}
\end{equation}
and in the following we check that these free energies (obtained from the refined topological recursion) are consistent with  $F_h^{\mathrm{even}}(\beta,\mu)$ and $F_h^{\mathrm{odd}}(\beta,\mu)$ found in \cite{GouHarJack} and summarized in (\ref{Fh_evenodd}) so that, in terms of $\gamma=\beta^{1/2}-\beta^{-1/2}$, we can write
\begin{equation}
F_h^{\mathrm{even}}(\beta,\mu)=\sum_{
\substack{
g,\ell=0\\
g+\ell=h
}}^{\infty}\gamma^{2\ell}
F^{\mathrm{P}/\mathrm{G}}_{g,2\ell},\ \ \ \ 
F_h^{\mathrm{odd}}(\beta,\mu)=\sum_{
\substack{
g,\ell=0\\
g+\ell=h
}}^{\infty}\gamma^{2\ell+1}
F^{\mathrm{P}/\mathrm{G}}_{g,2\ell+1}.
\label{penner_f_ag}
\end{equation}
To verify the above statement we use expressions for $\mathrm{Rec}^{(g,h)}_{\ell}(\mathrm{z}_1,\ldots,\mathrm{z}_h)$ given in appendix \ref{app:integrands_rec}, the redefinition (\ref{loop_var_ch_zhu}), and the reformulation of the refined topological recursion given in (\ref{ref_loop_eq_z}), to determine differentials $W^{(g,1)}_{\ell}(x)$ summarized in appendix \ref{ssec-compPenner}. Stable free energies are then computed from (\ref{free_ene_rec}), and we find the following results for (\ref{Penner-FPG})
\begin{align}
&\chi=-1:\qquad F^{\mathrm{P}/\mathrm{G}}_{1,1}=-\frac{{\mu}}{24(1+{\mu})},\qquad   F^{\mathrm{P}/\mathrm{G}}_{0,3}=0, \nonumber \\
&\chi=-2:\qquad F^{\mathrm{P}/\mathrm{G}}_{2,0}=\frac{{\mu}(2+{\mu})}{240(1+{\mu})^2},\qquad 
F^{\mathrm{P}/\mathrm{G}}_{1,2}=-\frac{{\mu}(2+{\mu})}{180(1+{\mu})^2}, \qquad  
F^{\mathrm{P}/\mathrm{G}}_{0,4}=-\frac{{\mu}(2+{\mu})}{720(1+{\mu})^2}, \nonumber \\
&\chi=-3: \qquad F^{\mathrm{P}/\mathrm{G}}_{2,1}=\frac{{\mu}(3+3{\mu}+{\mu}^2)}{240(1+{\mu})^3},\qquad
F^{\mathrm{P}/\mathrm{G}}_{1,3}=\frac{{\mu}(3+3{\mu}+{\mu}^2)}{720(1+{\mu})^3},\qquad
F^{\mathrm{P}/\mathrm{G}}_{0,5}=0,\nonumber \\
&\chi=-4:\qquad F^{\mathrm{P}/\mathrm{G}}_{3,0}=-\frac{{\mu}(2+{\mu})(2+2{\mu}+{\mu}^2)}{1008(1+{\mu})^4},\quad
F^{\mathrm{P}/\mathrm{G}}_{2,2}=\frac{11{\mu}(2+{\mu})(2+2{\mu}+{\mu}^2)}{10080(1+{\mu})^4},\nonumber \\
& \qquad \qquad \qquad F^{\mathrm{P}/\mathrm{G}}_{1,4}=\frac{{\mu}(2+{\mu})(2+2{\mu}+{\mu}^2)}{840(1+{\mu})^4},\quad
F^{\mathrm{P}/\mathrm{G}}_{0,6}=\frac{{\mu}(2+{\mu})(2+2{\mu}+{\mu}^2)}{5040(1+{\mu})^4}, \nonumber
\end{align}
where we ignored $\mu$-independent constants. From these results we find the agreements (\ref{penner_f_ag}).

Finally we determine unstable free energies, following the discussion in appendix \ref{sec-Funstable}. The spectral curve (\ref{Penner_curve}) has one cut $D=[a,b]$ with $a=-2\sqrt{\mu}-2\sqrt{\mu^2+\mu}$ and $b=-2\sqrt{\mu}+2\sqrt{\mu^2+\mu}$, so evaluating 
the integral in (\ref{F_00_xi}) we obtain
\begin{equation}
-\frac{1}{4\pi i}\int_a^b dz y(z)V(z)=-\frac{1}{2}\mu^2 + \frac{1}{4\pi i}\int_a^b dz y(z)\log(1-z).
\end{equation}
Setting $\Lambda=1-\epsilon$, $\epsilon\ll 1$, from (\ref{analytic_xi2}) we get
\begin{align}
\begin{split}
\frac{1}{4\pi i}\int_a^b dz y(z)\log(1-z)&=
\frac{1}{4}\xi
+\lim_{\epsilon\to 0}\Big(V(1-\epsilon)
-\int_b^{1-\epsilon}dz y(z)\Big) =  \\
&=
\frac{1}{4}\big(\xi
+(1+\mu)\log (1+\mu)-\mu \log\mu\big).
\end{split}
\end{align}
From (\ref{F_00_xi_f1}) the integration constant $\xi$ is now computed as
\begin{equation}
\xi=(1+\mu)\log (1+\mu)+\mu \log\mu-2\mu,
\end{equation}
and then the planar free energy (\ref{F_00_xi}) of the Penner model is determined as
\begin{equation}
F_{0,0}^{\mathrm{P}}=\frac{1}{2}(1+\mu)^2\log (1+\mu)-\frac{3}{4}(1+\mu)^2
+\frac{1}{2}\mu^2\log \mu-\frac{3}{4}\mu^2+\mu+\frac{3}{4}.
\end{equation}
Subtracting the Gaussian contribution (\ref{F00-gaussian}) we obtain the result $F^{\mathrm{P}/\mathrm{G}}_{0,0}=F_{0,0}^{\mathrm{P}}-F_{0,0}^{\mathrm{G}}$ that agrees with the planar free energy given in appendix \ref{ssec-penner}. Furthermore, computing (\ref{RP2_one_cut}) and subtracting the Gaussian contribution we get
\be
\partial_{{\mu}}F^{\mathrm{P}}_{0,1}=1+\frac12\log {\mu}-\frac12\log(1+{\mu}),  \qquad
\gamma\partial_{{\mu}}F^{\mathrm{P}/\mathrm{G}}_{0,1}=-\frac{\gamma}{2}\log(1+{\mu}),
\ee
in agreement with $\partial_{{\mu}}F_0^{\mathrm{odd}}(\beta,\mu)$ given in appendix \ref{ssec-penner}. Finally, from (\ref{torus_formula}) and (\ref{Klein_one_cut}) 
\be
F^{\mathrm{P}}_{1,0}=-\frac{1}{12}\log\big(16{\mu}(1+{\mu})\big), \qquad
F^{\mathrm{P}}_{0,2}=\frac{1}{12}\log\big(16{\mu}(1+{\mu})\big), 
\ee
so that
\begin{equation}
F^{\mathrm{P}/\mathrm{G}}_{1,0}+\gamma^2F^{\mathrm{P}/\mathrm{G}}_{0,2}=\frac{-1+\gamma^2}{12}\log(1+{\mu})
\end{equation}
coincides with $F_1^{\mathrm{even}}(\beta,\mu)$ given in appendix \ref{ssec-penner}.


\subsubsection{Quantum curves from the topological recursion}\label{sssec-Penner-gs}

We also verified that higher level quantum curves discussed in section \ref{sssec-Penner-quantum}, up to level 4, are consistent with quantum curves and wave-functions determined from the topological recursion. To this end we determined $W^{(g,h)}_\ell$ to several orders in $g_s$ following the discussion in section \ref{ssec-zhukovsky}; from these results we constructed wave-functions $\widehat{\psi}_{\alpha}(x)$ in the form (\ref{det_op_exp}) and checked that they satisfy quantum curve equations up to level 4. From $\widehat{\psi}_{\alpha}(x)$ obtained in this way, quantum curves can also be reconstructed as discussed in section \ref{ssec-reconstruct}. 


A modified wave-function $\overline{\psi}_{\beta=1,n=1}(x)$ defined as in (\ref{wave_involution}) can be also considered via multi-resolvents integrated from the reference point identified with the conjugate point $\overline{x}$. Such a wave-function should be related to the wave function (\ref{psi-n}) as in (\ref{psi-refpts-relation}) and should satisfy a modified quantum curve equation. We confirmed that this is the case, and found that the quantum curve at level 2 for the wave-function corresponding to $\overline{x}$ reference point takes form
\begin{equation}
\left(g_s^2\partial_x^2
-\frac{g_s^2}{1-x}\partial_x
-\frac{x^2+4{\mu}x-4{\mu}}{(1-x)^2}\right) \overline{\psi}_{\beta=1,n=1}(x)=0.
\label{Penner_unref2}
\end{equation}
In the Gaussian limit (\ref{Penner_Gaussian}) this equation reduces to (\ref{gaussian-limit-penner-qcurve}).


\subsubsection{$x$-expansion of the wave-function}    \label{sssec-Penner-x}

Finally, following the discussion in section \ref{ssec-x-expansion}, we illustrate how to extract a dependence of the wave-function $\psi_{\alpha}(x)$ on $x$ perturbatively, from the knowledge of a time-dependent differential equation in the $t$-deformed Penner model. For definiteness we choose the value $\alpha=-\epsilon_1/2$, and assume that the wave-function takes form as in (\ref{psi-x-expansion}) 
\begin{align}
\psi_{-\frac{\epsilon_1}{2}}(x)  = e^{S(x,t)}, \quad
S(x,t) = -\frac{2\mu}{\epsilon_2}\log x+\sum_{k=0}^{\infty}S_k(t)x^{-k}.
\end{align}
Substituting this form to the equation (\ref{BPZ2-t-Penner-ddt}) we obtain a hierarchy of differential equations for $S_k(t)$, which for $k \leq 3$ take form
\begin{equation}
\begin{split}
&
0 = \mu(1-t)+t^2\hbar^2S'_0(t),\\
&
0 = \mu\hbar  + \mu^2\sqrt{\beta} 
+ \hbar S_1(t) +t \hbar^2  \sqrt{\beta} S'_0(t)
+t^2 \hbar^2 \sqrt{\beta} S'_1(t),\\
&
0=2 \hbar S_1(t) + \sqrt{\beta} S_1(t) + 2 \mu \sqrt{\beta} S_1(t) 
+  2 \sqrt{\beta} S_2(t) + \hbar \beta S'_0(t) 
+  \hbar t \beta S'_1(t) + \hbar t^2\beta S'_2(t).   \nonumber
\end{split}
\end{equation}
Then in particular we find that the whole $t$-dependence of the partition function $Z$ of the $t$-deformed Penner model is given by
\be
Z=e^{S_{0}(t) + \{t\textrm{-independent terms}\}},\quad S_{0}(t) = \frac{\mu}{\hbar^2}\Big(\frac{1}{t} +\log t \Big).
\ee 
This structure of the partition function can also be seen upon the substitution of (\ref{asymptotics-psi}) into (\ref{L-1alpha-Penner-bis}). Note that this structure is in agreement with the statement that apart from the leading free energy $F_{0,0}$, all other free energies in the $t$-deformed Penner model are the same as in the undeformed model due to their invariance under symplectic transformation (\ref{x-symplectic-Penner}), and so the only dependence on $t$ can arise in the term $F_{0,0}$.


\subsection{Multi-Penner model and Liouville theory}   \label{ssec-multiPenner}

As the last example we consider the multi-Penner model with the potential
\begin{equation}
V(x)=\sum_{i=1}^{M}\alpha_i\log (x-x_i).
\label{m_penner_potential}
\end{equation}
It is well known \cite{Dotsenko:1984nm,Mironov:2010zs,Mironov:2010su,Dijkgraaf:2009pc} that the matrix model partition function $Z$ in (\ref{matrix_def}) with this potential computes, in minimal models or in Liouville theory, correlation functions of $(M+1)$ primary fields with momenta $\alpha_i$ and $\alpha_{\infty}$, inserted respectively at positions $x_i$ and at infinity on ${\IP}^1$. The primary field at infinity can also be removed by imposing the momentum conservation condition (\ref{m_penner_momemt_cons}). In this section we show that, more generally, various other objects that we introduced earlier -- such as the representation of Virasoro operators $\widehat{L}_{n}$, higher level quantum curves, etc. -- also reduce to familiar objects in minimal models or in Liouville theory for the above choice of potential. In this specialization the role of an infinite set of times $t_n$ in a general matrix model is played by parameters $x_i$ in (\ref{m_penner_potential}), i.e. positions of operator insertions. 

Before proceeding we also recall that, according to the AGT correspondence, the model with the potential (\ref{m_penner_potential}) describes a four-dimensional ${\cal N}=2$ $SU(2)^{M-2}$ superconformal linear quiver gauge theory \cite{Dijkgraaf:2009pc}; in particular for $M=2$ this model characterizes a theory of four free hypermultiplets. Therefore, via this link, various results that we present from the perspective of matrix models could also be linked to the realm of supersymmetric gauge theories and related topics, such as integrable models, topological string theory, Hitchin systems, etc.

We start our analysis with the observation, that for the choice (\ref{m_penner_potential}) the potential term in the matrix model integrand (\ref{matrix_def}) takes form
\be
e^{-\frac{\sqrt{\beta}}{\hbar}\sum_{a=1}^NV(z_a)} = \prod_{i=1}^M\prod_{a=1}^N (z_a-x_i)^{-\frac{2\alpha_i}{\epsilon_2}}.
\ee
Amusingly, each factor in this expression corresponding to fixed $i$ has the same form as the insertion of (\ref{fermion_alpha}) that defines wave-functions $\psi_{\alpha}(x)$ or $\widehat{\psi}_{\alpha}(x)$. Therefore for the potential (\ref{m_penner_potential}) these wave-functions can be interpreted as effectively representing correlation functions of $(M+2)$ primary fields in the presence of an additional field at infinity, or $(M+1)$ fields once the condition (\ref{m_penner_momemt_cons}) is imposed. This immediately suggests that quantum curve equations (\ref{Aalpha-psialpha}) in this case should reduce to differential equations for correlation functions of a number of primary fields, which include a distinguished field with a degenerate momentum $\alpha_{r,s}$ inserted at position $x$, represented by (\ref{fermion_alpha}). These are nothing but the BPZ equations, and in what follows we show that they indeed arise from our general formalism. Moreover, while in general these equations are time-dependent (i.e. for multi-Penner potential they include derivatives with respect to positions $x_i$), in some cases we will be able to rewrite them as time-independent equations by taking advantage of Virasoro constraints (\ref{vir_const_wave_2}), similarly as we did for the Penner model in section \ref{ssec-Penner}. In particular, at level 2 we obtain in this way familiar in minimal models or Liouville theory hypergeometric differential  equations for four-point correlation functions.

Let us discuss how various objects introduced earlier specialize upon the choice of potential (\ref{m_penner_potential}). First, we see that the operator $\widehat{f}(x)$ in (\ref{sp_hat_f_op})
takes form
\begin{equation}
\widehat{f}(x)=-\epsilon_1\epsilon_2\sum_{i=1}^{M}\frac{1}{x-x_i}\partial_{x_i}.  \label{fhat-multiPenner}
\end{equation}
In what follows it is useful to introduce an additional normalization factor and consider the following wave-function
\begin{equation}
\widetilde{\psi}_{\alpha}(x) = \widehat{\psi}_{\alpha}(x) \cdot \prod_{i\neq j}(x_i-x_j)^{-\frac{\alpha_i\alpha_j}{\epsilon_1\epsilon_2}}.
\label{m_penner_normalize}
\end{equation}
Then, rewriting (\ref{fhat-multiPenner}) accordingly, the representation of Virasoro algebra (\ref{Lm_fer_op}), when acting on $\widetilde{\psi}_{\alpha}(x)$, takes form
$\widetilde{L}_0=\Delta_{\alpha}, \widetilde{L}_{-1}=\partial_x$, and
\be
\widetilde{L}_{-n}=
\sum_{i=1}^M\Big(\frac{(n-1)\Delta_{\alpha_i}}{(x_i-x)^n}
-\frac{1}{(x_i-x)^{n-1}}\partial_{x_i}\Big),
\quad \textrm{for } n \ge 2.
\label{vir_m_penner}
\ee
Amusingly, these generators coincide with well-known expressions for Virasoro generators (\ref{tildeLp}) acting on correlation functions in conformal field theory.

Furthermore, consider the $SL(2,{\IC})$ subalgebra generated by Virasoro operators $\ell_{-1}^{\alpha}(x)$, $\ell_0^{\alpha}(x)$ and $\ell_{1}^{\alpha}(x)$ given in (\ref{l_m123_g}), 
which for the potential (\ref{m_penner_potential}) take form
\be
\begin{split}
\ell_{-1}^{\alpha}(x) & =\ell_{-1}^{\textrm{Witt}}(x)+
\sum_{i=1}^M \ell_{-1}^{\textrm{Witt}}(x_i),   \\
\ell_{0}^{\alpha}(x) & =\ell_{0}^{\textrm{Witt}}(x)+
\sum_{i=1}^M \ell_{0}^{\textrm{Witt}}(x_i)
-\frac{2\mu}{\epsilon_1\epsilon_2}\Big(2\mu+\epsilon_1+\epsilon_2
-2\alpha-2\sum_{i=1}^M\alpha_i\Big),  \\
\ell_{1}^{\alpha}(x) & =\ell_{1}^{\textrm{Witt}}(x)+
\sum_{i=1}^M \ell_{1}^{\textrm{Witt}}(x_i)
+\frac{4\mu}{\epsilon_1\epsilon_2}\Big(\alpha x+\sum_{i=1}^M \alpha_i
x_i\Big) +  \\
&\qquad
+\frac{2}{\epsilon_2}\Big(2\mu+\epsilon_1+\epsilon_2
-\alpha-\sum_{i=1}^M\alpha_i\Big)\sum_{a=1}^N z_a,
\label{l_1_m_penner}
\end{split}
\ee
where $\ell_n^{\textrm{Witt}}(x)=-x^{n+1}\partial_x$ are the generators of the Witt algebra (\ref{LWitt}), and as usual $\mu=\beta^{1/2}\hbar N$. These operators impose the constraints (\ref{vir_const_wave_2}) 
\begin{equation}
\ell_{-1}^{\alpha}(x)\psi_{\alpha}(x)=\ell_{0}^{\alpha}(x)\psi_{\alpha}(x)
=\ell_{1}^{\alpha}(x)\psi_{\alpha}(x)=0.
\label{sl2c_vir}
\end{equation}
Note that $\ell_{-1}^{\alpha}(x)$ and $\ell_{0}^{\alpha}(x)$ can be written as differential operators, however in general it is not possible to do so for $\ell_{1}^{\alpha}(x)$ due to its last summand. In conformal field theory language, the presence of this term signals the presence of a primary field at $x=\infty$ with the momentum
\begin{equation}
\alpha_{\infty}=2\mu+\epsilon_1+\epsilon_2-\alpha-\sum_{i=1}^M\alpha_i,
\label{m_penner_mom_inf}
\end{equation}
and in terms of this momentum we can introduce
\begin{align}
\begin{split}
\widetilde{\ell}_{-1}^{\alpha}(x) & =\ell_{-1}^{\textrm{Witt}}(x)+
\sum_{i=1}^M \ell_{-1}^{\textrm{Witt}}(x_i),   \\
\widetilde{\ell}_{0}^{\alpha}(x) & =\ell_{0}^{\textrm{Witt}}(x)+
\sum_{i=1}^M \ell_{0}^{\textrm{Witt}}(x_i)
-\Delta(\alpha)-\sum_{i=1}^M\Delta_{\alpha}+\Delta_{\alpha_{\infty}},
\label{m_penner_mvi}
\end{split}
\end{align}
that act on the normalized field $\widetilde{\psi}_{\alpha}(x)$ in (\ref{m_penner_normalize}). Now, for $M=2$, first two constraints in  (\ref{sl2c_vir}) are rewritten as $\widetilde{\ell}_{-1}^{\alpha}(x)\widetilde{\psi}_{\alpha}(x)=0$ and $\widetilde{\ell}_{0}^{\alpha}(x)\widetilde{\psi}_{\alpha}(x)=0$ and explicitly take form
\begin{align}
\begin{split}
&
\big((x_1-x_2)\partial_{x_1}
+ (x-x_2)\partial_{x}
+ \Delta_{\alpha}+\Delta_{\alpha_1} 
+ \Delta_{\alpha_2}-\Delta_{\alpha_{\infty}}\big)\widetilde{\psi}_{\alpha}(x)=0,
\\
&
\big((x_2-x_1)\partial_{x_2}
+ (x-x_1)\partial_{x}
+ \Delta_{\alpha} 
+ \Delta_{\alpha_1}
+ \Delta_{\alpha_2}-\Delta_{\alpha_{\infty}}\big)\widetilde{\psi}_{\alpha}(x)=0.  \label{ddx1ddx2-ddx}
\end{split}
\end{align}
Using these relations we can express derivatives $\partial_{x_1}$ and $\partial_{x_2}$ in terms of $\partial_x$, and in consequence write all the actions of Virasoro operators $\widetilde{L}_{-n}$ on $\widetilde{\psi}_{\alpha}(x)$ simply as ordinary differential operators in $\partial_x$. For example
\begin{equation}
\widetilde{L}_{-2}\widetilde{\psi}_{\alpha}(x)=
\bigg[-\sum_{i=1}^2 \frac{1}{x-x_i}\partial_x
+\sum_{i=1}^2 \frac{\Delta_{\alpha_i}}{(x-x_i)^2}
+\frac{\Delta_{\alpha_{\infty}}-\Delta_{\alpha_{1}}-\Delta_{\alpha_{2}}
-\Delta_{\alpha}}{(x-x_{1})(x-x_{2})}\bigg]\widetilde{\psi}_{\alpha}(x),
\label{m_penner_m2_2}
\end{equation}
while $\widetilde{L}_{-n}$ with $n>2$ can be found either by a direct substitution of (\ref{ddx1ddx2-ddx}) in (\ref{vir_m_penner}).

We can be also more specific and set the last term of $\ell_{1}^{\alpha}(x)$ in (\ref{l_1_m_penner}) to zero, simply by imposing the condition
\begin{equation}
2\mu+\epsilon_1+\epsilon_2-\alpha-\sum_{i=1}^M\alpha_i=0,
\label{m_penner_momemt_cons}
\end{equation}
which in conformal field theory is interpreted as the momentum conservation and demanding that there is no primary field at infinity. Taking this condition into account and including the normalization (\ref{m_penner_normalize}), all three operators (\ref{l_1_m_penner}) are transformed into expressions familiar in conformal field theory too
\begin{align}
\begin{split}
\widetilde{\ell}_{-1}^{\alpha}(x) & =\ell_{-1}^{\textrm{Witt}}(x)+
\sum_{i=1}^M \ell_{-1}^{\textrm{Witt}}(x_i),  \\
\widetilde{\ell}_{0}^{\alpha}(x) & =\ell_{0}^{\textrm{Witt}}(x)+
\sum_{i=1}^M \ell_{0}^{\textrm{Witt}}(x_i)
-\Delta_{\alpha}-\sum_{i=1}^M\Delta_{\alpha},  \\
\widetilde{\ell}_{1}^{\alpha}(x) & =\ell_{1}^{\textrm{Witt}}(x)+
\sum_{i=1}^M \ell_{1}^{\textrm{Witt}}(x_i)
-2\Delta_{\alpha}x-2\sum_{i=1}^M\Delta_{\alpha_i}x_i.
\label{m_penner_mv}
\end{split}
\end{align}
In terms of these operators the constraints (\ref{sl2c_vir}) take form $\widetilde{\ell}_{j}^{\alpha}(x)\widetilde{\psi}_{\alpha}(x)=0$ for $j=-1,0,1$, which implies that three among $M$ partial derivatives $\partial_{x_i}$ in Virasoro generators $\widetilde{L}_{-n}$ can be expressed in terms $\partial_x$. In particular, for $M=2$, with the momentum conservation (\ref{m_penner_momemt_cons}) condition imposed, we find that
\begin{equation}
\widetilde{L}_{-2}\widetilde{\psi}_{\alpha}(x)=
\bigg[-\sum_{i=1}^2 \frac{1}{x-x_i}\partial_x
+\sum_{i=1}^2 \frac{\Delta_{\alpha_i}}{(x-x_i)^2}
-\frac{\Delta_{\alpha_{1}}+\Delta_{\alpha_{2}}
+\Delta_{\alpha}}{(x-x_{1})(x-x_{2})}\bigg]\widetilde{\psi}_{\alpha}(x).   \label{L-2-mPenner-M2}
\end{equation}
For $M=3$, also imposing (\ref{m_penner_momemt_cons}), the constraints $\widetilde{\ell}_{j}^{\alpha}(x)\widetilde{\psi}_{\alpha}(x)=0$ for $j=-1,0,1$ take form
\begin{align}
\begin{split}
&
\Big((x_I-x_{I+1})(x_I-x_{I+2})\partial_{x_I}+(x-x_{I+1})(x-x_{I+2})\partial_x\Big)\widetilde{\psi}_{\alpha}(x) = \\
& =\Big( -2\Delta_{\alpha}x-2\sum_{i=1}^3\Delta_{\alpha_i}x_i+
\Big(\Delta_{\alpha}+\sum_{i=1}^3\Delta_{\alpha_i}\Big)
(x_{I+1}+x_{I+2})\Big)\widetilde{\psi}_{\alpha}(x),
\end{split}
\end{align}
where $I = 1,2,3$ (mod $3$), and in consequence the action of the Virasoro generator $\widetilde{L}_{-2}$ can take form
\begin{equation}
\widetilde{L}_{-2}\widetilde{\psi}_{\alpha}(x)=
\bigg[-\sum_{i=1}^3 \frac{1}{x-x_i}\partial_x
+\sum_{i=1}^3 \frac{\Delta_{\alpha_i}}{(x-x_i)^2}
+\sum_{I=1}^3 \frac{\Delta_{\alpha_I}-\Delta_{\alpha_{I+1}}-\Delta_{\alpha_{I+2}}
-\Delta_{\alpha}}{(x-x_{I+1})(x-x_{I+2})}\bigg]\widetilde{\psi}_{\alpha}(x).   \label{L-2-mPenner-M3}
\end{equation}
For both (\ref{L-2-mPenner-M2}) and (\ref{L-2-mPenner-M3}), the actions of other generators $\widetilde{L}_{-n}$ with $n>2$, as ordinary differential operators, can be easily obtained either by substituting the relevant constraints into (\ref{vir_m_penner}).
Note that taking the limit $x_{3}\to \infty$ and identifying $\alpha_3=\alpha_{\infty}$, (\ref{L-2-mPenner-M3}) reduces to (\ref{m_penner_m2_2}), and analogous statement holds for other $\widetilde{L}_{-n}$.

Finally we can consider specialization of our construction of quantum curves to the case of multi-Penner model, which brings us to another amusing point. In general quantum curves for multi-Penner model are expressed as in (\ref{chat-Lhat}), and substituting the representation of Virasoro operators given in (\ref{vir_m_penner}) we obtain differential equations for wave-functions $\widetilde{\psi}_{\alpha}(x)$ which have the same form as the original BPZ equations in conformal field theory (\ref{BPZ-cft}). Recall that these equations make sense only for momenta of the operator (\ref{fermion_alpha}) taking degenerate values $\alpha=\alpha_{r,s}$ given in (\ref{alpha-rs-2}). Including an additional insertion (\ref{fermion_alpha}) that defines the wave-function $\widetilde{\psi}_{\alpha}(x)$ itself, this wave-function is identified with a correlation function of $(M+1)$ fields with momenta $\alpha=\alpha_{r,s}$ and $\alpha_i$, $i=1,\ldots,M$, that are parameters of the potential (\ref{m_penner_potential}).

Furthermore, for $M=3$ quantum curves can be built from Virasoro operators (\ref{L-2-mPenner-M3}) and all other corresponding $\widetilde{L}_{-n}$. These quantum curves are written entirely in terms of $\partial_x$ and no other derivatives, and including an additional insertion (\ref{fermion_alpha}) that defines the wave-function itself, they impose conditions on four-point functions. Therefore for $M=3$ we obtain time-independent equations for four-point functions that include a field with a degenerate momentum. At level 2 these turn out to be hypergeometric differential equations, and take the same form as the original BPZ equations at level 2 in conformal field theory \cite{Belavin:1984vu}
\be
\big( \partial_x^2 + b^{\pm 2} \widetilde{L}_{-2} \big) \widetilde{\psi}_{\alpha}(x) = 0,
\ee
respectively for the choice of momentum $\alpha=\alpha_{2,1}$ and $\alpha=\alpha_{1,2}$, and
with $\widetilde{L}_{-2}$ given in (\ref{L-2-mPenner-M3}). In this case $\widetilde{\psi}_{\alpha}(x)$ is identified with the four-point function of fields with momenta $\alpha$ and $\alpha_i$ for $i=1,2,3$ that are parameters of the potential (\ref{m_penner_potential}). One can also immediately write down higher level BPZ equations, which take form (\ref{Aalpha-psialpha}) with actions of Virasoro operators given by (\ref{L-2-mPenner-M3}) and corresponding $\widetilde{L}_{-n}$. As usual, one can further simplify these equations by setting $x_1=0$, $x_1=1$, and $x_3=\infty$.



\newpage

\section{Epilogue}

To sum up, we have come a long, albeit hopefully enjoyable way, starting from a review of the structure of Virasoro singular vectors and BPZ equations in conformal field theory in section \ref{sec-virasoro}, and then in the end rederiving these results from the multi-Penner model. From this latter perspective all conformal field theory results arise just as a special case of a much more general formalism that involves matrix models with arbitrary potentials, depending on an infinite number of times. Specializing these potentials to various cases of interest we obtain families of quantum curves which are in one-to-one correspondence with singular vectors, and which can be thought of as infinite hierarchies of, in general, partial differential equations. These differential equations should capture interesting information in all systems -- be it topological string theory, supersymmetric gauge theories, knot theory, moduli spaces of Riemann surfaces, etc. -- which are related to matrix models by various dualities. We believe this information will be fascinating to reveal.


\acknowledgments{We thank Hidetoshi Awata, Hiroyuki Fuji,  Kohei Iwaki, Zbigniew Jask\'{o}lski, Hiroaki Kanno, Ivan Kostov, Motohico Mulase, and Marcin Pi{\k{a}}tek for insightful discussions and comments on the manuscript. We very much appreciate hospitality of the Simons Center for Geometry and Physics where parts of this work were done. This work is supported by the ERC Starting Grant no. 335739 \emph{``Quantum fields and knot homologies''} funded by the European Research Council under the European Union's Seventh Framework Programme, and the Ministry of Science and Higher Education in Poland.}

\newpage 

\appendix

\section{Unstable refined free energies}   \label{sec-Funstable}

Refined free energies $F_{g,\ell}$ are defined via the expansion (\ref{free_energy_beta}). Stable free energies, i.e. those with $2-2g-\ell<0$, are determined from $W^{(g,1)}_{\ell}$ as given in (\ref{free_ene_rec}). In this appendix we provide formulas for the remaining, unstable free energies, i.e. $F_{g,\ell}$ with $2-2g-\ell\ge 0$. We use the same notation as in section \ref{sec-refined}.

First, the sphere contribution to the free energy is given by \cite{Brezin:1977sv, Marino:2004eq}
\begin{equation}
F_{0,0}=-\mu\int_{D}dz \rho(z)V(z)+\mu^2\int_{D\times D}dz dz'\rho(z)\rho(z')\log|z-z'|,
\end{equation}
where $\rho(z)=\lim_{N \to \infty}\frac{1}{N}\sum_{a=1}^{N}\delta(z-z_a)$ is the eigenvalue density which is given by
\begin{equation}
\rho(z)=\frac{1}{2\pi i \mu}\big(W^{(0,1)}_{0}(z-i\epsilon)-W^{(0,1)}_{0}(z+i\epsilon)\big)
=\frac{1}{2\pi i \mu}y(z),\ \ \ z\in D.
\end{equation}
Using the integration of the saddle point equation in the $\beta$-deformed eigenvalue integral (\ref{matrix_def})
\begin{equation}
\frac{1}{2\mu}V(z)=\int_{D}dz'\rho(z')\log|z-z'|-\frac{1}{2\mu}\xi,\ \ \ z\in D,
\label{int_saddle_eq_const}
\end{equation}
with the integration constant $\xi$ which is to be fixed, one obtains
\begin{equation}
F_{0,0}=-\frac{1}{4\pi i}\int_{D}dz y(z)V(z)+\frac{\mu}{2}\xi.
\label{F_00_xi}
\end{equation}
In the one-cut case with $D=[a,b]$, the integration constant $\xi$ can be determined by the analytic continuation of $z$ in (\ref{int_saddle_eq_const}) to the value $z=-\Lambda<a$, which gives
\begin{equation}
2\mu\int_a^b dz'\rho(z')\log|z-z'|=
\int_{-\Lambda}^a dz' y(z')+
2\mu\int_a^b dz'\rho(z')\log (\Lambda+z').
\end{equation}
In the limit $\Lambda \to \infty$ we find
\begin{equation}
\xi
=\lim_{\Lambda \to \infty}
\Big(\int_{-\Lambda}^a dz y(z)-V(-\Lambda)+2\mu \log \Lambda\Big),
\label{F_00_xi_f1}
\end{equation}
where we assumed that the integration of $y(z)$ for $z<a$ is well-defined. Alternatively, by considering the analytic continuation to $z=\Lambda>b$ and using
\begin{equation}
2\mu\int_a^b dz'\rho(z')\log|z-z'|=
\int_b^{\Lambda} dz' y(z')+
2\mu\int_a^b dz'\rho(z')\log (\Lambda-z'),
\label{analytic_xi2}
\end{equation}
we obtain
\begin{equation}
\xi
=\lim_{\Lambda \to \infty}\Big(\int_b^{\Lambda}dz y(z)-V(\Lambda)+2\mu \log \Lambda\Big).
\end{equation}

Second, the $RP^2$ free energy is given by \cite{Chekhov:2006rq, Chekhov:2010xj}
\begin{equation}
F_{0,1}=\frac{1}{2\pi}\int_Ddz |y(z)|\log|y(z)|.
\end{equation}
In the one-cut case (\ref{one_cut_ab}) with the moment function (\ref{spectral_gen}), using $\partial_{\mu}y(x)=-2/\sqrt{\sigma(x)}$, one obtains \cite{Brini:2010fc}
\begin{equation}
\partial_{\mu}F_{0,1}=1+\log |c|+\frac12\log\Big(\frac{a-b}{4}\Big)^2+\sum_{i=1}^fm_i\log\bigg[\frac12\Big(\alpha_i-\frac{a+b}{2}+\sqrt{\sigma(\alpha_i)}\Big)\bigg].
\label{RP2_one_cut}
\end{equation}

Third, the torus free energy is given by \cite{Chekhov:2004vx} (see also \cite{Akemann:1996zr} for two-cut case)
\begin{equation}
F_{1,0}=-\frac12\log|\det A|-\frac{1}{24}\log\biggl|\prod_{i=1}^{2s}M(q_i)\biggr|-\frac{1}{12}\log|\mathsf{\Delta}(q)|,
\label{torus_formula}
\end{equation}
where
\begin{equation}
A_{i,j}=\oint_{\mathcal{A}_{i}}\frac{z^{i-1}}{\sqrt{\sigma(z)}}dz,\ \ \ i,j=1,\ldots,s-1
\end{equation}
is the period matrix, and $\mathsf{\Delta}(q)=\prod_{i<j}(q_i-q_j)^2$ is the discriminant of the spectral curve.

Furthermore, the free energy $F_{0,2}$ is given by \cite{Chekhov:2006rq, Chekhov:2010xj} (see also \cite{Borot:2010tr} for cases with a hard edge)
\begin{equation}
F_{0,2}=-\frac{1}{8\pi^2}\oint_{\mathcal{A}}\frac{dy(z')}{y(z')}
\int_DdS(z,z')\log |y(z)|-\frac{1}{24}\log \bigg|\mathsf{\Delta}(q)\prod_{i=1}^{2s}M(q_i)^2\bigg|.
\end{equation}
In the one-cut case (\ref{one_cut_ab}) one can prove the following formula \cite{Andersen:2016ida}
\be
\begin{split}
F_{0,2}&
=-\frac{1}{2}\sum_{i=1}^fm_i\log\big(1-{s}_i^{-2}\big)-\frac12\sum_{i,j=1}^fm_im_j\log\big(1-{s}_i^{-1} {s}_j^{-1}\big)  + \\
&\ \ \ \
+\frac{1}{24}\log\big|M(a)M(b)(a-b)^4\big|,
\label{Klein_one_cut}
\end{split}
\ee
where ${s}_i$ for $i=1,\ldots,f$ are defined in (\ref{alpha_s}). Note that the Klein bottle contribution, in general given by $F_{1,0}+F_{0,2}$,  in the one-cut case takes the following explicit form
\begin{equation}
F_{1,0}+F_{0,2}=-\frac{1}{2}\sum_{i=1}^fm_i\log\big(1-{s}_i^{-2}\big)-\frac12\sum_{i,j=1}^fm_im_j\log\big(1-{s}_i^{-1} {s}_j^{-1}\big).
\end{equation}

\section{Integrands in the refined topological recursion}       \label{app:integrands_rec}

In this appendix we write down explicitly the integrands $\mathrm{Rec}^{(g,h)}_{\ell}(\mathrm{z}_1,\ldots,\mathrm{z}_h)$ in the refined topological recursion (\ref{ref_loop_eq_z}). In particular, these results are used in the computation (\ref{free_ene_rec}) of stable free energies $F_{g,\ell}$ with $\chi=2-2g-\ell<0$. Throughout this section we use the notation
\begin{equation}
D_{\zeta}=
\frac{\partial}{\partial\zeta}
+\frac{\partial^2\zeta}{\partial{w}^2}\left(\frac{\partial{w}}{\partial\zeta}\right)^2,
\end{equation}
where ${w}=\frac{a+b}{2}-\frac{a-b}{4}(\zeta+\zeta^{-1})$ with $|\zeta|>1$.

\subsection*{Integrands at order $\chi=-1$}\label{app:integrands_rec1}

At this order we find the following results
\begin{align}
\label{rec_111}
&
\mathrm{Rec}^{(1,1)}_{1}(\zeta)=\widehat{W}^{(0,2)}_{1}(\zeta,\zeta)+2\widehat{W}^{(0,1)}_{1}(\zeta)\widehat{W}^{(1,1)}_{0}(\zeta)
+d\zeta^2D_{\zeta}\frac{\widehat{W}^{(1,1)}_{0}(\zeta)}{d\zeta},\\
\label{rec_013}
&
\mathrm{Rec}^{(0,1)}_{3}(\zeta)=2\widehat{W}^{(0,1)}_{1}(\zeta)\widehat{W}^{(0,1)}_{2}(\zeta)
+d\zeta^2D_{\zeta}\frac{\widehat{W}^{(0,1)}_{2}(\zeta)}{d\zeta},
\end{align}
which are necessary in particular to determine the free energies $F_{1,1}$ and $F_{0,3}$ with $\chi=-1$. Here $\widehat{W}^{(1,1)}_0$, $\widehat{W}^{(0,2)}_1$, and $\widehat{W}^{(0,1)}_2$ are obtained from
\begin{align}
\label{rec_110}
&
\mathrm{Rec}^{(1,1)}_{0}(\zeta)=\widehat{W}^{(0,2)}_{0}(\zeta,\zeta),\\
\label{rec_021}
&
\mathrm{Rec}^{(0,2)}_{1}(\zeta,{\mathrm{z}})=2\widehat{W}^{(0,1)}_{1}(\zeta)\widehat{\mathcal{W}}^{(0,2)}_{0}(\zeta,{\mathrm{z}})
+d\zeta^2D_{\zeta}\frac{\widehat{W}^{(0,2)}_{0}(\zeta,{\mathrm{z}})}{d\zeta},\\
\label{rec_012}
&
\mathrm{Rec}^{(0,1)}_{2}(\zeta)=\widehat{W}^{(0,1)}_{1}(\zeta)^2
+d\zeta^2D_{\zeta}\frac{\widehat{W}^{(0,1)}_{1}(\zeta)}{d\zeta}.
\end{align}

\subsection*{Integrands at order $\chi=-2$}\label{app:integrands_rec2}

At this order we find
\begin{align}
\label{rec_210}
\mathrm{Rec}^{(2,1)}_{0}(\zeta)&=\widehat{W}^{(1,2)}_{0}(\zeta,\zeta)+\widehat{W}^{(1,1)}_{0}(\zeta)^2,\\
\label{rec_112}
\mathrm{Rec}^{(1,1)}_{2}(\zeta)&=\widehat{W}^{(0,2)}_{2}(\zeta,\zeta)+2\widehat{W}^{(1,1)}_{0}(\zeta)\widehat{W}^{(0,1)}_{2}(\zeta)+2\widehat{W}^{(1,1)}_{1}(\zeta)\widehat{W}^{(0,1)}_{1}(\zeta) + \nonumber\\
&\ \ \
+d\zeta^2D_{\zeta}\frac{\widehat{W}^{(1,1)}_{1}(\zeta)}{d\zeta},\\
\label{rec_014}
\mathrm{Rec}^{(0,1)}_{4}(\zeta)&=2\widehat{W}^{(0,1)}_{1}(\zeta)\widehat{W}^{(0,1)}_{3}(\zeta)+\widehat{W}^{(0,1)}_{2}(\zeta)^2
+d\zeta^2D_{\zeta}\frac{\widehat{W}^{(0,1)}_{3}(\zeta)}{d\zeta},
\end{align}
which determine the free energies $F_{2,0}$, $F_{1,2}$, and $F_{0,4}$ with $\chi=-2$. Here $\widehat{W}^{(1,1)}_1$, $\widehat{W}^{(0,1)}_3$, $\widehat{W}^{(1,1)}_0$, $\widehat{W}^{(0,1)}_2$, $\widehat{W}^{(1,2)}_0$, and $\widehat{W}^{(0,2)}_2$ are obtained from (\ref{rec_111}), (\ref{rec_013}), (\ref{rec_110}) and (\ref{rec_012}), and
\begin{align}
\label{rec_120}
\mathrm{Rec}^{(1,2)}_{0}(\zeta,{\mathrm{z}})&=\widehat{W}^{(0,3)}_{0}(\zeta,\zeta,{\mathrm{z}})+2\widehat{W}^{(1,1)}_{0}(\zeta)\widehat{\mathcal{W}}^{(0,2)}_{0}(\zeta,{\mathrm{z}}),\\
\label{rec_022}
\mathrm{Rec}^{(0,2)}_{2}(\zeta,{\mathrm{z}})&=2\widehat{W}^{(0,1)}_{2}(\zeta)\widehat{\mathcal{W}}^{(0,2)}_{0}(\zeta,{\mathrm{z}})+2\widehat{W}^{(0,1)}_{1}(\zeta)\widehat{W}^{(0,2)}_{1}(\zeta,{\mathrm{z}}) + \nonumber\\
&\ \ \
+d\zeta^2D_{\zeta}\frac{\widehat{W}^{(0,2)}_{1}(\zeta,{\mathrm{z}})}{d\zeta}.
\end{align}
In addition $W^{(0,3)}_0$ that appears in (\ref{rec_120}) can be obtained from
\begin{equation}
\mathrm{Rec}^{(0,3)}_{0}(\zeta,{\mathrm{z}}_1,{\mathrm{z}}_2)=2\widehat{\mathcal{W}}^{(0,2)}_{0}(\zeta,{\mathrm{z}}_1)\widehat{\mathcal{W}}^{(0,2)}_{0}(\zeta,{\mathrm{z}}_2).
\label{rec_030}
\end{equation}

\subsection*{Integrands at order $\chi=-3$}\label{app:integrands_rec3}

Similarly we find
\begin{align}
\label{rec_211}
\mathrm{Rec}^{(2,1)}_{1}(\zeta)&=\widehat{W}^{(1,2)}_{1}(\zeta,\zeta)+2\widehat{W}^{(0,1)}_{1}(\zeta)\widehat{W}^{(2,1)}_{0}(\zeta)+2\widehat{W}^{(1,1)}_{0}(\zeta)\widehat{W}^{(1,1)}_{1}(\zeta) 
+d\zeta^2D_{\zeta}\frac{\widehat{W}^{(2,1)}_{0}(\zeta)}{d\zeta},\\
\label{rec_113}
\mathrm{Rec}^{(1,1)}_{3}(\zeta)&=\widehat{W}^{(0,2)}_{3}(\zeta,\zeta)+2\widehat{W}^{(1,1)}_{0}(\zeta)\widehat{W}^{(0,1)}_{3}(\zeta)+2\widehat{W}^{(1,1)}_{1}(\zeta)\widehat{W}^{(0,1)}_{2}(\zeta) + \nonumber\\
&\ \ \ 
+2\widehat{W}^{(0,1)}_{1}(\zeta)\widehat{W}^{(1,1)}_{2}(\zeta)
+d\zeta^2D_{\zeta}\frac{\widehat{W}^{(1,1)}_{2}(\zeta)}{d\zeta},\\
\label{rec_015}
\mathrm{Rec}^{(0,1)}_{5}(\zeta)&=2\widehat{W}^{(0,1)}_{1}(\zeta)\widehat{W}^{(0,1)}_{4}(\zeta)+2\widehat{W}^{(0,1)}_{2}(\zeta)\widehat{W}^{(0,1)}_{3}(\zeta)
+d\zeta^2D_{\zeta}\frac{\widehat{W}^{(0,1)}_{4}(\zeta)}{d\zeta},
\end{align}
which are necessary to determine the free energies $F_{2,1}$, $F_{1,3}$, and $F_{0,5}$ with $\chi=-3$. Here $\widehat{W}^{(1,1)}_1$, $\widehat{W}^{(0,1)}_3$, $\widehat{W}^{(1,1)}_0$, $\widehat{W}^{(0,1)}_2$, $\widehat{W}^{(2,1)}_0$, $\widehat{W}^{(1,1)}_2$, $\widehat{W}^{(0,1)}_4$, $\widehat{W}^{(1,2)}_1$, and $\widehat{W}^{(0,2)}_3$ are obtained from (\ref{rec_111}), (\ref{rec_013}), (\ref{rec_110}), (\ref{rec_012}), (\ref{rec_210}), (\ref{rec_112}), (\ref{rec_014}), and
\begin{align}
\label{rec_121}
\mathrm{Rec}^{(1,2)}_{1}(\zeta,{\mathrm{z}})&=\widehat{W}^{(0,3)}_{1}(\zeta,\zeta,{\mathrm{z}})+2\widehat{W}^{(1,1)}_{0}(\zeta)\widehat{W}^{(0,2)}_{1}(\zeta,{\mathrm{z}})+2\widehat{W}^{(1,1)}_{1}(\zeta)\widehat{\mathcal{W}}^{(0,2)}_{0}(\zeta,{\mathrm{z}}) + \nonumber\\
&\ \ \ 
+2\widehat{W}^{(0,1)}_{1}(\zeta)\widehat{W}^{(1,2)}_{0}(\zeta,{\mathrm{z}})
+d\zeta^2D_{\zeta}\frac{\widehat{W}^{(1,2)}_{0}(\zeta,{\mathrm{z}})}{d\zeta},\\
\label{rec_023}
\mathrm{Rec}^{(0,2)}_{3}(\zeta,{\mathrm{z}})&=2\widehat{W}^{(0,1)}_{3}(\zeta)\widehat{\mathcal{W}}^{(0,2)}_{0}(\zeta,{\mathrm{z}})+2\widehat{W}^{(0,1)}_{2}(\zeta)\widehat{W}^{(0,2)}_{1}(\zeta,{\mathrm{z}}) + \nonumber\\
&\ \ \ 
+2\widehat{W}^{(0,1)}_{1}(\zeta)\widehat{W}^{(0,2)}_{2}(\zeta,{\mathrm{z}})
+d\zeta^2D_{\zeta}\frac{\widehat{W}^{(0,2)}_{2}(\zeta,{\mathrm{z}})}{d\zeta}.
\end{align} 
In particular $\widehat{W}^{(0,2)}_1$, $\widehat{W}^{(1,2)}_0$ and $\widehat{W}^{(0,2)}_2$ 
are obtained from (\ref{rec_021}), (\ref{rec_120}), and (\ref{rec_022}).

\subsection*{Integrands at order $\chi=-4$}\label{app:integrands_rec4}

Finally we find
\begin{align}
\label{rec_310}
\mathrm{Rec}^{(3,1)}_{0}(\zeta)&=\widehat{W}^{(2,2)}_{0}(\zeta,\zeta)+2\widehat{W}^{(1,1)}_{0}(\zeta)\widehat{W}^{(2,1)}_{0}(\zeta),\\
\label{rec_212}
\mathrm{Rec}^{(2,1)}_{2}(\zeta)&=\widehat{W}^{(1,2)}_{2}(\zeta,\zeta)+2\widehat{W}^{(1,1)}_{0}(\zeta)\widehat{W}^{(1,1)}_{2}(\zeta)+2\widehat{W}^{(2,1)}_{0}(\zeta)\widehat{W}^{(0,1)}_{2}(\zeta)+\widehat{W}^{(1,1)}_{1}(\zeta)^2 + \nonumber\\
&\ \ \ 
+2\widehat{W}^{(0,1)}_{1}(\zeta)\widehat{W}^{(2,1)}_{1}(\zeta)
+d\zeta^2D_{\zeta}\frac{\widehat{W}^{(2,1)}_{1}(\zeta)}{d\zeta},\\
\label{rec_114}
\mathrm{Rec}^{(1,1)}_{4}(\zeta)&=\widehat{W}^{(0,2)}_{4}(\zeta,\zeta)+2\widehat{W}^{(1,1)}_{0}(\zeta)\widehat{W}^{(0,1)}_{4}(\zeta)+2\widehat{W}^{(1,1)}_{1}(\zeta)\widehat{W}^{(0,1)}_{3}(\zeta) + \nonumber\\
&\ \ \ 
+2\widehat{W}^{(0,1)}_{2}(\zeta)\widehat{W}^{(1,1)}_{2}(\zeta)
+2\widehat{W}^{(0,1)}_{1}(\zeta)\widehat{W}^{(1,1)}_{3}(\zeta)
+d\zeta^2D_{\zeta}\frac{\widehat{W}^{(1,1)}_{3}(\zeta)}{d\zeta},\\
\label{rec_016}
\mathrm{Rec}^{(0,1)}_{6}(\zeta)&=2\widehat{W}^{(0,1)}_{1}(\zeta)\widehat{W}^{(0,1)}_{5}(\zeta)+2\widehat{W}^{(0,1)}_{2}(\zeta)\widehat{W}^{(0,1)}_{4}(\zeta)+\widehat{W}^{(0,1)}_{3}(\zeta)^2 + \nonumber\\
&\ \ \ 
+d\zeta^2D_{\zeta}\frac{\widehat{W}^{(0,1)}_{5}(\zeta)}{d\zeta},
\end{align}
which are necessary to determine the free energies $F_{3,0}$, $F_{2,2}$, $F_{1,4}$, and $F_{0,6}$ with $\chi=-4$. Here $\widehat{W}^{(1,1)}_1$, $\widehat{W}^{(0,1)}_3$, $\widehat{W}^{(1,1)}_0$, $\widehat{W}^{(0,1)}_2$, $\widehat{W}^{(2,1)}_0$, $\widehat{W}^{(1,1)}_2$, $\widehat{W}^{(0,1)}_4$, $\widehat{W}^{(2,1)}_1$, $\widehat{W}^{(1,1)}_3$, $\widehat{W}^{(0,1)}_5$, $\widehat{W}^{(2,2)}_0$, $\widehat{W}^{(1,2)}_2$, and $\widehat{W}^{(0,2)}_4$ are obtained respectively from (\ref{rec_111}), (\ref{rec_013}), (\ref{rec_110}), (\ref{rec_012}), (\ref{rec_210}), (\ref{rec_112}), (\ref{rec_014}), (\ref{rec_211}), (\ref{rec_113}), (\ref{rec_015}), and
\begin{align}
\label{rec_220}
\mathrm{Rec}^{(2,2)}_{0}(\zeta,{\mathrm{z}})&=\widehat{W}^{(1,3)}_{0}(\zeta,\zeta,{\mathrm{z}})+2\widehat{W}^{(1,1)}_{0}(\zeta)\widehat{W}^{(1,2)}_{0}(\zeta,{\mathrm{z}})+2\widehat{W}^{(2,1)}_{0}(\zeta)\widehat{\mathcal{W}}^{(0,2)}_{0}(\zeta,{\mathrm{z}}),\\
\label{rec_122}
\mathrm{Rec}^{(1,2)}_{2}(\zeta,{\mathrm{z}})&=\widehat{W}^{(0,3)}_{2}(\zeta,\zeta,{\mathrm{z}})+2\widehat{W}^{(1,1)}_{0}(\zeta)\widehat{W}^{(0,2)}_{2}(\zeta,{\mathrm{z}})+2\widehat{W}^{(0,1)}_{2}(\zeta)\widehat{W}^{(1,2)}_{0}(\zeta,{\mathrm{z}}) + \nonumber\\
&\ \ \ 
+2\widehat{W}^{(1,1)}_{1}(\zeta)\widehat{W}^{(0,2)}_{1}(\zeta,{\mathrm{z}})
+2\widehat{W}^{(0,1)}_{1}(\zeta)\widehat{W}^{(1,2)}_{1}(\zeta,{\mathrm{z}}) + \nonumber\\
&\ \ \ 
+2\widehat{W}^{(1,1)}_{2}(\zeta)\widehat{\mathcal{W}}^{(0,2)}_{0}(\zeta,{\mathrm{z}})
+d\zeta^2D_{\zeta}\frac{\widehat{W}^{(1,2)}_{1}(\zeta,{\mathrm{z}})}{d\zeta},\\
\label{rec_024}
\mathrm{Rec}^{(0,2)}_{4}(\zeta,{\mathrm{z}})&=2\widehat{W}^{(0,1)}_{4}(\zeta)\widehat{\mathcal{W}}^{(0,2)}_{0}(\zeta,{\mathrm{z}})+2\widehat{W}^{(0,1)}_{3}(\zeta)\widehat{W}^{(0,2)}_{1}(\zeta,{\mathrm{z}}) + \nonumber\\
&\ \ \ 
+2\widehat{W}^{(0,1)}_{2}(\zeta)\widehat{W}^{(0,2)}_{2}(\zeta,{\mathrm{z}})
+2\widehat{W}^{(0,1)}_{1}(\zeta)\widehat{W}^{(0,2)}_{3}(\zeta,{\mathrm{z}})
+d\zeta^2D_{\zeta}\frac{\widehat{W}^{(0,2)}_{3}(\zeta,{\mathrm{z}})}{d\zeta}.
\end{align}
In addition $\widehat{W}^{(0,2)}_1$, $\widehat{W}^{(1,2)}_0$, $\widehat{W}^{(0,2)}_2$, $\widehat{W}^{(1,2)}_1$, $\widehat{W}^{(0,2)}_3$, $\widehat{W}^{(1,3)}_0$, and $\widehat{W}^{(0,3)}_2$, which appear in the recursions (\ref{rec_220}), (\ref{rec_122}), and (\ref{rec_024}), can be obtained respectively from (\ref{rec_021}), (\ref{rec_120}), (\ref{rec_022}), (\ref{rec_121}), (\ref{rec_023}), and
\begin{align}
\label{rec_130}
\mathrm{Rec}^{(1,3)}_{0}(\zeta,{\mathrm{z}}_1,{\mathrm{z}}_2)&=\widehat{W}^{(0,4)}_{0}(\zeta,\zeta,{\mathrm{z}}_1,{\mathrm{z}}_2)+2\widehat{W}^{(1,1)}_{0}(\zeta)\widehat{W}^{(0,3)}_{0}(\zeta,{\mathrm{z}}_1,{\mathrm{z}}_2) + \nonumber\\
&\ \ \ 
+2\widehat{W}^{(1,2)}_{0}(\zeta,{\mathrm{z}}_1)\widehat{\mathcal{W}}^{(0,2)}_{0}(\zeta,{\mathrm{z}}_2)+2\widehat{W}^{(1,2)}_{0}(\zeta,{\mathrm{z}}_2)\widehat{\mathcal{W}}^{(0,2)}_{0}(\zeta,{\mathrm{z}}_1),\\
\label{rec_032}
\mathrm{Rec}^{(0,3)}_{2}(\zeta,{\mathrm{z}}_1,{\mathrm{z}}_2)&=2\widehat{W}^{(0,2)}_{2}(\zeta,{\mathrm{z}}_1)\widehat{\mathcal{W}}^{(0,2)}_{0}(\zeta,{\mathrm{z}}_2)+2\widehat{W}^{(0,2)}_{2}(\zeta,{\mathrm{z}}_2)\widehat{\mathcal{W}}^{(0,2)}_{0}(\zeta,{\mathrm{z}}_1) + \nonumber\\
&\ \ \ 
+2\widehat{W}^{(0,1)}_{1}(\zeta)\widehat{W}^{(0,3)}_{1}(\zeta,{\mathrm{z}}_1,{\mathrm{z}}_2)+2\widehat{W}^{(0,2)}_{1}(\zeta,{\mathrm{z}}_1)\widehat{W}^{(0,2)}_{1}(\zeta,{\mathrm{z}}_2) + \nonumber\\
&\ \ \ 
+2\widehat{W}^{(0,1)}_{2}(\zeta)\widehat{W}^{(0,3)}_{0}(\zeta,{\mathrm{z}}_1,{\mathrm{z}}_2)
+d\zeta^2D_{\zeta}\frac{\widehat{W}^{(0,3)}_{1}(\zeta,{\mathrm{z}}_1,{\mathrm{z}}_2)}{d\zeta}.
\end{align}
Finally $\widehat{W}^{(0,3)}_0$, $\widehat{W}^{(0,4)}_0$, and $\widehat{W}^{(0,3)}_1$ that appear in the recursions (\ref{rec_130}) and (\ref{rec_032}) are obtained from (\ref{rec_030}) and 
\begin{align}
\label{rec_040}
\mathrm{Rec}^{(0,4)}_{0}(\zeta,{\mathrm{z}}_1,{\mathrm{z}}_2,{\mathrm{z}}_3)&=2\widehat{W}^{(0,3)}_{0}(\zeta,{\mathrm{z}}_1,{\mathrm{z}}_2)\widehat{\mathcal{W}}^{(0,2)}_{0}(\zeta,{\mathrm{z}}_3)+2\widehat{W}^{(0,3)}_{0}(\zeta,{\mathrm{z}}_1,{\mathrm{z}}_3)\widehat{\mathcal{W}}^{(0,2)}_{0}(\zeta,{\mathrm{z}}_2) + \nonumber\\
&\ \ \ 
+2\widehat{W}^{(0,3)}_{0}(\zeta,{\mathrm{z}}_2,{\mathrm{z}}_3)\widehat{\mathcal{W}}^{(0,2)}_{0}(\zeta,{\mathrm{z}}_1),\\
\label{rec_031}
\mathrm{Rec}^{(0,3)}_{1}(\zeta,{\mathrm{z}}_1,{\mathrm{z}}_2)&=2\widehat{W}^{(0,2)}_{1}(\zeta,{\mathrm{z}}_1)\widehat{\mathcal{W}}^{(0,2)}_{0}(\zeta,{\mathrm{z}}_2)+2\widehat{W}^{(0,2)}_{1}(\zeta,{\mathrm{z}}_2)\widehat{\mathcal{W}}^{(0,2)}_{0}(\zeta,{\mathrm{z}}_1) + \nonumber\\
&\ \ \ 
+2\widehat{W}^{(0,1)}_{1}(\zeta)\widehat{W}^{(0,3)}_{0}(\zeta,{\mathrm{z}}_1,{\mathrm{z}}_2)
+d\zeta^2D_{\zeta}\frac{\widehat{W}^{(0,3)}_{0}(\zeta,{\mathrm{z}}_1,{\mathrm{z}}_2)}{d\zeta}.
\end{align}

\section{Free energies in $\beta$-deformed Gaussian and Penner matrix models}   \label{sec-GP}

In this appendix we review the form of the free energy and its asymptotic expansion in $\beta$-deformed Gaussian and Penner matrix models. In section \ref{sec-examples} we show that these results are correctly reproduced by the refined topological recursion. 

\subsection{Gaussian model}  \label{ssec-gaussian}

The $\beta$-deformed Gaussian model is defined by the integral (\ref{matrix_def}) with the potential 
\be
V(x)=\frac{x^2}{2},
\ee
and its partition function  $Z^{\mathrm{G}}$ can be evaluated and written as follows
\begin{align}
Z^{\mathrm{G}} & = 
\frac{1}{(2\pi)^NN!}\int_{{\IR}^N}  \Delta(z)^{2\beta} e^{-\frac{\sqrt{\beta}}{2\hbar}\sum_{a=1}^N z_a^2}  \prod_{a=1}^Ndz_a  = \nonumber \\
& = \big(\hbar^{\frac{1}{2}}\beta^{-\frac{1}{4}}\big)^{\beta N^2+(1-\beta)N}
\frac{\prod_{j=1}^N \Gamma(1+\beta j)}{N!(2\pi)^{N/2}\Gamma(1+\beta)^N} = \label{ZGaussian} \\
& = (2\pi)^{\frac{1}{2}}\big(\hbar^{\frac{1}{2}}\beta^{\frac{1}{4}}\big)^{\beta N^2+(1-\beta)N}
\beta^{\frac{1}{2}-(1-\beta)N}\Gamma(\beta)^{-N}e^{\chi'(0;1,\beta^{-1})+\chi'(0;-1,\beta^{-1})}\Gamma_2(N|-1,\beta^{-1}),  \nonumber
\end{align}
where $\Gamma_2(x|a,b)$ is the Barnes double Gamma function defined by
\begin{align}
\begin{split}
\Gamma_2(x|a,b)&=\exp\left(\left.\frac{d}{ds}\right|_{s=0}\zeta_2(s;a,b,x)
-\chi'(0;a,b)\right),
\label{Barnes} \\
\zeta_2(s;a,b,x)&=\frac{1}{\Gamma(s)}\int_0^\infty dt\;t^{s-1}\frac{e^{-tx}}{(1-e^{-at})(1-e^{-bt})},\\
\chi'(0;a,b) &= \lim_{x\to 0}\left(\left.\frac{d}{ds}\right|_{s=0}\zeta_2(s;a,b,x)+\log x\right).
\end{split}
\end{align}
The second line of (\ref{ZGaussian}) follows from the Mehta formula
\be
\int \Delta(z)^{2\beta}e^{-\frac{1}{2}\sum_{a=1}^Nz_a^2} \prod_{a=1}^N dz_a =(2\pi)^{N/2}\prod_{a=1}^N\frac{\Gamma(1+\beta j)}{\Gamma(1+\beta)}.
\ee
Furthermore, from the functional equation \cite{spreafico},
\begin{equation}
\Gamma_2(x+b|a,b)=\sqrt{2\pi}a^{\frac{1}{2}-x/a}\Gamma(x/a)^{-1}\Gamma_2(x|a,b),
\end{equation}
and recalling that $\Gamma_2(a|a,b)=\sqrt{2\pi/b}$, we obtain the identity 
\begin{equation}
\prod_{j=1}^N\Gamma(1+\beta j)=(2\pi)^{(N+1)/2}\beta^{1/2+\beta N^2/2+(1+\beta)N/2}N!\Gamma_2(N+1|1,\beta^{-1})^{-1},
\end{equation}
which together with the relation
\begin{equation}
\Gamma_2(N+1|1,\beta^{-1})^{-1}=e^{\chi'(0;1,\beta^{-1})+\chi'(0;-1,\beta^{-1})}\Gamma_2(N|-1,\beta^{-1})
\end{equation}
implies the third line of (\ref{ZGaussian}). From this third line we can determine the large $N$ asymptotic expansion of the free energy $F^{\mathrm{G}}(\hbar,\beta,\mu)=\log Z^{\mathrm{G}} $. To write it in a concise form we introduce
\be
\begin{split}
&
\widehat{F}_{0}^{\mathrm{even}}(\beta,\mu)=\frac12 \mu^2\log\mu-\frac34\mu^2,\\
&
\widehat{F}_{1}^{\mathrm{even}}(\beta,\mu)=\frac{1}{12}\big(-3+\beta+\beta^{-1}\big)\log\mu,\\
&
\widehat{F}_{h\ge2}^{\mathrm{even}}(\beta,\mu)
=-\frac{\sum_{r=0}^{h}\binom{2h}{2r}B_{2h-2r}B_{2r}\beta^{2r-h}}{2h(2h-1)(2h-2){\mu}^{2h-2}},
\end{split}
\ee
and
\be
\begin{split}
&
\widehat{F}_{0}^{\mathrm{odd}}(\beta,\mu)
=\frac{1}{2}\left(\mu\log\mu-\mu\right)\big(\beta^{1/2}-\beta^{-1/2}\big),\\
&
\widehat{F}_{h\ge1}^{\mathrm{odd}}(\beta,\mu)
=-\frac{B_{2h}\big(\beta^{h-1/2}-\beta^{-h+1/2}\big)}{4h(2h-1){\mu}^{2h-1}},
\end{split}
\ee
where $\mu=\beta^{1/2}\hbar N$ as in (\ref{tHooft}), $(-n)!=0$ for $n\ge 1$, and Bernoulli numbers $B_k$ are defined by
\begin{equation}
\frac{x}{e^x-1}=\sum_{k=0}^{\infty}\frac{B_k}{k!}x^k,
\end{equation}
so that the first few of them takes form
\begin{equation}
B_0=1,\ B_1=-\frac12,\ B_2=\frac16,\ B_4=-\frac{1}{30},\ B_6=\frac{1}{42},\quad \textrm{and}\  B_{2k+1}=0\ \textrm{for}\ k\ge 1.
\end{equation}
In terms of these quantities, ignoring some additive terms, the free energy of the $\beta$-deformed Gaussian model can be written as
\begin{equation}
\boxed{\ F^{\mathrm{G}}(\hbar,\beta,\mu)=\log Z^{\mathrm{G}} \simeq \sum_{h=0}^{\infty}\hbar^{2h-2}\widehat{F}_h^{\mathrm{even}}(\beta,\mu)+\sum_{h=0}^{\infty}\hbar^{2h-1}\widehat{F}_h^{\mathrm{odd}}(\beta,\mu).\ }
\end{equation}
This free energy can also be expressed as \cite{Brini:2010fc}
\begin{equation}
F^{\mathrm{G}}(\hbar,\beta,\mu)\simeq
\log \Gamma_2({\mu}|-\beta^{1/2}\hbar, \beta^{-1/2}\hbar).
\label{gauss_fr_en_exact}
\end{equation}


\subsection{Penner model}  \label{ssec-penner}

The $\beta$-deformed Penner model is defined by the integral (\ref{matrix_def}) with the potential 
\be
V(x)=-x-\log(1-x).
\ee
It is convenient to consider the partition function $Z^{\mathrm{P}}$ of the $\beta$-deformed Penner model  normalized by the partition function of the Gaussian model, and to define
\be
Z^{\mathrm{P}/\mathrm{G}}(\hbar,\beta,\mu)
=\exp \left(F^{\mathrm{P}/\mathrm{G}}(\hbar,\beta,\mu)  \right)
 = \frac{Z^{\mathrm{P}}(\hbar,\beta,\mu)}{Z^{\mathrm{G}}(\hbar,\beta,\mu)}.
\ee
One can show that the corresponding free energy can be expressed in terms of the Barnes double Gamma function (\ref{Barnes})
\be
F^{\mathrm{P}/\mathrm{G}}(\hbar,\beta,\mu)=
F^{\mathrm{P}}(\hbar,\beta,\mu)-F^{\mathrm{G}}(\hbar,\beta,\mu)
= \log \frac{\Gamma_2(1+{\mu}|\beta^{\frac{1}{2}}\hbar, -\beta^{-\frac{1}{2}}\hbar)}{\Gamma_2(1|\beta^{\frac{1}{2}}\hbar, -\beta^{-\frac{1}{2}}\hbar)}+\frac{\mu}{\hbar^2}.
\label{Gauss_Pen_norm}
\ee
Let us introduce 
\be
\begin{split}
&
\label{Fh_evenodd}
F_h^{\mathrm{even}}(\beta,\mu)=
\sum_{n=1}^{\infty}\frac{(-1)^{n+1}(2h+n-3)!}{(2h)!n!}{\mu}^n\Big\{2hB_{2h-1}+\sum_{r=0}^{2h}\binom{2h}{r}B_{2h-r}B_r\beta^{r-h}\Big\},\\
&
F_h^{\mathrm{odd}}(\beta,\mu)=
\sum_{n=1}^{\infty}\frac{(-1)^n(2h+n-2)!}{2(2h)!n!}B_{2h}{\mu}^n\big(\beta^{h-1/2}-\beta^{-h+1/2}\big).
\end{split}
\ee
The summations in these formulas can be performed and the results can also be expressed in terms of quantities introduced in appendix \ref{ssec-gaussian}
\begin{equation}
\begin{split}
&
F_{h}^{\mathrm{even}}(\beta,\mu)=\widehat{F}_{h}^{\mathrm{even}}(\beta^{-1},1+\mu)-\widehat{F}_{h}^{\mathrm{even}}(\beta^{-1},1)+\mu\delta_{h,0},\\
&
F_{h}^{\mathrm{odd}}(\beta,\mu)=\widehat{F}_{h}^{\mathrm{odd}}(\beta^{-1},1+\mu)-\widehat{F}_{h}^{\mathrm{odd}}(\beta^{-1},1).
\label{Fh_sumt_eo}
\end{split}
\end{equation}
Using this notation, the free energy of the (normalized) Penner model (\ref{Gauss_Pen_norm}) is expressed as \cite{GouHarJack}
\begin{equation}
\boxed{\ F^{\mathrm{P}/\mathrm{G}}(\hbar,\beta,\mu)=\sum_{h=0}^{\infty}\hbar^{2h-2}F_h^{\mathrm{even}}(\beta,\mu)+\sum_{h=0}^{\infty}\hbar^{2h-1}F_h^{\mathrm{odd}}(\beta,\mu) . \ }
\label{F_penner_asym}
\end{equation}

For completeness, let us recall that free energies in the $\beta$-deformed Penner model encode virtual Euler characteristics of moduli spaces $\mathcal{M}_{g,n}^{\mathrm{comp}}$ and $\mathcal{M}_{g,n}^{\mathrm{real}}$ of complex and real algebraic curves of genus $g$ and $n$ marked points \cite{P2,GouHarJack,Mulase:2002cr}, given respectively by 
\be
\begin{split}
\chi(\mathcal{M}_{g,n}^{\mathrm{comp}}) & =\frac{(-1)^n(2g+n-3)!(2g-1)}{(2g)!n!}B_{2g}, \\
\chi(\mathcal{M}_{g,n}^{\mathrm{real}}) & =\frac{(g+n-2)!(2^{g-1}-1)}{2(g)!n!}B_g,\ \ \ \frac{g}{2}\in {\IZ}_{\ge 0}.
\end{split}
\ee
To show this, the following formulas (from appendix B in \cite{Mulase:2002cr}) are useful
\be
(1-2n)B_{2n}  =\sum_{k=0}^n \binom{2n}{2k}B_{2n-2k}B_{2k} =\sum_{k=0}^n \binom{2n}{2k}B_{2n-2k}B_{2k}2^{2k} \ \qquad \textrm{for}\ n\neq 1.\label{BBB}
\ee
Indeed, plugging $\beta=1$ in (\ref{Fh_evenodd}) and using the first equality above, one finds
\begin{equation}
F_g^{\mathrm{even}}(1,{\mu})=\sum_{n=1}^{\infty}\chi(\mathcal{M}_{g,n}^{\mathrm{comp}}){\mu}^n.
\label{high_f_g0}
\end{equation}
Similarly, plugging $\beta=1/2$ in (\ref{Fh_evenodd}) and using the second equality in (\ref{BBB}), one finds
\begin{align}
2^hF_h^{\mathrm{even}}\Big(\frac12, {\mu}\Big) & =\sum_{n=1}^{\infty}\chi(\mathcal{M}_{h,n}^{\mathrm{comp}}){\mu}^n, \\
2^{h+1/2}F_h^{\mathrm{odd}}\Big(\frac12, {\mu}\Big) & =
-2\sum_{n=1}^{\infty}\chi(\mathcal{M}_{2h,n}^{\mathrm{real}})(-{\mu})^n.
\end{align}


\section{One-point differentials in the Penner model}    \label{ssec-compPenner}

In this appendix we use the refined topological recursion to compute one-point differentials $W^{(g,1)}_{\ell}(x)$, 
for the range $0\ge \chi=1-2g-\ell \ge -5$, for the spectral curve of the Penner model given in (\ref{Penner_curve}). Note that by the rescaling (\ref{Penner_Gaussian}) and taking the Gaussian limit $T\to 0$, the results below reduce to the results for the Gaussian model computed e.g. in \cite{Brini:2010fc, Mironov:2011jn, Witte:2013cea}.

\noindent\underline{$\chi=0$} :
\begin{align}
\frac{W^{(0,1)}_1(x)}{dx}&=\frac{1}{2(1-x)}\left(\frac{1}{\sigma_{\mathrm{P}}(x)^{1/2}}-1\right)-\frac{x+2{\mu}}{2\sigma_{\mathrm{P}}(x)}.
\end{align}

\noindent\underline{$\chi=-1$} :
\begin{align}
\frac{W^{(1,1)}_0(x)}{dx}&=\frac{{\mu}(1+{\mu})(1-x)}{\sigma_{\mathrm{P}}(x)^{5/2}},
\end{align}
\begin{align}
\frac{W^{(0,1)}_2(x)}{dx}&=-\frac{x+2{\mu}}{\sigma_{\mathrm{P}}(x)^2}+\frac{(1+2{\mu})x^2+{\mu}(1-3{\mu})(1-x)}{\sigma_{\mathrm{P}}(x)^{5/2}}.
\end{align}

\noindent\underline{$\chi=-2$} :
\begin{align} 
\frac{W^{(1,1)}_1(x)}{dx}&=\frac{1-x}{2\sigma_{\mathrm{P}}(x)^{7/2}}
\big(x^2+4{\mu}x+2{\mu}(3+5{\mu})\big)-\frac{1-x}{2\sigma_{\mathrm{P}}(x)^4}\big((1+2{\mu})x^3\nonumber\\
&
-6{\mu}(3+2{\mu})x^2+6{\mu}(5+{\mu}-2{\mu}^2)x+4{\mu}^2(5+3{\mu})\big),
\end{align}
\begin{align}
\frac{W^{(0,1)}_3(x)}{dx}&=\frac{1}{\sigma_{\mathrm{P}}(x)^{7/2}}
\big(-3x^3+(5-8{\mu})x^2+{\mu}(7-9{\mu})x+{\mu}(5+9{\mu})\big)\nonumber\\
&
-\frac{1}{\sigma_{\mathrm{P}}(x)^4}\big(-3(1+2{\mu})x^4+(5+12{\mu}-12{\mu}^2)x^3-6{\mu}(2-3{\mu}+2{\mu}^2)x^2\nonumber\\
&
+2{\mu}(5-{\mu}+12{\mu}^2)x-4{\mu}^2(1+3{\mu})\big).
\end{align}

\noindent\underline{$\chi=-3$} :
\begin{align}
\frac{W^{(2,1)}_0(x)}{dx}&
=\frac{{\mu}(1+{\mu})(1-x)}{\sigma_{\mathrm{P}}(x)^{11/2}}\big(8x^4-4(7-2{\mu})x^3+3(7-5{\mu}+3{\mu}^2)x^2-2{\mu}(7+9{\mu})x
\nonumber\\
&
+3{\mu}(7+3{\mu})\big),
\end{align}
\begin{align}
\frac{W^{(1,1)}_2(x)}{dx}&
=-\frac{1-x}{2\sigma_{\mathrm{P}}(x)^5}\big(-12x^4+(23-50{\mu})x^3-2{\mu}(23+98{\mu})x^2
\nonumber\\
&
+4{\mu}(45+20{\mu}-52{\mu}^2)x+8{\mu}^2(25+26{\mu})\big)
+\frac{1-x}{2\sigma_{\mathrm{P}}(x)^{11/2}}\big(-12(1+2{\mu})x^5\nonumber\\
&
+(23+184{\mu}+64{\mu}^2)x^4-4{\mu}(131+43{\mu}-14{\mu}^2)x^3\nonumber\\
&
+2{\mu}(227-49{\mu}-60{\mu}^2+12{\mu}^3)x^2+8{\mu}^2(15-19{\mu}-6{\mu}^2)x
\nonumber\\
&
+8{\mu}^2(22+27{\mu}+3{\mu}^2)\big),
\end{align}
\begin{align}
\frac{W^{(0,1)}_4(x)}{dx}&
=-\frac{1}{\sigma_{\mathrm{P}}(x)^5}\big(12x^5+15(3-2{\mu})x^4+(37-60{\mu}+60{\mu}^2)x^3\nonumber\\
&
-6{\mu}(9+18{\mu}-8{\mu}^2)x^2+4{\mu}(23-10{\mu}-24{\mu}^2)x+8{\mu}^2(11+6{\mu})\big)
\nonumber\\
&
+\frac{1}{\sigma_{\mathrm{P}}(x)^{11/2}}\big(12(1+2{\mu})x^6
-(45+100{\mu}+44{\mu}^2)x^5
\nonumber\\
&
+(37+166{\mu}-94{\mu}^2+84{\mu}^3)x^4
-{\mu}(205+62{\mu}+210{\mu}^2
-63{\mu}^3)x^3 \nonumber\\
&
+{\mu}(123+190{\mu}+46{\mu}^2-189{\mu}^3)x^2
-{\mu}^2(99-202{\mu}-189{\mu}^2)x
\nonumber\\
&
+{\mu}^2(21-122{\mu}-63{\mu}^2)\big).
\end{align}

\noindent\underline{$\chi=-4$} :
\begin{align}
\frac{W^{(2,1)}_1(x)}{dx}&=\frac{1-x}{2\sigma_{\mathrm{P}}(x)^{13/2}}\big(8x^6-4(7-10{\mu})x^5+(21+86{\mu}+286{\mu}^2)x^4
\nonumber\\
&
-4{\mu}(133+73{\mu}
-142{\mu}^2)x^3+2{\mu}(210-291{\mu}-334{\mu}^2+259{\mu}^3)x^2
\nonumber\\
&
+4{\mu}^2(77-190{\mu}-259{\mu}^2)x
+2{\mu}^2(147+430{\mu}+259{\mu}^2)\big)
\nonumber\\
&
-\frac{1-x}{2\sigma_{\mathrm{P}}(x)^7}\big(8(1+2{\mu})x^7-4(7+114{\mu}+86{\mu}^2)x^6
+3(7+758{\mu}+520{\mu}^2
\nonumber\\
&
-112{\mu}^3)x^5
-2{\mu}(1849+960{\mu}-168{\mu}^2+336{\mu}^3)x^4
+4{\mu}(471+739{\mu}
\nonumber\\
&
+970{\mu}^2+372{\mu}^3-120{\mu}^4)x^3
-24{\mu}^2(243+241{\mu}-76{\mu}^2-60{\mu}^3)x^2
\nonumber\\
&
+16{\mu}^2(225-31{\mu}-339{\mu}^2-90{\mu}^3)x
+96{\mu}^3(25+29{\mu}+5{\mu}^2)\big),
\end{align}
\begin{align}
\frac{W^{(1,1)}_3(x)}{dx}&=\frac{1-x}{2\sigma_{\mathrm{P}}(x)^{13/2}}\big(116x^6-6(81-70{\mu})x^5+(445+246{\mu}+2346{\mu}^2)x^4
\nonumber\\
&
-4{\mu}(1127+699{\mu}
-1098{\mu}^2)x^3+4{\mu}(1083-1507{\mu}-1706{\mu}^2+794{\mu}^3)x^2
\nonumber\\
&
+16{\mu}^2(341-204{\mu}-397{\mu}^2)x
+8{\mu}^2(189+712{\mu}+397{\mu}^2)\big)
\nonumber\\
&
-\frac{1-x}{2\sigma_{\mathrm{P}}(x)^7}\big(116(1+2{\mu})x^7-2(243+1052{\mu}+240{\mu}^2)x^6
+(445+7766{\mu}
\nonumber\\
&
+3132{\mu}^2+168{\mu}^3)x^5
-6{\mu}(2185+926{\mu}-68{\mu}^2-104{\mu}^3)x^4
+2{\mu}(3857
\nonumber\\
&
+2475{\mu}+266{\mu}^2-588{\mu}^3+264{\mu}^4)x^3
-4{\mu}^2(2247+835{\mu}-244{\mu}^2+396{\mu}^3)x^2
\nonumber\\
&
+8{\mu}^2(930-101{\mu}-115{\mu}^2+198{\mu}^3)x
+16{\mu}^3(190+31{\mu}-33{\mu}^2)\big),
\end{align}
\begin{align}
\frac{W^{(0,1)}_5(x)}{dx}&=\frac{1}{\sigma_{\mathrm{P}}(x)^{13/2}}\big(-60x^7+6(61-18{\mu})x^6-(651-310{\mu}+338{\mu}^2)x^5
\nonumber\\
&
+(353+794{\mu}
+934{\mu}^2-504{\mu}^3)x^4-{\mu}(2507-938{\mu}-1386{\mu}^2+315{\mu}^3)x^3
\nonumber\\
&
+{\mu}(1527-3316{\mu}+14{\mu}^2
+945{\mu}^3)x^2+{\mu}^2(1383-2170{\mu}-945{\mu}^2)x
\nonumber\\
&
+7{\mu}^2(57+182{\mu}+45{\mu}^2)\big)
-\frac{1}{\sigma_{\mathrm{P}}(x)^7}\big(-60(1+2{\mu})x^8
+6(61+136{\mu}
\nonumber\\
&
-24{\mu}^2)x^7
-21(31+102{\mu}-12{\mu}^2+24{\mu}^3)x^6
+(353+3660{\mu}+2016{\mu}^2
\nonumber\\
&
+1584{\mu}^3-720{\mu}^4)x^5
-2{\mu}(1982+2799{\mu}-534{\mu}^2-1260{\mu}^3
+216{\mu}^4)x^4
\nonumber\\
&
+2{\mu}(883+2675{\mu}-3408{\mu}^2-636{\mu}^3+864{\mu}^4)x^3
-4{\mu}^2(681-1549{\mu}
\nonumber\\
&
+1026{\mu}^2+648{\mu}^3)x^2
+8{\mu}^2(106-139{\mu}+693{\mu}^2+216{\mu}^3)x
\nonumber\\
&
-16{\mu}^3(2+9{\mu})(13+3{\mu})\big).
\end{align}

\noindent\underline{$\chi=-5$} :
\begin{align}
\frac{W^{(3,1)}_0(x)}{dx}&=\frac{{\mu}(1+{\mu})(1-x)}{\sigma_{\mathrm{P}}(x)^{17/2}}\big(180x^8-16(89-2{\mu})x^7+8(465+38{\mu}+66{\mu}^2)x^6
\nonumber\\
&
-24(165
+211{\mu}+87{\mu}^2-30{\mu}^3)x^5+5(297+3018{\mu}-316{\mu}^2-540{\mu}^3+90{\mu}^4)x^4\nonumber\\
&
-20{\mu}(825
-601{\mu}-26{\mu}^2+90{\mu}^3)x^3+6{\mu}(1023-1975{\mu}+1240{\mu}^2+450{\mu}^3)x^2
\nonumber\\
&
+8{\mu}^2(154
-1155{\mu}-225{\mu}^2)x+2{\mu}^2(869+1630{\mu}+225{\mu}^2)\big),
\end{align}
\begin{align}
\frac{W^{(2,1)}_2(x)}{dx}&=-\frac{1-x}{\sigma_{\mathrm{P}}(x)^8}\big(-114x^8+(677-470{\mu})x^7-(1173+1350{\mu}+4640{\mu}^2)x^6
\nonumber\\
&
+2(309+7316{\mu}+4674{\mu}^2-6164{\mu}^3)x^5
-4{\mu}(6459-3567{\mu}-5802{\mu}^2
\nonumber\\
&
+4804{\mu}^3)x^4
+16{\mu}(819-1640{\mu}+1926{\mu}^2+2869{\mu}^3-774{\mu}^4)x^3
\nonumber\\
&
-16{\mu}^2(666+4324{\mu}+175{\mu}^2-2322{\mu}^3)x^2
+16{\mu}^2(1125+668{\mu}
\nonumber\\
&
-3453{\mu}^2-2322{\mu}^3)x+32{\mu}^3(525+980{\mu}+387{\mu}^2)\big)\nonumber\\
&
+\frac{1-x}{\sigma_{\mathrm{P}}(x)^{17/2}}\big(-114(1+228{\mu})x^9+(677+4833{\mu}+2781{\mu}^2)x^8
\nonumber\\
&
-(1173+29656{\mu}+20682{\mu}^2-1044{\mu}^3)x^7
+2(309+37335{\mu}
\nonumber\\
&
+27498{\mu}^2+1186{\mu}^3+2420{\mu}^4)x^6
-2{\mu}(40788+53619{\mu}+31484{\mu}^2
\nonumber\\
&
+6705{\mu}^3-3126{\mu}^4)x^5
+{\mu}(32043+202026{\mu}
+137448{\mu}^2-50071{\mu}^3
\nonumber\\
&
-21522{\mu}^4+3246{\mu}^5)x^4
-8{\mu}^2(27887+3387{\mu}-19833{\mu}^2+1000{\mu}^3
\nonumber\\
&
+1623{\mu}^4)x^3
+{\mu}^2(91299-112454{\mu}-84681{\mu}^2+90612{\mu}^3+19476{\mu}^4)x^2\nonumber\\
&
+4{\mu}^3(11205-17084{\mu}-25599{\mu}^2-3246{\mu}^3)x
+2{\mu}^3(8417+26497{\mu}
\nonumber\\
&
+17527{\mu}^2+1623{\mu}^3)\big),
\end{align}
\begin{align}
\frac{W^{(1,1)}_4(x)}{dx}&=-\frac{1-x}{2\sigma_{\mathrm{P}}(x)^8}\big(-1104x^8+16(463-178{\mu})x^7-32(452+142{\mu}+765{\mu}^2)x^6
\nonumber\\
&
+(8567+87562{\mu}+55188{\mu}^2-61128{\mu}^3)x^5
-2{\mu}(89521-50126{\mu}-70884{\mu}^2
\nonumber\\
&
+40968{\mu}^3)x^4
+8{\mu}(12661-29864{\mu}+14604{\mu}^2+26538{\mu}^3-5772{\mu}^4)x^3
\nonumber\\
&
+16{\mu}^2(1048-23244{\mu}-2271{\mu}^2+8658{\mu}^3)x^2
+16{\mu}^2(5850+5762{\mu}
\nonumber\\
&
-14781{\mu}^2-8658{\mu}^3)x
+32{\mu}^3(2570+4452{\mu}+1443{\mu}^2)\big)
\nonumber\\
&
+\frac{1-x}{2\sigma_{\mathrm{P}}(x)^{17/2}}\big(-1104(1+2{\mu})x^9
+4(1852+6315{\mu}+1347{\mu}^2)x^8
-64(226
\nonumber\\
&
+1815{\mu}+846{\mu}^2+115{\mu}^3)x^7
+(8567+281358{\mu}+200798{\mu}^2+16736{\mu}^3
\nonumber\\
&
-17392{\mu}^4)x^6
-4{\mu}(83345+91560{\mu}-4398{\mu}^2-11568{\mu}^3+5808{\mu}^4)x^5
\nonumber\\
&
+4{\mu}(36889+125103{\mu}-4627{\mu}^2+7194{\mu}^3+19383{\mu}^4-3219{\mu}^5)x^4
\nonumber\\
&
-8{\mu}^2(65871-4278{\mu}+18133{\mu}^2+5480{\mu}^3-6438{\mu}^4)x^3
+4{\mu}^2(60795
\nonumber\\
&
-32096{\mu}+32129{\mu}^2-25338{\mu}^3-19314{\mu}^4)x^2
+16{\mu}^3(3404-3997{\mu}+8766{\mu}^2
\nonumber\\
&
+3219{\mu}^3)x
+4{\mu}^3(7809+5711{\mu}-12341{\mu}^2-3219{\mu}^3)\big),
\end{align}
\begin{align}
\frac{W^{(0,1)}_6(x)}{dx}&=-\frac{1}{\sigma_{\mathrm{P}}(x)^8}\big(360x^9-12(259-22{\mu})x^8+24(367-4{\mu}+84{\mu}^2)x^7
-9(1125
\nonumber\\
&
+1678{\mu}+860{\mu}^2-472{\mu}^3)x^6
+(4081+54084{\mu}-14616{\mu}^2-15408{\mu}^3
\nonumber\\
&
+5040{\mu}^4)x^5
-2{\mu}(32755-36378{\mu}+6468{\mu}^2+9360{\mu}^3-1296{\mu}^4)x^4
\nonumber\\
&
+8{\mu}(3299-7928{\mu}+10764{\mu}^2+888{\mu}^3-1296{\mu}^4)x^3
-32{\mu}^2(249+2441{\mu}
\nonumber\\
&
-1269{\mu}^2-486{\mu}^3)x^2+16{\mu}^2(1186+74{\mu}-3303{\mu}^2
-648{\mu}^3)x+32{\mu}^3(466
\nonumber\\
&
+588{\mu}+81{\mu}^2)\big)
+\frac{1}{\sigma_{\mathrm{P}}(x)^{17/2}}\big(360(1+2{\mu})x^{10}-12(259+587{\mu}-13{\mu}^2)x^9
\nonumber\\
&
+4(2202+6677{\mu}+879{\mu}^2+820{\mu}^3)x^8-(10125+61802{\mu}+42466{\mu}^2\nonumber\\
&
+13040{\mu}^3-6600{\mu}^4)x^7+(4081+94968{\mu}+132406{\mu}^2-23436{\mu}^3-27210{\mu}^4
\nonumber\\
&
+7596{\mu}^5)x^6
-{\mu}(82143+188337{\mu}-137570{\mu}^2+12448{\mu}^3+34182{\mu}^4-3798{\mu}^5)x^5\nonumber\\
&
+{\mu}(28625+156083{\mu}-167614{\mu}^2+172280{\mu}^3+30126{\mu}^4-18990{\mu}^5)x^4
\nonumber\\
&
-2{\mu}^2(44095-22968{\mu}+123285{\mu}^2-34698{\mu}^3-18990{\mu}^4)x^3+4{\mu}^2(6708
\nonumber\\
&
+8722{\mu}+25015{\mu}^2-40266{\mu}^3-9495{\mu}^4)x^2-2{\mu}^3(9661-14885{\mu}-59385{\mu}^2
\nonumber\\
&
-9495{\mu}^3)x+2{\mu}^3(869-11241{\mu}-15321{\mu}^2-1899{\mu}^3)\big).
\end{align}

\newpage

\bibliographystyle{JHEP}
\bibliography{abmodel}

\end{document}